\documentclass{article}

\usepackage{amsmath}
\usepackage{amssymb}
\usepackage{authblk}
\usepackage{hyperref}
\usepackage{graphicx}
\usepackage{setspace}
\usepackage{tikz}
\usepackage{tikz-cd}
\usepackage{fullpage}

\newtheorem{definition}{Definition}

\title{A Functorial Perspective on (Multi)computational Irreducibility}
\author[1,2]{Jonathan Gorard}
\affil[1]{Cardiff University, Cardiff, UK\footnote{\url{gorardj@cardiff.ac.uk}}}
\affil[2]{University of Cambridge, Cambridge, UK\footnote{\url{jg865@cantab.ac.uk}}}

\begin{document}
\maketitle

\begin{abstract}
This article aims to provide a novel formalization of the concept of computational irreducibility in terms of the exactness of functorial correspondence between a category of data structures and elementary computations and a corresponding category of (1-dimensional) cobordisms. We proceed to demonstrate that, by equipping both categories with a symmetric monoidal structure and considering the case of higher-dimensional cobordism categories, we obtain a natural extension of this formalism that serves also to encompass non-deterministic or ``multiway'' computations, in which one quantifies not only the irreducibility in the behavior of a single (deterministic) computation path, but in the branching and merging behavior of an entire ``multiway system'' of such paths too. We finally outline how, in the most general case, the resulting symmetric monoidal functor may be considered to be adjoint to the functor characterizing the Atiyah-Segal axiomatization of a functorial quantum field theory. Thus, we conclude by arguing that the irreducibility of (multi)computations may be thought of as being dual to the locality of time evolution in functorial approaches to quantum mechanics and quantum field theory. In the process, we propose an extension of the methods of standard (monoidal) category theory, in which morphisms are effectively equipped with intrinsic computational complexity data, together with an algebra for how those complexities compose (both in sequence and in parallel, subject to the monoidal structure). Some possible extensions of this formalism (for instance to encompass notions of causality, space complexity, irreversibility, complexity of index manipulation in tensor networks, etc.), as well as potential implications for physics (for instance by providing an ability to distinguish formally between certain computational and multicomputational definitions of entropy) are briefly discussed.
\end{abstract}

\section{Introduction}

Computational irreducibility, as first proposed by Stephen Wolfram in the 1980s\cite{wolfram}, explored empirically within \textit{A New Kind of Science}\cite{wolfram2} (NKS) and subsequently refined and formalized by the author in later work\cite{gorard}, refers to the general phenomenon in which the outcome of any sufficiently sophisticated computational process cannot be predicted (or ``shortcut'') using any less computational effort than the system itself requires for its own explicit evolution. In this way, computational irreducibility may be considered to be an extension of the standard recursion-theoretic concepts of universality and undecidability\cite{turing}\cite{garey}\cite{hopcroft} (indeed, it is straightforward to prove by diagonalization that any Turing-complete computation must be irreducible, and undecidability may be thought of as corresponding to a limiting case of irreducibility in which certain computations become infinite in length). Within the NKS paradigm (in which all natural processes are conceived of as corresponding to computations), traditional methods of theoretical science thus correspond to those cases in which computations are reducible, and can therefore be preempted by means of sufficiently sophisticated tools of scientific and mathematical analysis. The abstract phenomenon of \textit{complexity} in natural systems may, in turn, be thought of as corresponding to those cases in which computations are fundamentally \textit{irreducible}, and therefore in which the only reasonable course of action available to the theoretical scientist is to simulate the behavior of the system explicitly\cite{zenil}\cite{zwirn}\cite{reisinger}.

The present article seeks to recast the author's previous formal definition of computational irreducibility within the broader conceptual framework of \textit{functoriality}\cite{maclane}. In Section \ref{sec:Section1}, by considering a category whose objects are data structures and whose morphisms are elementary computations (in the first instance, the article employs Turing machines as its chosen formal model of computation, whose corresponding category has as its objects configurations of the Turing machine's tape/head and head position, and as its morphisms applications of the Turing machine's partial transition function), we are able to reformulate the definition of reducibility/irreducibility in terms of the compositional properties of a certain map from this category to a category whose objects are time steps/step numbers and whose morphisms are discrete intervals between these step numbers. This has the effect of equipping the morphisms of our original category of Turing machine states and computations with intrinsic computational complexity data, along with a natural algebra (encoded by the action of the map on the composition operation) describing how the corresponding time complexities compose. Computational irreducibility then corresponds to the case where these complexities compose purely additively, and therefore in which this map is a pure functor; this allows us to say, in a rather precise sense, that irreducibility \textit{is} functoriality. Computational reducibility is then a measure of the extent to which this map distorts pure additive composition of time complexities, i.e. it is a measure of deviation away from pure functoriality. The codomain of this map (i.e. the category of step numbers and discrete intervals) may be interpreted as being a certain discretization of a (1-dimensional) category of cobordisms/``gluings of boundaries'' between (0-dimensional) manifolds\cite{milnor}\cite{adams}.

In Section \ref{sec:Section2}, we proceed to demonstrate that, when our category of data structures and elementary computations is also equipped with a symmetric tensor product structure (thereby promoting it to a \textit{symmetric monoidal category}\cite{maclane2}), which allows us to describe a \textit{multiway system} of many different branching and merging (singleway) computation paths, with the \textit{branchial graphs} of that multiway system describing the tensor product structure of the corresponding monoidal category, then there exists a very natural extension of the previous irreducibility formalism for ordinary categories and singleway computations to encompass the \textit{multicomputational} case too. Just as (singleway) computational reducibility may be formulated as a measure of distortion of the additivity of time complexity under ordinary (sequential) composition of computations, \textit{multicomputational reducibility} may therefore be formulated as a measure of distortion of the additivity of time complexity under parallel (tensor product) composition of computations, and hence, a multicomputationally irreducible computation is one under which the map from the (monoidal) category of data structures and elementary computations to the (monoidal) category consisting of parallel compositions of step numbers and parallel compositions of discrete intervals (effectively describing the coordinatization of branchial graphs), is a pure \textit{symmetric monoidal} functor. This idea formally encodes and concretizes the intuition that a singleway computation is irreducible if one must explicitly trace every intermediate step in the composition in order to determine the final result, whereas a multiway computation is irreducible if one must explicitly trace every singleway computation path in order to determine the system's overall branchial structure. The codomain of this map between symmetric monoidal categories is now interpretable as a certain discretization of a higher-dimensional category of cobordisms between higher-dimensional manifolds (equipped with a Riemannian structure). We discuss how ordinary computational irreducibility is a byproduct of the complexity of the \textit{state evolution function} used in the specification of a multiway system, whilst multicomputational irreducibility is a byproduct of the complexity of its \textit{state equivalence function}, and emphasize that the two concepts are therefore essentially orthogonal (e.g. one can easily build multicomputationally irreducible systems out of tensor products of many computationally reducible singleway paths). To illustrate this fact, we also analyze the case of multiway hypergraph rewriting systems, in which the state evolution function has comparable computational complexity to that of Turing machine evolution, but the state equivalence function (based on hypergraph isomorphism) is now far more non-trivial.

In Section \ref{sec:Section3}, we illustrate how, in a certain general sense, the functor from data structures and computations to step numbers and intervals that characterizes an irreducible computation may be considered to be the \textit{left adjoint} of the functor encoding the passage from moments of time and intervals to vector spaces and linear isomorphisms in the context of categorical quantum mechanics\cite{baez}. Thus, the same functoriality that characterizes the irreducibility of a computation may be thought of as being formally dual/adjoint to the functoriality that characterizes the locality of time evolution in non-relativistic quantum mechanics (in which any global unitary evolution over an interval may be decomposed into several local unitary evolutions over subintervals). In a fairly natural extension of this idea, we show how the symmetric monoidal functor encoding an irreducible multicomputation can be viewed as the left adjoint of the symmetric monoidal functor from manifolds and cobordisms to vector spaces and linear isomorphisms that defines a functorial quantum field theory, with the functoriality of multicomputational irreducibility now being dual/adjoint to the \textit{Atiyah-Segal sewing laws}\cite{atiyah}\cite{atiyah2}\cite{segal}, in which the path integral over any global interval may be decomposed into local path integrals over subintervals. Hence, we argue for the existence of a general adjunction relationship between computational complexity theory and categorical quantum mechanics, and between multicomputational complexity theory and functorial quantum field theory. We discuss how some of the other algebraic structure inherent to categorical quantum mechanics and functorial quantum field theory models (in particular, the involutive \textit{dagger} and \textit{compact} structures of the corresponding categories) might therefore have potential complexity-theoretic interpretations (for instance, in terms of the irreducibility of reversing computations, or of swapping arguments and values in a composition of multi-argument, multi-valued functions as described by a tensor network or monoidal string diagram, etc.), although a complete analysis of these interpretations lies beyond the scope of the present work.

Finally, in Section \ref{sec:Section4}, we outline some of the broader implications of the formalism developed herein, including the generalization of techniques of ordinary (monoidal) category theory to accommodate the case where morphisms designate explicit computations with associated computational complexity classes, and thus in which composition operations must respect the underlying algebra for how the complexities of those various computations interact. We also discuss the possibility of using these new algebraic techniques to conduct a more systematic investigation of computational complexity classes and (multi)computational complexity classes that are in some way ``wilder'' or less structured (in the sense that their algebra of composition is less well-behaved) than those studied within the setting of traditional computational complexity theory; we, moreover, indicate the various ways in which multicomputational complexity classes differ from traditional non-deterministic complexity classes (namely by considering the inherent computational complexity of the branching and merging operations of the multiway system, rather than solely considering the result yielded non-deterministically by a single multiway branch). We summarize some of the potential implications for physics, including a plausible disambiguation of two different computational definitions of entropy - one in which microstates are treated as being possible instantaneous states of a system (and thus based on computational irreducibility), and one in which microstates are treated as being possible paths of evolution history for the system (and thus based on multicomputational irreducibility) - that are often otherwise conflated. We propose some possible future directions of investigation, including the extension of the relevant maps/functors to also encompass preservation of dagger structure and/or compact structure, the extension of the overall formalism to encompass the additional information encoded within the structure of causal relations via certain higher categories/higher functors and the extension to multi-argument computations as encoded through \textit{glocal} multiway systems and their description as tensor networks/string diagrams.

Note that the majority (though by no means all) of the categories considered within this article are \textit{small} (and all of the ones which are not are sufficiently widely studied that their behavior is known to be at least somewhat well-behaved). For this reason, we shall henceforth neglect all considerations of set-theoretic issues, and shall use terms such as ``object set'' and ``hom set'' irrespective of whether the structures involved are sets or proper classes. Moreover, note that the code necessary to reproduce the results, proofs and figures from this article is open source and freely exposed through the \textit{Wolfram Function Repository}, for instance via \href{https://resources.wolframcloud.com/FunctionRepository/resources/MultiwayTuringMachine}{MultiwayTuringMachine}, \href{https://resources.wolframcloud.com/FunctionRepository/resources/TuringMachineGlocalMultiwaySystem/}{TuringMachineGlocalMultiwaySystem} and \href{https://resources.wolframcloud.com/FunctionRepository/resources/MultiwaySystem}{MultiwaySystem} for system evolution, \href{https://resources.wolframcloud.com/FunctionRepository/resources/AbstractCategory/}{AbstractCategory}, \href{https://resources.wolframcloud.com/FunctionRepository/resources/AbstractFunctor/}{AbstractFunctor} and \href{https://resources.wolframcloud.com/FunctionRepository/resources/AbstractStrictMonoidalCategory}{AbstractStrictMonoidalCategory} for representation of the underlying category-theoretic structures (and for automating the process of theorem-proving over them), etc.

\section{(Singleway) Irreducibility as Functoriality}
\label{sec:Section1}

We begin by adopting the formal definition of computational irreducibility proposed by the author in \cite{gorard}:

\begin{definition}
If ${f : \mathbb{N} \to \mathbb{N}}$ is a function on natural numbers and $T$ is a Turing machine that computes the value of ${f \left( i \right)}$ for some fixed input $i$ in $n$ steps, then $T$'s computation is \textit{reducible} if and only if there exists a Turing machine ${T^{*}}$ that computes ${f \left( i \right)}$ in $m$ steps, where ${m < n}$\cite{garey}\cite{hopcroft}.
\end{definition}
The ``degree'' of reducibility of the computation may thus be quantified in terms of the discrepancy, i.e. the value of ${n - m}$ (the converse value, i.e. ${m - n}$ for the case of irreducible computations in which ${m \geq n}$, is known as the \textit{slowdown} of ${T^{*}}$'s simulation of $T$). Note that we are here and henceforth assuming that all Turing machines are 1-tape Turing machines,\cite{zenil} which we can do without loss of generality since any $k$-tape Turing machine $M$ operating in time ${f \left( n \right)}$ may be simulated by a 1-tape Turing machine ${M^{\prime}}$ operating in time ${O \left( \left[ f \left( n \right) \right]^2 \right)}$, i.e. for any input $x$, ${M^{\prime} \left( x \right) = M \left( x \right)}$\cite{goldreich}\cite{papadimitriou}. We have also (implicitly) adopted the Hopcroft-Ullman formalization\cite{hopcroft} of a 1-tape Turing machine $T$ as a 7-tuple ${T = \left\langle Q, \Gamma, b, \Sigma, \delta, q_0, F \right\rangle}$, with finite alphabet set ${\Gamma \neq \varnothing}$, blank symbol ${b \in \Gamma}$, input symbols ${\Sigma \subseteq \Gamma \setminus \left\lbrace b \right\rbrace}$, finite state set ${Q \neq \varnothing}$, initial state ${q_0 \in Q}$, accepting states ${F \subseteq Q}$ and (partial) transition function:

\begin{equation}
\delta : \left( Q \setminus F \right) \times \Gamma \nrightarrow Q \times \Gamma \times \left\lbrace L, F \right\rbrace.
\end{equation}

\begin{figure}[ht]
\centering
\includegraphics[width=0.395\textwidth]{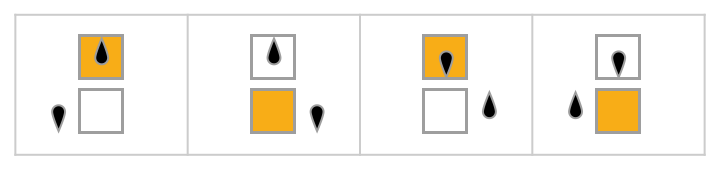}
\caption{A graphical representation of the 2-state, 2-color Turing machine rule number 2506, with the black icon representing the location and state of the head.}
\label{fig:Figure1}
\end{figure}

\begin{figure}[ht]
\centering
\includegraphics[width=0.245\textwidth]{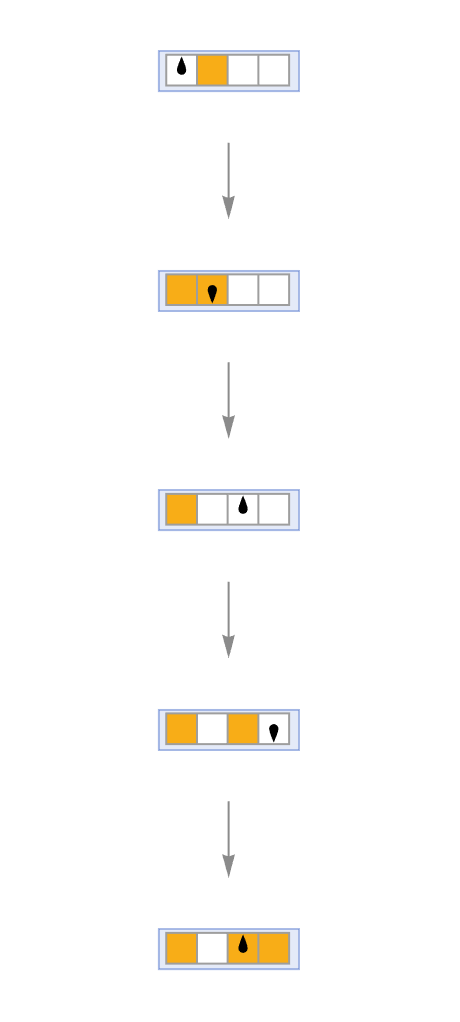}\hspace{0.1\textwidth}
\includegraphics[width=0.495\textwidth]{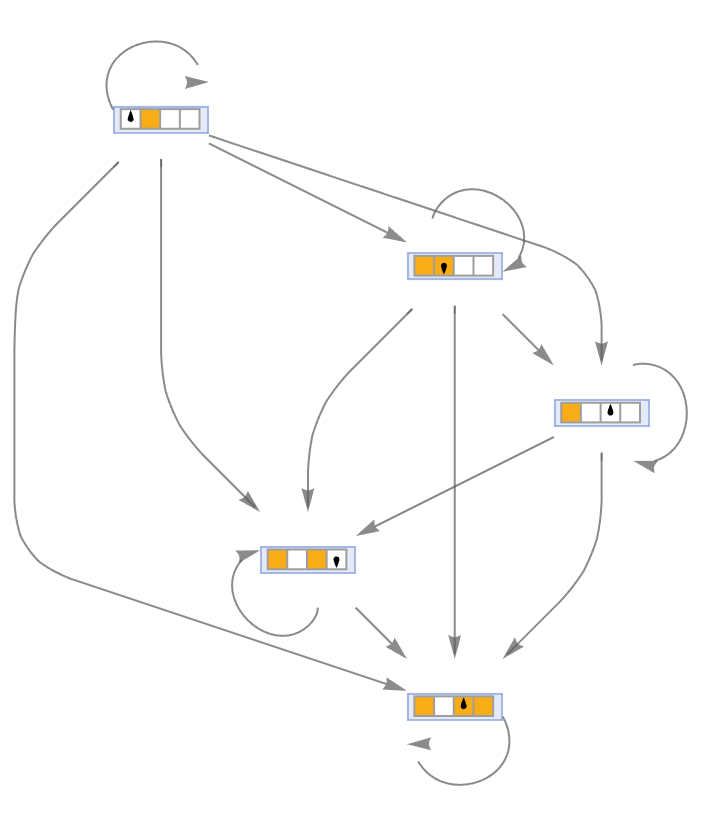}
\caption{On the left, a graph-theoretic representation of the evolution of the 2-state, 2-color Turing machine rule 2506, starting from the tape state ${\left\lbrace 0, 1, 0, 0 \right\rbrace}$, for 4 evolution steps; each arrow/edge of the quiver represents a single application of the Turing machine's transition function. On the right, a graph-theoretic representation of the category that is freely generated from this quiver.}
\label{fig:Figure2}
\end{figure}

We are now in a position to be able to construct a \textit{category}\cite{maclane} representing a particular formal (abstract) model of computation, whose objects are data structures and whose morphisms are elementary/primitive computations. For the specific case considered here of the category generated by the Turing machine $T$, which we shall denote ${\mathcal{T}}$, we choose as its \textit{object set} ${\mathrm{ob} \left( \mathcal{T} \right)}$ the set ${\Gamma^{\aleph_0} \times Q \times \mathbb{N}}$ of possible ordered triples consisting of tape state, head state and head position (assuming an infinite-length tape), and we wish for its \textit{morphism set} ${\mathrm{hom} \left( \mathcal{T} \right)}$ to consist of all possible valid transitions of the Turing machine $T$. In order to construct this morphism set, we therefore begin by constructing a \textit{quiver} (i.e. a directed multigraph) whose arrows/edges correspond to applications of the (partial) transition function ${\delta}$; for our present purposes, it suffices to think of the set of all possible such arrows as being ${N \times \left( \Gamma^{\aleph_0} \times Q \times \mathbb{N} \right)^2}$, i.e. each arrow is treated as an ordered triple ${\left( f, X, Y \right)}$, denoted ${f : X \to Y}$, consisting of a name ${f \in N}$, and a pair of elements ${X, Y \in \mathrm{ob} \left( \mathcal{T} \right)}$ ($X$ being the arrow's source/domain and $Y$ being its target/codomain). This quiver freely generates a category if we populate the morphism set ${\mathrm{hom} \left( \mathcal{T} \right)}$ initially with the set of arrows/transitions, and proceed to introduce a binary composition operator ${\circ}$ such that:

\begin{equation}
\forall \left( f : X \to Y \right), \left( g : Y \to Z \right) \in \mathrm{hom} \left( \mathcal{T} \right), \qquad \left( g \circ f : X \to Z \right) \in \mathrm{hom} \left( \mathcal{T} \right),
\end{equation}
i.e. any pair of morphisms with matching codomain and domain can be composed by means of the operator ${\circ}$. If $X$, $Y$ and $Z$ represent possible Turing machine states (including tape state, head state and head position), and $f$ and $g$ represent possible Turing machine transitions, this procedure of generating a category from the underlying quiver can be illustrated diagrammatically as follows:

\begin{equation}
\begin{tikzcd}
& Y \arrow[swap, dr, "g"'{name=g}]  & & & & & & Y \arrow[dr, "g"] &\\
X \arrow[ur, "f"] & & Z & & & & X \arrow[swap, ur, "f"'{name=f}] \arrow[swap, rr, "g \circ f"] & & Z.
\arrow[mapsto, from=g, to=f, shorten=6em]
\end{tikzcd}
\end{equation}
The resulting algebraic structure is indeed a category, since the operator ${\circ}$ inherits associativity:

\begin{multline}
\forall \left( f : X \to Y \right), \left( g : Y \to Z \right), \left( h : Z \to W \right) \in \mathrm{hom} \left( \mathcal{T} \right),\\
\left( \left( h \circ g \right) \circ f : X \to W \right) = \left( h \circ \left( g \circ f \right) : X \to W \right),
\end{multline}
from the fact that (partial) function composition is associative, and the identity axiom, namely:

\begin{equation}
\forall X \in \mathrm{ob} \left( \mathcal{T} \right), \qquad \exists \left( id_X : X \to X \right) \in \mathrm{hom} \left( \mathcal{T} \right),
\end{equation}
such that:

\begin{equation}
\forall \left( f : X \to Y \right) \in \mathrm{hom} \left( \mathcal{T} \right), \qquad \left( f \circ id_X : X \to Y \right) = \left( f : X \to Y \right),
\end{equation}
and:

\begin{equation}
\forall \left( g : Y \to X \right) \in \mathrm{hom} \left( \mathcal{T} \right), \qquad \left( id_X \circ g : Y \to X \right) = \left( g : Y \to X \right),
\end{equation}
holds by virtue of the fact that the (partial) transition function may be augmented by a neutral (``no shift'') operation ${id}$ as follows:

\begin{equation}
\delta : \left( Q \setminus F \right) \times \Gamma \nrightarrow Q \times \Gamma \times \left\lbrace L, R, id \right\rbrace.
\end{equation}
Note that, since the Turing machines considered here are all deterministic/classical, meaning that the (partial) transition function is always single-valued, it follows that the resulting quiver must take the form of either a path graph, a cycle graph or a union of path and cycle graphs. An explicit example of this construction, for the 2-state, 2-color Turing machine shown in Figure \ref{fig:Figure1} (known  as rule number 2506 in the canonical Turing machine enumeration) is shown in Figure \ref{fig:Figure2}.

\begin{figure}[ht]
\centering
\includegraphics[width=0.495\textwidth]{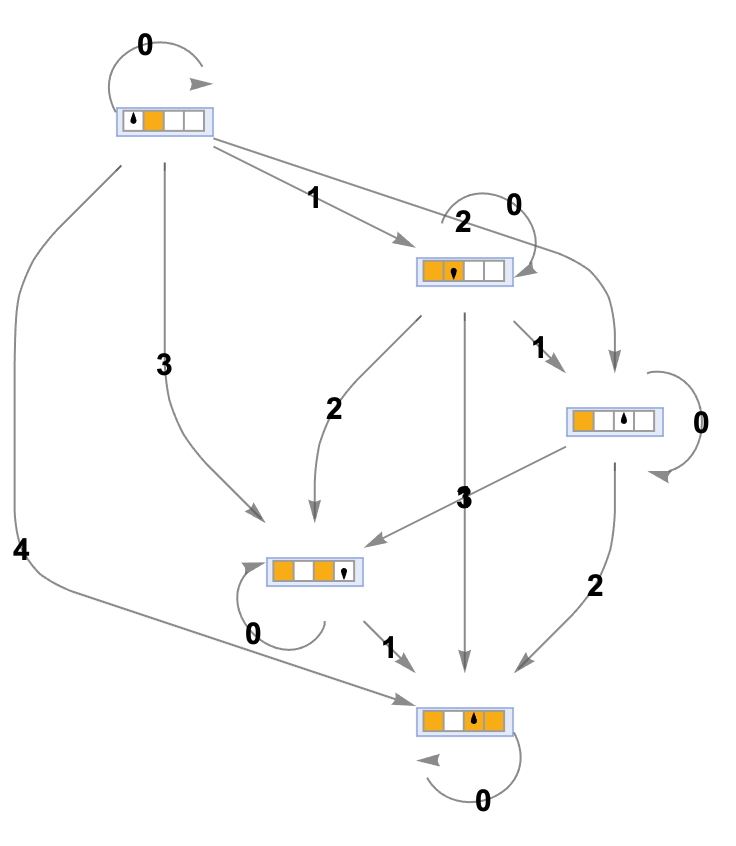}
\caption{A graph-theoretic representation of the category that is freely generated from the quiver representing the evolution of the 2-state, 2-color Turing machine rule 2506, with edges/morphisms tagged with additional metadata corresponding to the number of ``steps'' (i.e. transitions) required to perform the requisite computation.}
\label{fig:Figure3}
\end{figure}

The process of obtaining the full category ${\mathcal{T}}$ representing the Turing machine $T$ may consequently be thought of graph-theoretically as taking a reflexive transitive closure of the underlying transition quiver. However, this operation of taking transitive closures appears, at least conceptually, to be very much against the spirit of computational irreducibility - it seems intuitively to imply that whenever there exists a computation from $X$ to $Y$, and another computation from $Y$ to $Z$, then one can necessarily always ``jump ahead'' to get directly from $X$ to $Z$ with the same amount of computational effort. Ideally, we would like to consider some form of ``decorated'' category in which morphisms are tagged with some additional metadata corresponding to the computational complexity of the underlying computation that the morphism signifies; we could then proceed to define an algebra to describe formally how these complexities behave under the action of the composition operator ${\circ}$. Under such an algebra, irreducible computations would correspond to those computations whose complexities behave purely additively under composition, i.e. if the computation underlying morphism ${f : X \to Y}$ requires at least $n$ steps to execute, and the computation underlying morphism ${g : Y \to Z}$ requires at least $m$ steps to execute, then the composite computation ${g \circ f : X \to Z}$ is irreducible (under the definition given above) if and only if it cannot be executed in fewer than ${n + m}$ steps. Conversely, if the complexities of the computations behave strictly \textit{subadditively} under composition, then the composite computation is reducible (with the ``degree'' of subadditivity corresponding to the ``degree'' of reducibility). This intuition is illustrated in Figure \ref{fig:Figure3} for the case of the 2-state, 2-color Turing machine rule considered above, with each edge/morphism tagged with the number of transitions necessary to perform the corresponding computation; the composition operation here is purely additive, indicating that all computations shown are irreducible.

\begin{figure}[ht]
\centering
\includegraphics[width=0.495\textwidth]{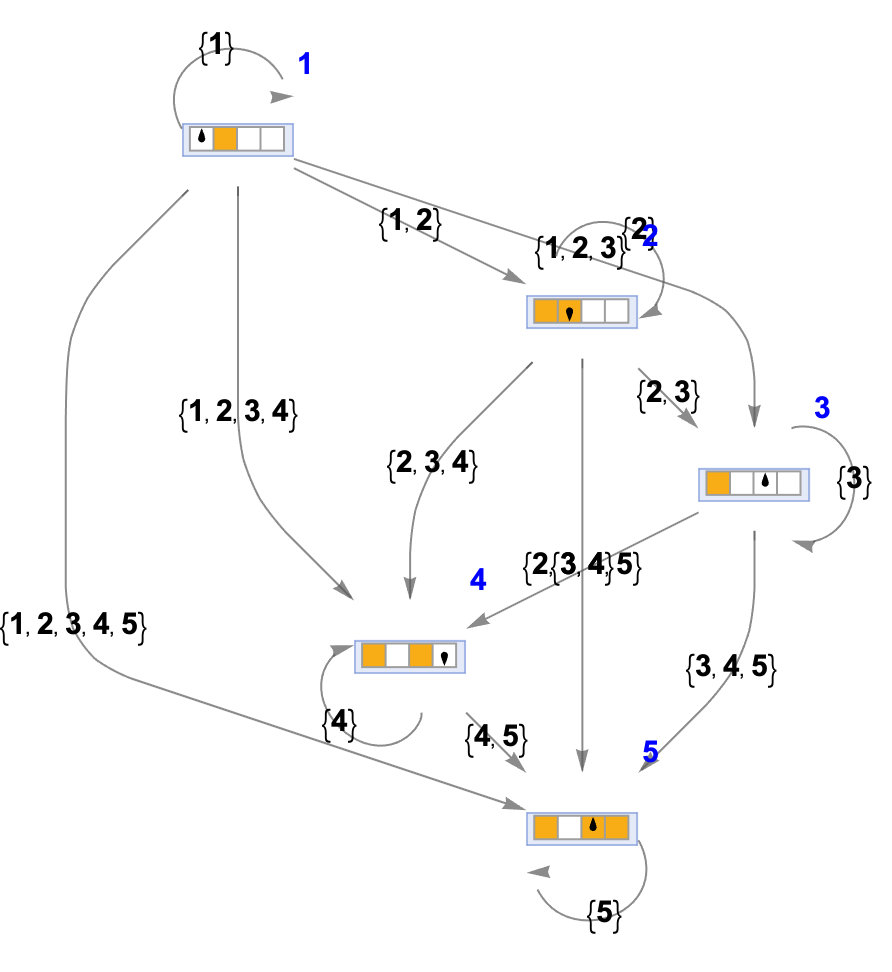}
\caption{A graph-theoretic representation of the category that is freely generated from the quiver representing the evolution of the 2-state, 2-color Turing machine rule 2506, with vertices/objects tagged with additional metadata corresponding to the step number on which they occur (shown in blue) and with edges/morphisms tagged with additional metadata corresponding to all intermediate step numbers traversed as part of the requisite computation (shown in black).}
\label{fig:Figure4}
\end{figure}

In order to formalize this intuition, let us proceed on the assumption that all computations are fully irreducible. We can now consider a function ${Z^{\prime}}$ mapping from our category of data structures and computations (for the case of Turing machines, this is the category ${\mathcal{T}}$ defined above) to a category ${\mathcal{B}}$ whose objects are natural numbers (i.e. ${\mathrm{ob} \left( \mathcal{B} \right) = \mathbb{N}}$) corresponding to step numbers/moments of time, and whose morphisms are discrete intervals between these step numbers/moments of time, i.e:

\begin{equation}
\mathrm{hom} \left( \mathcal{B} \right) = \left\lbrace \left[ n, m \right] \cap \mathbb{N} \left\vert n, m \in \mathbb{N} \text{ and } n \leq m \right. \right\rbrace.
\end{equation}
If we now equip ${\mathcal{B}}$ with a composition operation given by the union of discrete (contiguous) intervals ${\cup}$, then it trivially forms a category (with ${\left[ n, n \right] \cap \mathbb{N} = \left\lbrace n \right\rbrace}$ being the identity morphism for any ${n \in \mathbb{N}}$), and, moreover, since all computations are irreducible (and therefore computational complexities are purely additive under composition) by hypothesis, the function ${Z^{\prime} : \mathcal{T} \to \mathcal{B}}$ trivially forms a \textit{functor}\cite{jacobson}. More precisely, ${Z^{\prime}}$ is a map between categories ${\mathcal{T}}$ and ${\mathcal{B}}$, i.e:

\begin{equation}
\forall X \in \mathrm{ob} \left( \mathcal{T} \right), \qquad \exists Z^{\prime} \left( X \right) \in \mathrm{ob} \left( \mathcal{B} \right),
\end{equation}
and:

\begin{multline}
\forall \left( f : X \to Y \right) \in \mathrm{hom} \left( \mathcal{T} \right), \qquad \exists \left( Z^{\prime} \left( f \right) : Z^{\prime} \left( X \right) \to Z^{\prime} \left( Y \right) \right) \in \mathrm{hom} \left( \mathcal{B} \right),\\
\text{ with } \left( Z^{\prime} \left( f \right) : Z^{\prime} \left( X \right) \to Z^{\prime} \left( Y \right) \right) = \left[ Z^{\prime} \left( X \right), Z^{\prime} \left( Y \right) \right] \cap \mathbb{N},
\end{multline}
such that the structure of composition is preserved:

\begin{multline}
\forall \left( f : X \to Y \right), \left( g : Y \to Z \right) \in \mathrm{hom} \left( \mathcal{T} \right),\\
\left( Z^{\prime} \left( g \circ f \right) : Z^{\prime} \left( X \right) \to Z^{\prime} \left( Z \right) \right) = \left( Z^{\prime} \left( g \right) \cup Z^{\prime} \left( f \right) : Z^{\prime} \left( X \right) \to Z^{\prime} \left( Z \right) \right) = \left[ Z^{\prime} \left( X \right), Z^{\prime} \left( Z \right) \right] \cap \mathbb{N},
\end{multline}
and all identity morphisms are preserved:

\begin{multline}
\forall X \in \mathrm{ob} \left( \mathcal{T} \right),\\
\left( Z^{\prime} \left( id_X \right) : Z^{\prime} \left( X \right) \to Z^{\prime} \left( X \right) \right) = \left( id_{Z^{\prime} \left( X \right)} : Z^{\prime} \left( X \right) \to Z^{\prime} \left( X \right) \right) = \left[ Z^{\prime} \left( X \right), Z^{\prime} \left( X \right) \right] \cap \mathbb{N} = \left\lbrace Z^{\prime} \left( X \right) \right\rbrace.
\end{multline}
If $X$, $Y$ and $Z$ represent Turing machine states reached at step numbers ${t_1}$, ${t_2}$ and ${t_3}$ respectively, and ${f : X \to Y}$ and ${g : Y \to Z}$ represent the transitions between states $X$ and $Y$ and states $Y$ and $Z$ respectively, then this functoriality between categories ${\mathcal{T}}$ and ${\mathcal{B}}$ can be illustrated diagrammatically as follows:

\begin{equation}
\begin{tikzcd}
& & Y \arrow[swap, ddrr, "g"'{name=g}] & & & & & & & t_2 \arrow[ddrr, "{\left[ t_2, t_3 \right] \cap \mathbb{N}}"] & &\\\\
X \arrow[uurr, "f"] \arrow[swap, rrrr, "g \circ f"] & & & & Z & & & t_1 \arrow[swap, uurr, "{\left[ t_1, t_2 \right] \cap \mathbb{N}}"'{name=f}] \arrow[swap, rrrr, "\substack{\left( \left[ t_2, t_3 \right] \cup \left[ t_1, t_2 \right] \right) \cap \mathbb{N} \\ = \left[ t_1, t_3 \right] \cap \mathbb{N}}"] & & & & t_3.
\arrow[mapsto, from=g, to=f, shorten=4em, "Z^{\prime}"]
\end{tikzcd}
\end{equation}
In this way, the cardinality of any morphism in ${\mathcal{B}}$ (e.g. ${\left\lvert \left[ t_1, t_2 \right] \cap \mathbb{N} \right\rvert}$) represents, up to an additive constant, the time complexity of the corresponding computation in ${\mathcal{T}}$ (e.g. ${f : X \to Y}$). Since this functorial relationship holds only in the case where all computations in ${\mathcal{T}}$ are irreducible, we are consequently able to conclude that, in some precise sense, computational irreducibility \textit{is} functoriality, and, moreover, that computational reducibility corresponds precisely to a deformation of the map ${Z^{\prime}}$ away from being a pure functor. This construction is demonstrated in Figure \ref{fig:Figure4} for the 2-state, 2-color Turing machine described previously, with each vertex/object tagged with its step number and each edge/morphism tagged with a list of step numbers traversed throughout the course of its corresponding computation.

Note that the axioms obeyed by this ``algebra of complexity'' are formally almost identical to those of a metric space: the complexity of applying the identity transition ${id}$ mapping from a state to itself is always 0 (or possibly 1, depending upon precisely how the intervals are defined), the complexity of applying any transition between two distinct states is always strictly positive, and the complexities always obey the triangle inequality (i.e. non-strict subadditivity) under composition; the only metric space axiom for which there exists no immediate analog is the symmetry axiom, since transitions need not necessarily be reversible, and the complexity of reversing a transition (when such a reversal exists) need not necessarily be equal to the complexity of performing the transition. Note, moreover, that, if we modify the definition of the category ${\mathcal{B}}$ such that its objects are not merely natural numbers but real ones (i.e. ${\mathrm{ob} \left( \mathcal{B} \right) = \mathbb{R}}$), and its morphisms are not merely discrete intervals but continuous ones, i.e:

\begin{equation}
\mathrm{hom} \left( \mathcal{B} \right) = \left\lbrace \left[ x, y \right] \left\lvert x, y \in \mathbb{R} \text{ and } x \leq y \right. \right\rbrace,
\end{equation}
with the composition operation still being the standard union of contiguous intervals ${\cup}$, then nothing in the above argument is substantively modified: the category ${\mathcal{T}}$ is still in functorial correspondence with the new version of category ${\mathcal{B}}$ if and only if it was in functorial correspondence with the old category ${\mathcal{B}}$. The only material difference that this modification makes is that it guarantees that the functor/map ${Z^{\prime}}$ (functor in the irreducible case, map in the reducible one) cannot be surjective on objects, but there was no requirement for it to be so in the first place. However, this slightly generalized definition of the category ${\mathcal{B}}$ does carry the definite advantage of making the underlying topological intuition of this construction manifest; the objects of ${\mathbb{R}}$ (real numbers, representing moments of time) can be thought of as being 0-dimensional manifolds, and its morphisms can be thought of as being 1-dimensional cobordisms between those manifolds. Thus, ${\mathcal{B}}$ is really a \textit{cobordism category}\cite{stong}.

Here and henceforth, in relation to category ${\mathcal{B}}$ and its descendants, we employ the formal definition of a cobordism category as an ordered triple ${\left\langle \mathcal{B}, \partial, i \right\rangle}$, where ${\mathcal{B}}$ has all finite \textit{coproducts} and is equipped with an \textit{initial object} ${\varnothing}$, ${\partial : \mathcal{B} \to \mathcal{B}}$ is an \textit{additive endofunctor} that preserves all coproducts and ${i : \partial \Rightarrow \mathrm{Id}_{\mathcal{B}}}$ (with ${\mathrm{Id}_{\mathcal{B}}}$ denoting the identity functor on ${\mathcal{B}}$ that sends every object/morphism to itself) is a \textit{natural transformation} satisfying:

\begin{equation}
\forall M \in \mathrm{ob} \left( \mathcal{B} \right), \qquad \partial \left( \partial \left( M \right) \right) = \varnothing.
\end{equation}
In the above, the coproduct of a pair of objects ${M_1, M_2 \in \mathrm{ob} \left( \mathcal{B} \right)}$ is taken to be an object ${M_1 \oplus M_2 \in \mathrm{ob} \left( \mathcal{B} \right)}$ equipped with a pair of \textit{canonical injection} morphisms:

\begin{equation}
\left( i_1 : M_1 \to M_1 \oplus M_2 \right), \left( i_2 : M_2 \to M_1 \oplus M_2 \right) \in \mathrm{hom} \left( \mathcal{B} \right),
\end{equation}
that are \textit{universal}, in the sense that, for any object ${M^{*} \in \mathrm{ob} \left( \mathcal{B} \right)}$ equipped with morphisms:

\begin{equation}
\left( i_{1}^{*} : M_1 \to M^{*} \right), \left( i_{2}^{*} : M_2 \to M^{*} \right) \in \mathrm{hom} \left( \mathcal{B} \right),
\end{equation}
there exists a unique morphism ${\left( u : M_1 \oplus M_2 \to M^{*} \right) \in \mathrm{hom} \left( \mathcal{B} \right)}$ such that:

\begin{equation}
\left( i_{1}^{*} : M_1 \to M^{*} \right) = \left( u \circ i_1 : M_1 \to M^{*} \right), \qquad \text{ and } \qquad \left( i_{2}^{*} : M_2 \to M^{*} \right) = \left( u \circ i_2 : M_2 \to M^{*} \right).
\end{equation}
This condition may be restated concisely by means of the following commutative diagram:

\begin{equation}
\begin{tikzcd}
& \forall M^{*} &\\
M_1 \arrow[ur, "\forall i_{1}^{*}"] \arrow[swap, r, "i_1"] & M_1 \oplus M_2 \arrow[u, dashed, "\exists ! u"] & M_2 \arrow[swap, ul, "\forall i_{2}^{*}"] \arrow[l, "i_2"].
\end{tikzcd}
\end{equation}
This definition can be extended in the obvious way to any finite collection of objects ${M_j \in \mathrm{ob} \left( \mathcal{B} \right)}$ to yield a finite coproduct ${\bigoplus\limits_{j} M_j \in \mathrm{ob} \left( \mathcal{B} \right)}$ equipped with a finite collection of (universal) injection morphisms:

\begin{equation}
\left( i_j : M_j \to \bigoplus\limits_{k} M_k \right) \in \mathrm{hom} \left( \mathcal{B} \right).
\end{equation}
An initial object ${\varnothing \in \mathrm{ob} \left( \mathcal{B} \right)}$ is a distinguished object such that, for any object ${M \in \mathrm{ob} \left( \mathcal{B} \right)}$, there exists a unique morphism ${\left( u : \varnothing \to M \right) \in \mathrm{hom} \left( \mathcal{B} \right)}$, i.e:

\begin{equation}
\begin{tikzcd}
\varnothing \arrow[r, dashed, "\exists ! u"] & \forall M.
\end{tikzcd}
\end{equation}
An endofunctor is any functor from a category to itself, and an additive functor is any functor that preserves finite coproducts (or, in certain contexts, finite \textit{biproducts}, though this is not the case considered here); in other words, zero objects are preserved (up to isomorphism):

\begin{equation}
\partial \left( \varnothing \right) \cong \varnothing \in \mathrm{ob} \left( \mathcal{B} \right),
\end{equation}
and, for any pair of objects ${M_1, M_2 \in \mathrm{ob} \left( \mathcal{B} \right)}$, there exists an isomorphism:

\begin{equation}
\partial \left( M_1 \oplus M_2 \right) \cong \partial \left( M_1 \right) \oplus \partial \left( M_2 \right),
\end{equation}
that preserves the canonical injection morphisms of the coproduct construction:

\begin{equation}
\begin{tikzcd}
& M_1 \oplus M_2 & & & & \partial \left( M_1 \oplus M_2 \right) \cong \partial \left( M_1 \right) \oplus \partial \left( M_2 \right) & \\
M_1 \arrow[ur, "i_1"] & & M_2 \arrow[ul, "i_2"'{name=g}] & & \partial \left( M_1 \right) \arrow[swap, ur, "\partial \left( i_1 \right)"'{name=f}] & & \partial \left( M_2 \right) \arrow[swap, ul, "\partial \left( i_2 \right)"].
\arrow[mapsto, from=g, to=f, shorten=4em, "\partial"]
\end{tikzcd}
\end{equation}
An isomorphism (indicated by ${\cong}$) here refers to any morphism ${\left( f : X \to Y \right) \in \mathrm{hom} \left( \mathcal{B} \right)}$ for which there exists a corresponding morphism ${\left( f^{-1} : Y \to X \right) \in \mathrm{hom} \left( \mathcal{B} \right)}$ that acts as both a left and right inverse of $f$, i.e:

\begin{equation}
\left( f^{-1} \circ f : X \to X \right) = \left( id_X : X \to X \right), \qquad \text{ and } \qquad \left( f \circ f^{-1} : Y \to Y \right) = \left( id_Y : Y \to Y \right).
\end{equation}
Finally, a natural transformation ${\eta : F \Rightarrow G}$ between functors ${F : \mathcal{C} \to \mathcal{D}}$ and ${G : \mathcal{C} \to \mathcal{D}}$ is characterized by a family of \textit{component} morphisms:

\begin{equation}
\forall X \in \mathrm{ob} \left( \mathcal{C} \right), \qquad \exists \left( \eta_X : F \left( X \right) \to G \left( X \right) \right) \in \mathrm{hom} \left( \mathcal{D} \right),
\end{equation}
such that one has:

\begin{equation}
\forall \left( f : X \to Y \right) \in \mathrm{hom} \left( \mathcal{C} \right), \qquad \left( \eta_Y \circ F \left( f \right) : F \left( X \right) \to G \left( Y \right) \right) = \left( G \left( f \right) \circ \eta_X : F \left( X \right) \to G \left( Y \right) \right),
\end{equation}
or, represented in the form of a commutative diagram:

\begin{equation}
\begin{tikzcd}
F \left( X \right) \arrow[r, "F \left( f \right)"] \arrow[swap, d, "\eta_X"] & F \left( Y \right) \arrow[d, "\eta_Y"]\\
G \left( X \right) \arrow[swap, r, "G \left( f \right)"] & G \left( Y \right).
\end{tikzcd}
\end{equation}

The intuition lying behind this formalization of cobordism categories can be articulated as follows. Every object ${M \in \mathrm{ob} \left( \mathcal{B} \right)}$ is interpreted as a (generalized) manifold; every morphism ${\left( f : M_1 \to M_2 \right) \in \mathrm{hom} \left( \mathcal{B} \right)}$ is interpreted as a (generalized) cobordism between those manifolds, i.e. a ``gluing'' together of those manifolds along their boundaries. The initial object ${\varnothing}$ plays the role of the empty set/empty manifold, and the additive endofunctor ${\partial}$ represents the boundary relation: ${\partial \left( M \right)}$ (for some ${M \in \mathrm{ob} \left( \mathcal{B} \right)}$) corresponds to the (generalized) boundary of manifold $M$, with the condition that ${\partial \left( \partial \left( M \right) \right) = \varnothing}$ thus designating that the boundary of a boundary is always empty. Coproducts ${\oplus}$ in the cobordism category ${\mathcal{B}}$ play the role of the direct sum/disjoint union of manifolds. A pair of objects/manifolds ${M_1, M_2 \in \mathrm{ob} \left( \mathcal{B} \right)}$ may be said to be \textit{cobordant} (in the sense of their disjoint union forming the boundary of a manifold in one dimension higher), written ${M_1 \sim M_2}$, if and only if:

\begin{equation}
\exists V_1, V_2 \in \mathrm{ob} \left( \mathcal{B} \right), \qquad \text{ such that } \qquad M_1 \oplus \partial \left( V_1 \right) \cong M_2 \oplus \partial \left( V_2 \right),
\end{equation}
where ${\cong}$ is an isomorphism in ${\mathcal{B}}$; the cobordism relation ${\sim}$ is hence an equivalence relation in which all isomorphic objects/manifolds are cobordant:

\begin{equation}
\forall M_1, M_2 \in \mathrm{ob} \left( \mathcal{B} \right), \qquad M_1 \cong M_2 \implies M_1 \sim M_2,
\end{equation}
and where:

\begin{equation}
\forall M \in \mathrm{ob} \left( \mathcal{B} \right), \qquad \partial \left( M \right) \sim \varnothing.
\end{equation}
This connection to cobordism categories may appear to be a rather trivial technical point, but it will proceed to play a crucial role in the forthcoming discussion on formal correspondences with categorical quantum mechanics and functorial quantum field theory\cite{baez}, and the relationship between computational irreducibility and the locality of time evolution.

\section{Multicomputational Irreducibility as Monoidal Functoriality}
\label{sec:Section2}

\begin{figure}[ht]
\centering
\includegraphics[width=0.395\textwidth]{ComputationalIrreducibility3}\hspace{0.1\textwidth}
\includegraphics[width=0.395\textwidth]{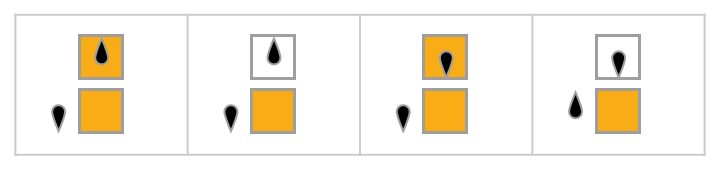}
\caption{A graphical representation of two 2-state, 2-color Turing machine rules (rule numbers 2506 and 3506), with the black icons representing the locations and states of the heads.}
\label{fig:Figure5}
\end{figure}

\begin{figure}[ht]
\centering
\includegraphics[width=0.445\textwidth]{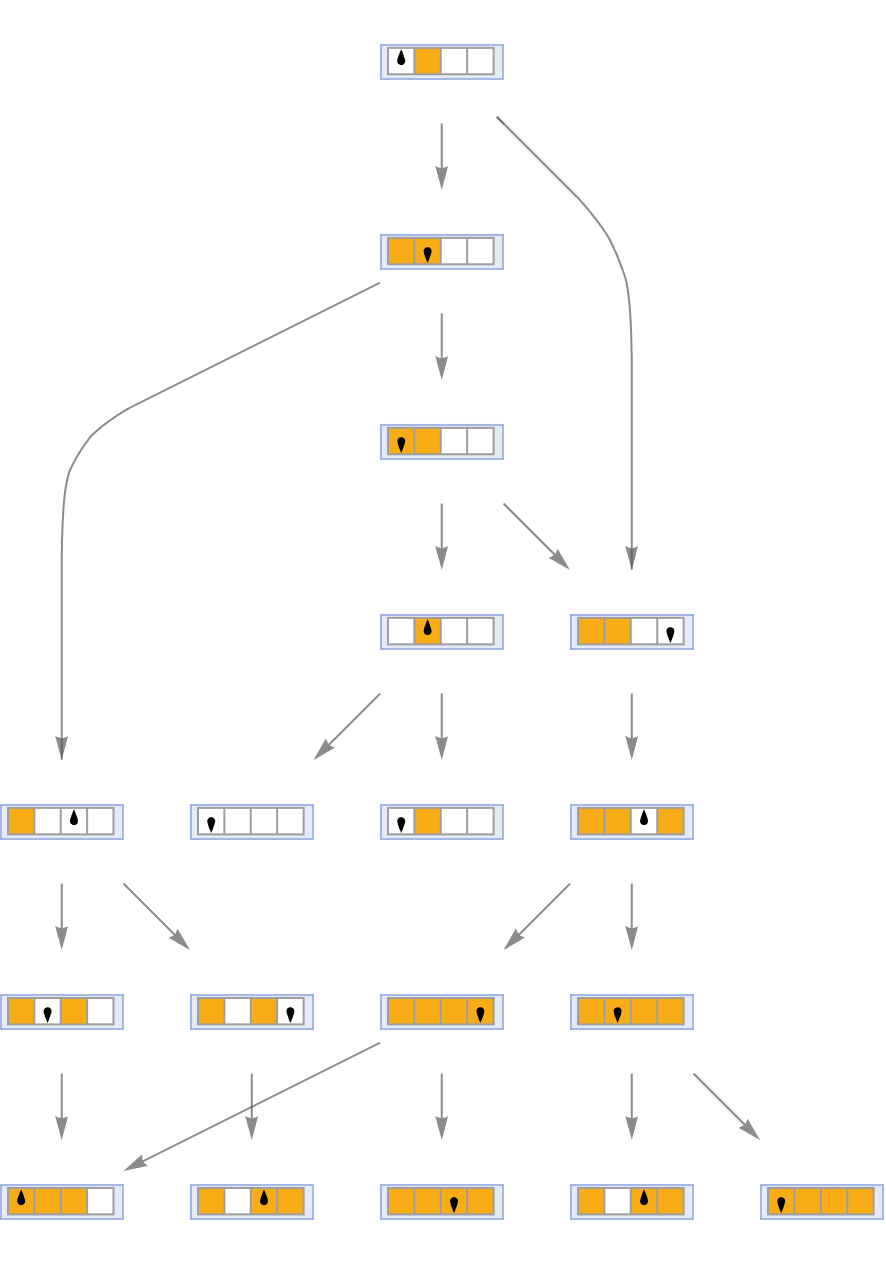}
\caption{A multiway evolution graph corresponding to the non-deterministic evolution of a 2-state, 2-color Turing machine constructed from the (parallel) composition of the Turing machine transition functions for rules 2506 and 3506, starting from the single tape state ${\left\lbrace 0, 1, 0, 0 \right\rbrace}$, for 4 steps; each edge represents a single application of one of the two Turing machine transition functions.}
\label{fig:Figure6}
\end{figure}

\begin{figure}[ht]
\centering
\includegraphics[width=0.795\textwidth]{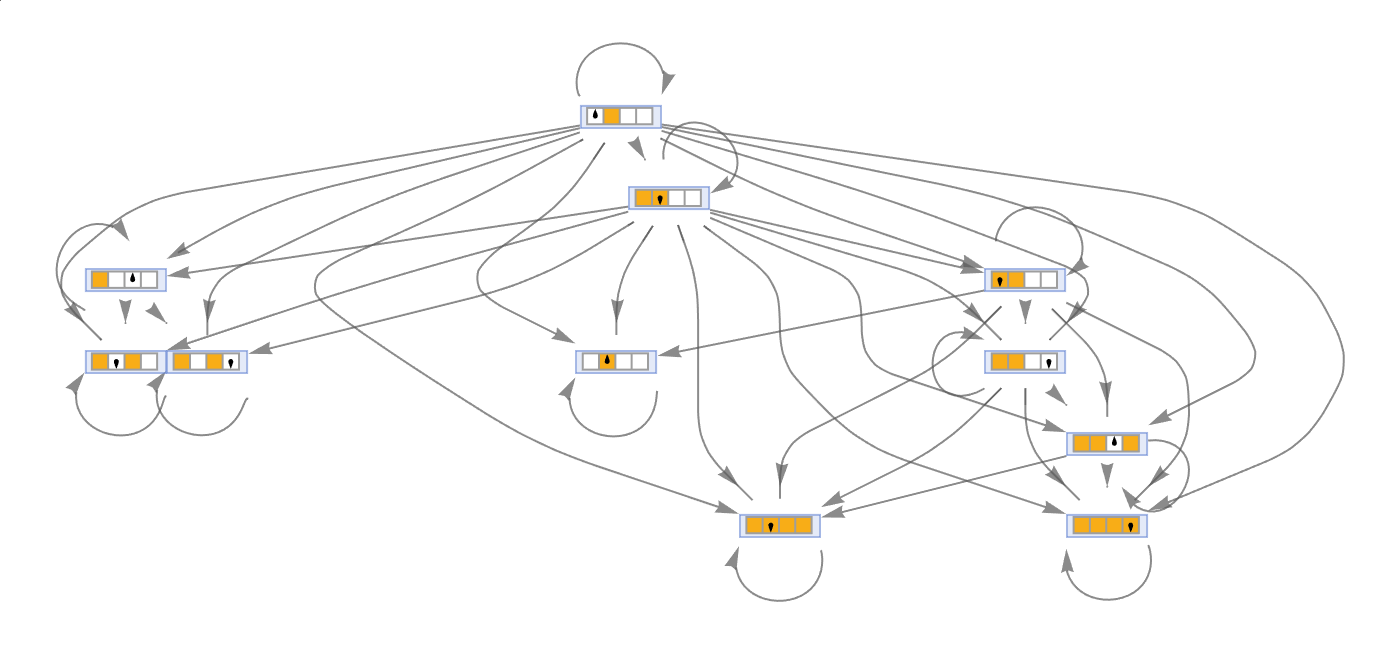}
\caption{A graph-theoretic representation of the category that is freely generated by the multiway evolution graph corresponding to the non-deterministic evolution of a 2-state, 2-color Turing machine constructed from the (parallel) composition of the Turing machine transition functions for rules 2506 and 3506 (considered as a quiver), starting from the single tape state ${\left\lbrace 0, 1, 0, 0 \right\rbrace}$, for 3 steps.}
\label{fig:Figure7}
\end{figure}

We now proceed to consider the case of non-deterministic (multiway) computations, in which the (partial) transition function acting on data structures/computational states is not necessarily single-valued. Such computations may be parameterized by means of a \textit{multiway system}\cite{wolfram3}, or, more precisely, a \textit{multiway evolution graph}\cite{gorard2}\cite{gorard3}, namely a directed acyclic graph whose vertices represent computational states and whose directed edges represent single-step transitions between those states. An explicit example of the multiway system construction, for a pair of 2-state, 2-color Turing machine rules shown in Figure \ref{fig:Figure5} (rule numbers 2506 and 3506 in the canonical Turing machine enumeration) is shown in Figure \ref{fig:Figure6}.  By considering the resulting multiway evolution graph as a quiver, we are able to construct the category that this quiver freely generates via the same procedure outlined in the previous section, as shown in Figure \ref{fig:Figure7}.

\begin{figure}[ht]
\centering
\includegraphics[width=0.695\textwidth]{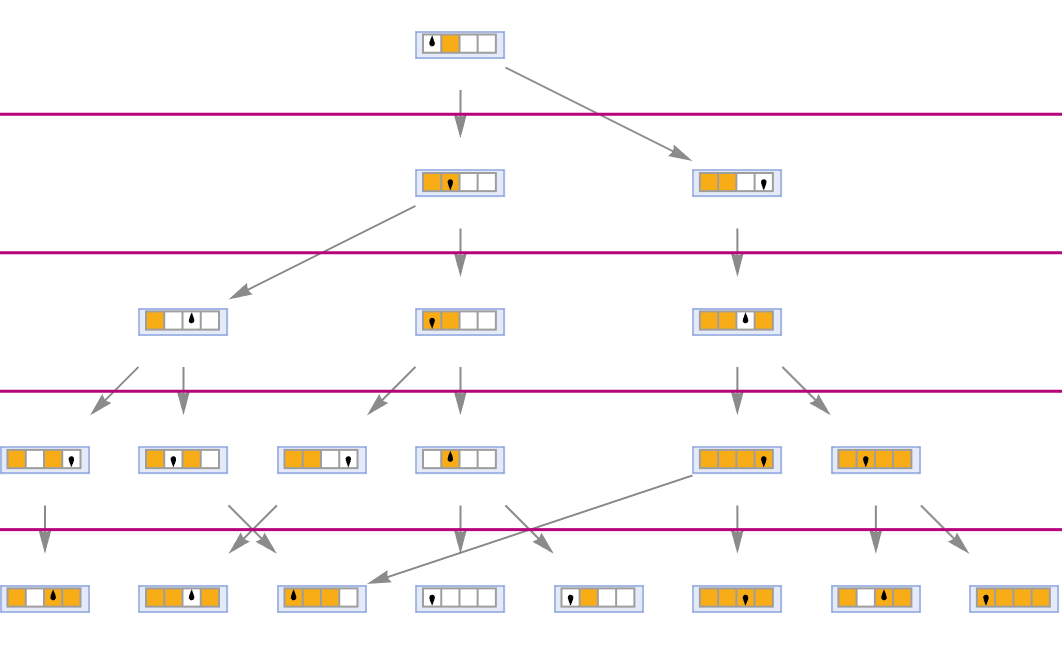}
\caption{The default ``foliation'' of the multiway evolution graph for the non-deterministic evolution of the 2-state, 2-color Turing machine constructed from parallel composition of the transition functions for rules 2506 and 3506, starting from the single tape state ${\left\lbrace 0, 1, 0, 0 \right\rbrace}$, for 4 steps.}
\label{fig:Figure8}
\end{figure}

\begin{figure}[ht]
\centering
\includegraphics[width=0.195\textwidth]{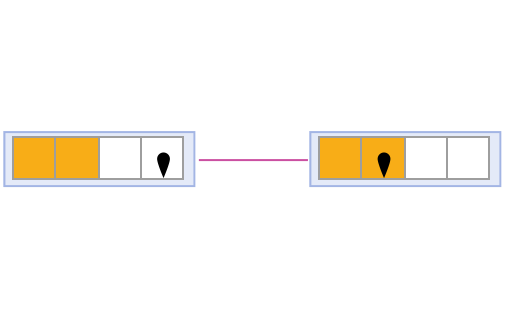}\hspace{0.05\textwidth}
\includegraphics[width=0.245\textwidth]{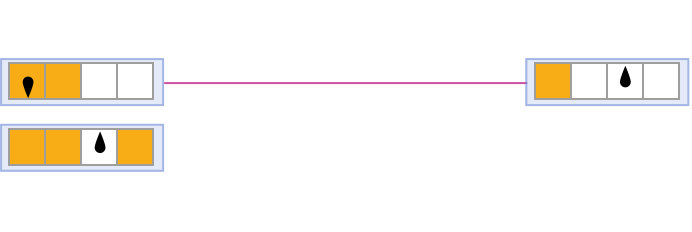}\hspace{0.05\textwidth}
\includegraphics[width=0.195\textwidth]{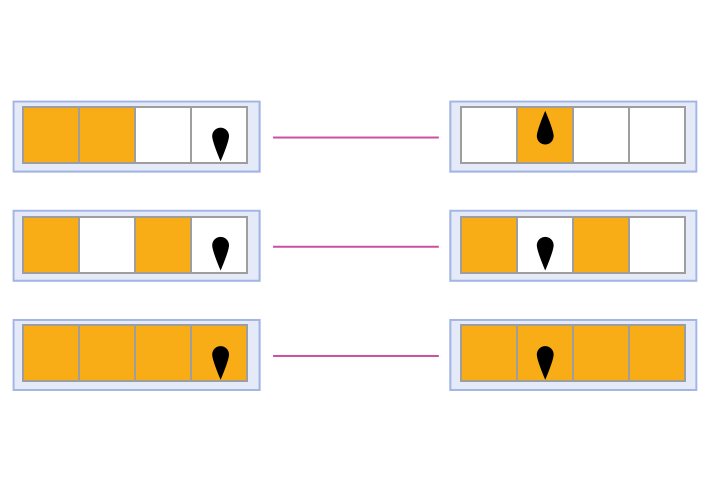}\hspace{0.05\textwidth}
\includegraphics[width=0.195\textwidth]{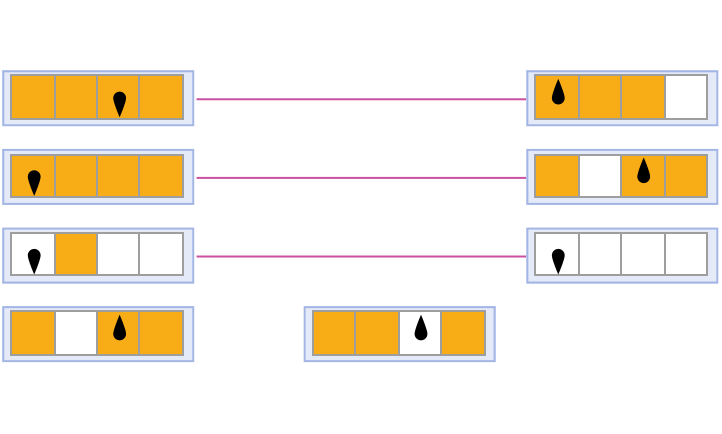}
\caption{The corresponding sequence of ``branchlike hypersurfaces'' associated to the default ``foliation'' of the multiway evolution graph for the non-deterministic evolution of the 2-state, 2-color Turing machine constructed from parallel composition of the transition functions for rules 2506 and 3506.}
\label{fig:Figure9}
\end{figure}

However, we are now able to construct a deterministic/singleway computation from this non-deterministic/multiway one by exploiting the formalism of \textit{branchial graphs}, through a procedure that is more-or-less directly analogous to the Rabin-Scott powerset/subset construction\cite{rabin} for converting non-deterministic finite automata into deterministic ones. More concretely, we ``foliate'' the multiway evolution graph into an ordered sequence of non-intersecting ``branchlike hypersurfaces'' ${\Sigma_t}$ that cover the entire multiway evolution graph, with the ordering relation defined by a universal time function:

\begin{equation}
t : V \to \mathbb{Z}, \qquad \text{ such that } \qquad \Delta t \neq 0 \text{ everywhere},
\end{equation}
where $V$ is the vertex set of the multiway evolution graph (i.e. the set of reachable Turing machine states), and where each ``branchlike hypersurface'' is now a level set of this function, satisfying:

\begin{equation}
\forall t_1, t_2 \in \mathbb{Z}, \qquad \Sigma_{t_1} = \left\lbrace p \in V : t \left( p \right) = t_1 \right\rbrace, \qquad \text{ and } \qquad \Sigma_1 \cap \Sigma_2 = \varnothing \Longleftrightarrow t_1 \neq t_2.
\end{equation}
Branchial graphs constitute a discrete/combinatorial representation of these abstract branchlike hypersurfaces, in which the common ancestry distance between state vertices is represented for any given value of the universal time function, i.e. vertices $X$ and $Y$ in the branchial graph for a given time step are connected by an undirected edge in the branchial graph if and only if they share a common ancestor state $Z$ in the multiway evolution graph. The default choice of foliation for the multiway evolution graph corresponding to the evolution of the non-deterministic 2-state, 2-color Turing machine discussed above is shown in Figure \ref{fig:Figure8}, with the associated sequence of branchial graphs shown in Figure \ref{fig:Figure9}. We can therefore think of every multiway system as being equipped with a certain \textit{tensor product} structure, in which certain pairs of states occur in parallel (as defined by the simultaneity surfaces of the universal time function, i.e. the branchlike hypersurfaces), and are therefore considered to be ``tensored'' together, such that the entire multiway system can be decomposed into a tensor product of single branches (i.e. of individual deterministic/singleway computations). The branchial graphs thus provide a combinatorial description of the tensor product structure of the multiway computation at each time step, and the overall non-deterministic/multiway computation can be recast as a deterministic/singleway computation over these tensor products/branchial graphs.

We can make this intuition mathematically rigorous by noting that our category ${\mathcal{T}}$ of Turing machine states and transitions is now (in the non-deterministic/multiway case) a \textit{monoidal category}, i.e. a category equipped with a tensor product structure, as proved for the case of generic multiway systems based on symbolic rewriting in \cite{gorard4} and \cite{gorard5}; concretely, this entails that ${\mathcal{T}}$ is really an ordered triple ${\left\langle T, \otimes, I \right\rangle}$ consisting of an underlying category, a tensor product operation ${\otimes}$ and a distinguished \textit{unit object} ${I \in \mathrm{ob} \left( \mathcal{T} \right)}$. The tensor product operation is encoded as a \textit{bifunctor} of the form:

\begin{equation}
\otimes : \mathcal{T} \times \mathcal{T} \to \mathcal{T},
\end{equation}
in other words a functor whose domain is the \textit{product category} ${\mathcal{T} \times \mathcal{T}}$, whose object set is given by ordered pairs of objects in ${\mathcal{T}}$:

\begin{equation}
\mathrm{ob} \left( \mathcal{T} \times \mathcal{T} \right) = \left\lbrace \left( X, Y \right) \left\lvert X, Y \in \mathrm{ob} \left( \mathcal{T} \right) \right. \right\rbrace,
\end{equation}
whose morphism set is given by ordered pairs of morphisms in ${\mathcal{T}}$:

\begin{equation}
\mathrm{hom} \left( \mathcal{T} \times \mathcal{T} \right) = \left\lbrace \left( \left( f, g \right) : \left( X_1, X_2 \right) \to \left( Y_1, Y_2 \right) \right) \left\lvert \left( f : X_1 \to Y_1 \right), \left( g : X_2 \to Y_2 \right) \in \mathrm{hom} \left( \mathcal{T} \right) \right. \right\rbrace,
\end{equation}
and in which composition and identity are defined component-wise, i.e:

\begin{multline}
\forall \left( \left( f_1, g_1 \right) : \left( X_1, X_2 \right) \to \left( Y_1, Y_2 \right) \right),  \left( \left( f_2, g_2 \right) : \left( Y_1, Y_2 \right) \to \left( Z_1, Z_2 \right) \right) \in \mathrm{hom} \left( \mathcal{T} \times \mathcal{T} \right),\\
\left( \left( f_2, g_2 \right) \circ \left( f_1, g_1 \right) : \left( X_1, X_2 \right) \to \left( Z_1, Z_2 \right) \right) = \left( \left( f_2 \circ f_1, g_2 \circ g_1 \right) : \left( X_1, X_2 \right) \to \left( Z_1, Z_2 \right) \right),
\end{multline}
and:

\begin{equation}
\forall \left( X, Y \right) \in \mathrm{ob} \left( \mathcal{T} \times \mathcal{T} \right), \qquad \left( id_{\left( X, Y \right)} : \left( X, Y \right) \to \left( X, Y \right) \right) = \left( \left( id_X, id_Y \right) : \left( X, Y \right) \to \left( X, Y \right) \right),
\end{equation}
respectively. In order to encode the fact that the tensor product operation should be (weakly) associative, there should exist a \textit{natural isomorphism} (i.e. a natural transformation whose components are all isomorphisms) ${\alpha}$, which we call the \textit{associator}\cite{kelly}\cite{kelly2}, of the general form:

\begin{equation}
\alpha : \left( - \right) \otimes \left( \left( - \right) \otimes \left( - \right) \right) \cong \left( \left( - \right) \otimes \left( - \right) \right) \otimes \left( - \right),
\end{equation}
with components:

\begin{equation}
\forall X, Y, Z \in \mathrm{ob} \left( \mathcal{T} \right), \qquad \alpha_{X, Y, Z} : X \otimes \left( Y \otimes Z \right) \cong \left( X \otimes Y \right) \otimes Z,
\end{equation}
such that the following diagram (the \textit{associator coherence}) commutes for all ${X, Y, Z, W \in \mathrm{ob} \left( \mathcal{T} \right)}$:

\begin{equation}
\begin{tikzcd}
X \otimes \left( Y \otimes \left( Z \otimes W \right) \right) \arrow[rr, "\alpha_{X, Y, Z \otimes W}"] \arrow[swap, dd, "id_{X} \otimes \alpha_{Y, Z, W}"] & &  \left( X \otimes Y \right) \otimes \left( Z \otimes W \right) \arrow[rr, "\alpha_{X \otimes Y, Z, W}"] & & \left( \left( X \otimes Y \right) \otimes Z \right) \otimes W\\\\
X \otimes \left( \left( Y \otimes Z \right) \otimes W \right) \arrow[swap, rrrr, "\alpha_{X, Y \otimes Z, W}"] & & & & \left( X \otimes \left( Y \otimes Z \right) \right) \otimes W \arrow[swap, uu, "\alpha_{X, Y, Z} \otimes id_{Z}"],
\end{tikzcd}
\end{equation}
i.e:

\begin{equation}
\forall X, Y, Z, W \in \mathrm{ob} \left( \mathcal{T} \right), \qquad \alpha_{X, Y, Z} \otimes id_Z \circ \left( \alpha_{X, Y \otimes Z, W} \circ id_X \otimes \alpha_{Y, Z, W} \right) = \alpha_{X \otimes Y, Z, W} \circ \alpha_{X, Y, Z \otimes W}.
\end{equation}
Moreover, in order to encode the fact that the tensor product is (weakly) unital, i.e. that the distinguished unit object $I$ acts as both a left and right identity for ${\otimes}$, there should exist a further pair of natural isomorphisms ${\lambda}$ and ${\rho}$, which we call the left and right \textit{unitor} isomorphisms respectively, of the general form:

\begin{equation}
\lambda : I \otimes \left( - \right) \cong \left( - \right), \qquad \text{ and } \qquad \rho : \left( - \right) \otimes I \cong \left( - \right),
\end{equation}
with components:

\begin{equation}
\forall X \in \mathrm{ob} \left( \mathcal{T} \right), \qquad \lambda_X : I \otimes X \cong X, \qquad \text{ and } \qquad \rho_X : X \otimes I \cong X,
\end{equation}
such that the following diagram (the \textit{unitor coherence}) commutes for all ${X, Y \in \mathrm{ob} \left( \mathcal{T} \right)}$:

\begin{equation}
\begin{tikzcd}
X \otimes \left( I \otimes Y \right) \arrow[rr, "\alpha_{X, I, Y}"] \arrow[swap, ddr, "id_X \otimes \lambda_Y"] & & \left( X \otimes I \right) \otimes Y \arrow[ddl, "\rho_X \otimes id_Y"]\\\\
& X \otimes Y &,
\end{tikzcd}
\end{equation}
i.e:

\begin{equation}
\forall X, Y \in \mathrm{ob} \left( \mathcal{T} \right), \qquad \rho_X \otimes id_Y \circ \alpha_{X, I, Y} = id_X \otimes \lambda_Y.
\end{equation}

Our monoidal category ${\left\langle \mathcal{T}, \otimes, I \right\rangle}$ inherits the associativity and unitality of its tensor product structure ${\otimes}$ from the associativity and unitality of the disjoint union operation ${\sqcup}$ (with the halt state ${HALT}$ of the Turing machine playing the role of the unit object $I$, since, by definition, parallel composition with the halt state does not substantively modify the structure of any multiway computation because the halt state does not evolve). Furthermore, since the disjoint union operation is also commutative, it follows that our monoidal category is, in fact, \textit{symmetric}\cite{baez2}\cite{joyal}, in the sense that it is also equipped with an additional natural isomorphism ${\sigma}$, called the \textit{symmetry} or \textit{braiding} isomorphism, of the general form:

\begin{equation}
\sigma : \left( - \right) \otimes \left( - \right) \cong \left( - \right) \otimes \left( - \right),
\end{equation}
with components:

\begin{equation}
\forall X, Y \in \mathrm{ob} \left( \mathcal{T} \right), \qquad \sigma_{X, Y} : X \otimes Y \cong Y \otimes X,
\end{equation}
such that the symmetry/braiding isomorphism is compatible with the associator, meaning that the following diagram (essentially an additional associator coherence condition) commutes for all ${X, Y, Z \in \mathrm{ob} \left( \mathcal{T} \right)}$:

\begin{equation}
\begin{tikzcd}
\left( X \otimes Y \right) \otimes Z \arrow[rr, "\sigma_{X, Y} \otimes id_Z"] \arrow[swap, dd, "\alpha_{X, Y, Z}^{-1}"] & & \left( Y \otimes X \right) \otimes Z \arrow[dd, "\alpha_{Y, X, Z}^{-1}"]\\\\
X \otimes \left( Y \otimes Z \right) \arrow[swap, dd, "\sigma_{X, Y \otimes Z}"] & & Y \otimes \left( X \otimes Z \right) \arrow[dd, "id_Y \otimes \sigma_{X, Z}"]\\\\
\left( Y \otimes Z \right) \otimes X \arrow[swap, rr, "\alpha_{Y, Z, X}^{-1}"] && Y \otimes \left( Z \otimes X \right),
\end{tikzcd}
\end{equation}
i.e:

\begin{equation}
\forall X, Y, Z \in \mathrm{ob} \left( \mathcal{T} \right), \qquad id_Y \otimes \sigma_{X, Z} \circ \left( \alpha_{Y, X, Z}^{-1} \circ \sigma_{X, Y} \otimes id_Z \right) = \alpha_{Y, Z, X}^{-1} \circ \left( \sigma_{X, Y \otimes Z} \circ \alpha_{X, Y, Z}^{-1} \right).
\end{equation}
Additionally, the symmetry/braiding isomorphism should be compatible with the left and right unitor isomorphisms, meaning that the following diagram (an additional unitor coherence condition) commutes for all ${X \in \mathrm{ob} \left( \mathcal{T} \right)}$:

\begin{equation}
\begin{tikzcd}
X \otimes I \arrow[rr, "\sigma_{X, I}"]  \arrow[swap, ddr, "\rho_X"] & & I \otimes X \arrow[ddl, "\lambda_X"]\\\\
& X &,
\end{tikzcd}
\end{equation}
i.e:

\begin{equation}
\forall X \in \mathrm{ob} \left( \mathcal{T} \right), \qquad \lambda_X \circ \sigma_{X, I} = \rho_X.
\end{equation}
Finally, the symmetry/braiding isomorphism should be involutive/self-inverse, meaning that the following diagram commutes for all ${X, Y \in \mathrm{ob} \left( \mathcal{T} \right)}$:

\begin{equation}
\begin{tikzcd}
& Y \otimes X \arrow[ddr, "\sigma_{Y, X}"]\\\\
X \otimes Y \arrow[swap, equal, rr, "id_{X \otimes Y}"] \arrow[uur, "\sigma_{X, Y}"] & & X \otimes Y,
\end{tikzcd}
\end{equation}
i.e:

\begin{equation}
\forall X, Y \in \mathrm{ob} \left( \mathcal{T} \right), \qquad \sigma_{Y, X} \circ \sigma_{X, Y} = id_{X \otimes Y}.
\end{equation}
Note that, if we relax the last condition (namely that the natural isomorphism ${\sigma}$ should be involutive/self-inverse), then we obtain not a symmetric monoidal category but the weaker notion of a \textit{braided} monoidal category\cite{joyal2}\cite{chari}, in which the action of ${\sigma}$ on an $n$-fold tensor product factors not through the symmetric group but through the braid group. For the case of braided monoidal categories, we must impose one further associator coherence condition, wherein the following diagram commutes for all ${X, Y, Z \in \mathrm{ob} \left( \mathcal{T} \right)}$:

\begin{equation}
\begin{tikzcd}
X \otimes \left( Y \otimes Z \right) \arrow[rr, "id_X \otimes \sigma_{Y, Z}"] \arrow[swap, dd, "\alpha_{X, Y, Z}"] & & X \otimes \left( Z \otimes Y \right) \arrow[dd, "\alpha_{X, Z, Y}"]\\\\
\left( X \otimes Y \right) \otimes Z \arrow[swap, dd, "\sigma_{X \otimes Y, Z}"] & & \left( X \otimes Z \right) \otimes Y \arrow[dd, "\sigma_{X, Z} \otimes id_Y"]\\\\
Z \otimes \left( X \otimes Y \right) \arrow[swap, rr, "\alpha_{Z, X, Y}"] & & \left( Z \otimes X \right) \otimes Y,
\end{tikzcd}
\end{equation}
i.e:

\begin{equation}
\forall X, Y, Z \in \mathrm{ob} \left( \mathcal{T} \right), \qquad \sigma_{X, Z} \otimes id_Y \circ \left( \alpha_{X, Z, Y} \circ id_X \otimes \sigma_{Y, Z} \right) = \alpha_{Z, X, Y} \circ \left( \sigma_{X \otimes Y, Z} \circ \alpha_{X, Y, Z} \right);
\end{equation}
however, this coherence condition is unnecessary for the case of symmetric monoidal categories, since it can be derived from a combination of the first associator coherence law and the involutive/self-inverse property of the symmetry/braiding isomorphism.\

\begin{figure}[ht]
\centering
\includegraphics[width=0.795\textwidth]{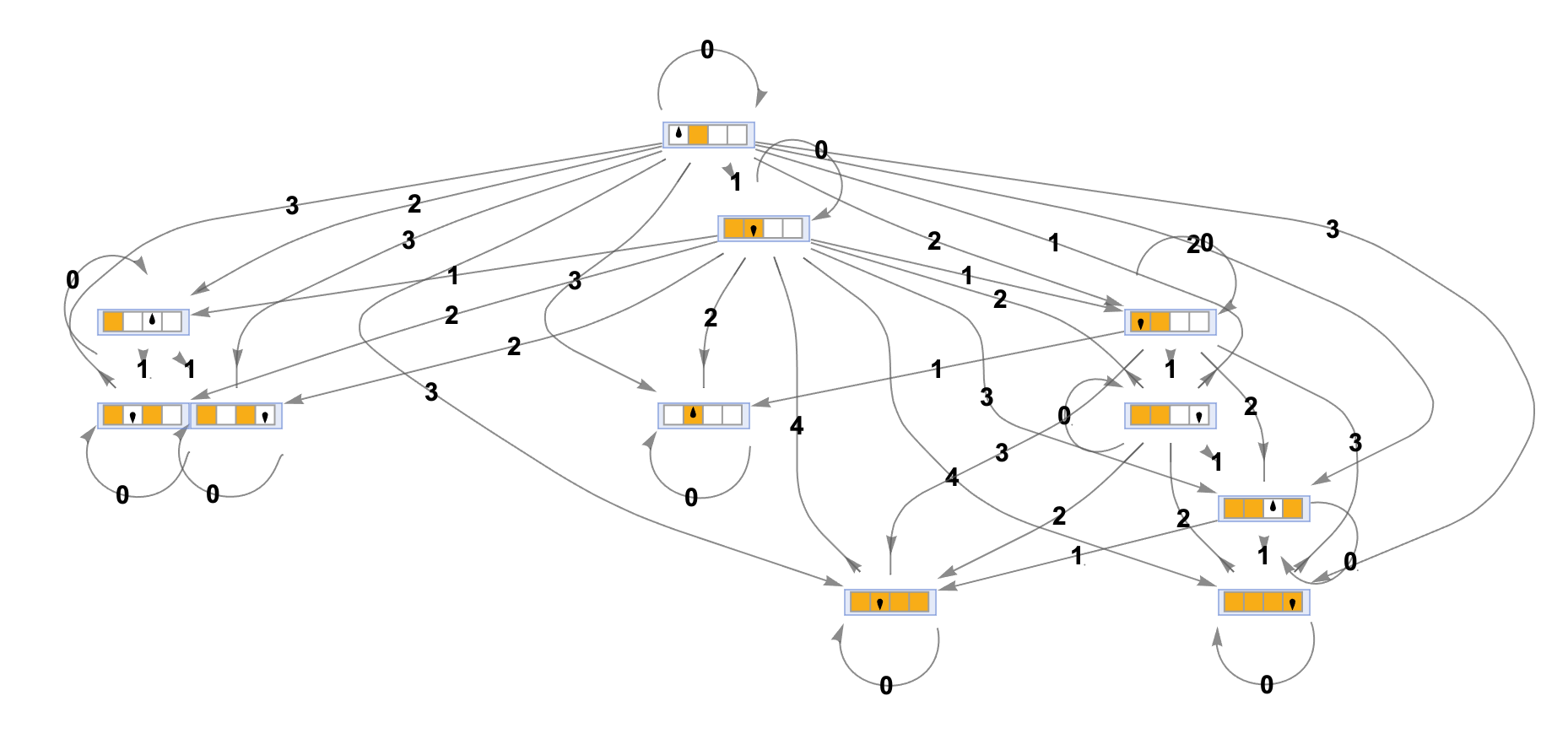}
\caption{A graph-theoretic representation of the category that is freely generated by the multiway evolution graph corresponding to the non-deterministic evolution of a 2-state, 2-color Turing machine constructed from the (parallel) composition of the Turing machine transition functions for rules 2506 and 3506 (considered as a quiver), with edges/morphisms tagged with additional metadata corresponding to the number of ``steps'' (i.e. transitions) required to perform the requisite computation.}
\label{fig:Figure10}
\end{figure}

As before, we can now consider equipping the morphisms of the category ${\mathcal{T}}$ with additional semantic metadata specifying the computational complexities of the corresponding computations that the morphisms signify symbolically. However, since we are now able to compose computations not only in sequence (using the standard composition operator ${\circ}$) but also in parallel (using the tensor product operation ${\otimes}$), the algebra describing how these complexities compose is now somewhat more complicated. As previously described, (singleway) computational irreducibility is characterized by additivity of sequential composition ${\circ}$, i.e. if morphism ${f : X \to Y}$ requires at least $n$ computational steps to enact and morphism ${g : Y \to Z}$ requires at least $m$ computational steps to enact, then their sequential composition ${g \circ f : X \to Z}$ requires at least ${n + m}$ computational steps to enact. On the other hand, due to the presence of the tensor product structure, we can now characterize a form of \textit{multicomputational irreducibility} in terms of additivity of parallel composition ${\otimes}$, i.e. if morphism ${f : X_1 \to Y_1}$ requires at least $n$ computational steps to enact and morphism ${g : X_2 \to Y_2}$ requires at least $m$ computational steps to enact, then their parallel composition ${f \otimes g : X_1 \otimes X_2 \to Y_1 \otimes Y_2}$ (acting on the branchial graph consisting initially of states ${X_1}$ and ${X_2}$) is multicomputationally irreducible if and only if it cannot be enacted in fewer than ${n + m}$ computational steps. Likewise, if the complexities of computations behave subadditively under parallel composition/tensor product, then the composite (multiway) computation is multicomputationally reducible (with the degree of subadditivity quantifying the degree of multicomputational reducibility). The intuition behind this characterization is that, much as ``ordinary'' computational irreducibility is intended to describe the case in which the outcome of a deterministic computation cannot be preempted without explicitly tracing out all intermediate steps, multicomputational irreducibility is intended to describe the case in which the outcome of a non-deterministic/multiway computation cannot be preempted without explicitly tracing out all parallel deterministic/singleway computations of which it is the tensor product. This intuition is illustrated in Figure \ref{fig:Figure10} for the case of the non-deterministic 2-state, 2-color Turing machine discussed above, with each edge/morphism tagged with the number of transitions necessary to perform the corresponding computation; neither the sequential composition operation ${\circ}$ nor the parallel tensor product operation ${\otimes}$ is purely additive, indicating that the system is, at least in part, multicomputationally reducible.

We can consequently extend our previous formalization of computational irreducibility in terms of the functoriality of the map ${Z^{\prime} : \mathcal{T} \to \mathcal{B}}$ to deal with the new case of multicomputational irreducibility too. If we suppose, in the first instance, that all singleway computation paths through the multiway system for the non-deterministic Turing machine $T$ are computationally irreducible, then ${Z^{\prime}}$ is a functor (as proved above). Furthermore, the (discrete) cobordism category ${\mathcal{B}}$ of step numbers/moments of time and (discrete) intervals between them can trivially be equipped with a commutative tensor product structure (namely the disjoint union of sets and intervals ${\sqcup}$) with a unit object given by the empty set ${\varnothing}$; recall that, for more abstract and general cobordism categories, the coproduct ${\oplus}$ plays the role of the disjoint union ${\sqcup}$ (i.e. the direct sum of manifolds), with the initial object ${\varnothing}$ playing the role of the empty set/empty manifold. The objects in the symmetric monoidal category ${\left\langle \mathcal{B}, \oplus, \varnothing \right\rangle}$ therefore represent not merely the time steps associated to individual computational states (as in the singleway case considered within the previous section), but parallel compositions of time steps associated to multiple states on the same ``branchlike hypersurface''; we can, as such, think of the category ${\mathcal{B}}$ as providing a ``coordinatization'' of the time-ordered sequence of branchial graphs computed by the multiway system. Thus, both ${\left\langle \mathcal{T}, \otimes, I = HALT \right\rangle}$, and ${\left\langle \mathcal{B}, \oplus, \varnothing \right\rangle}$ are symmetric monoidal categories, and therefore, subject to the additional hypothesis that all parallel compositions of singleway computations are multicomputationally irreducible (and therefore all computational complexities behave purely additively under tensor products), the map ${Z^{\prime} : \mathcal{T} \to \mathcal{B}}$ forms a \textit{symmetric monoidal functor}\cite{aguiar}. More precisely, ${Z^{\prime}}$ is a monoidal functor in the sense that it is a functor between monoidal categories:

\begin{equation}
Z^{\prime} : \left\langle \mathcal{T}, \otimes, I = HALT \right\rangle \to \left\langle \mathcal{B}, \oplus, \varnothing \right\rangle,
\end{equation}
that preserves the tensor product structure, meaning concretely that ${Z^{\prime}}$ is equipped with a morphism ${\left( \varepsilon : \varnothing \to Z^{\prime} \left( I \right) \right) \in \mathrm{hom} \left( \mathcal{B} \right)}$, along with a natural transformation ${\mu}$ between functors from ${\mathcal{T} \times \mathcal{T}}$ to ${\mathcal{B}}$, with components:

\begin{equation}
\forall X, Y \in \mathrm{ob} \left( \mathcal{T} \right), \qquad \mu_{X, Y} : Z^{\prime} \left( X \right) \oplus Z^{\prime} \left( Y \right) \to Z^{\prime} \left( X \otimes Y \right).
\end{equation}
Together, ${\varepsilon}$ and ${\mu}$ are known as the \textit{coherence maps} of the monoidal functor ${Z^{\prime}}$, and satisfy coherence conditions with the associators ${\alpha^{\mathcal{T}}, \alpha^{\mathcal{B}}}$, left unitor isomorphisms ${\lambda^{\mathcal{T}}, \lambda^{\mathcal{B}}}$ and right unitor isomorphisms ${\rho^{\mathcal{T}}, \rho^{\mathcal{B}}}$. The associator coherence condition is represented by the assertion that the following diagram commutes for all ${X, Y, Z \in \mathrm{ob} \left( \mathcal{T} \right)}$:

\begin{equation}
\begin{tikzcd}
\left( Z^{\prime} \left( X \right) \oplus Z^{\prime} \left( Y \right) \right) \oplus Z^{\prime} \left( Z \right) \arrow[rrr, "\alpha_{Z^{\prime} \left( X \right), Z^{\prime} \left( Y \right), Z^{\prime} \left( Z \right)}^{\mathcal{B}}"] \arrow[swap, ddd, "\mu_{X, Y} \oplus id_{Z^{\prime} \left( Z \right)}^{\mathcal{B}}"] & & & Z^{\prime} \left( X \right) \oplus \left( Z^{\prime} \left( Y \right) \oplus Z^{\prime} \left( Z \right) \right) \arrow[ddd, "id_{Z^{\prime} \left( Z \right)}^{\mathcal{B}}"]\\\\\\
Z^{\prime} \left( X \otimes Y \right) \oplus Z^{\prime} \left( Z \right) \arrow[swap, ddd, "\mu_{X \otimes Y, Z}"] & & & Z^{\prime} \left( X \right) \oplus Z^{\prime} \left( Y \otimes Z \right) \arrow[ddd, "\mu_{X, Y \otimes Z}"]\\\\\\
Z^{\prime} \left( \left( X \otimes Y \right) \otimes Z \right) \arrow[swap, rrr, "Z^{\prime} \left( \alpha_{X, Y, Z}^{\mathcal{T}} \right)"] & & & Z^{\prime} \left( X \otimes \left( Y \otimes Z \right) \right),
\end{tikzcd}
\end{equation}
i.e. (recalling that the composition operation in the cobordism category ${\mathcal{B}}$ is given by the ordinary union of contiguous intervals/cobordisms ${\cup}$):

\begin{multline}
\forall X, Y, Z \in \mathrm{ob} \left( \mathcal{T} \right), \\
\mu_{X, Y \otimes Z} \cup \left( id_{Z^{\prime} \left( Z \right)} \cup \alpha_{Z^{\prime} \left( X \right), Z^{\prime} \left( Y \right), Z^{\prime} \left( Z \right)}^{\mathcal{B}} \right) = Z^{\prime} \left( \alpha_{X, Y, Z}^{\mathcal{T}} \right) \cup \left( \mu_{X \otimes Y, Z} \cup \mu_{X, Y} \oplus id_{Z^{\prime} \left( Z \right)}^{\mathcal{B}} \right),
\end{multline}
while the left and right unitor coherence conditions are represented by the assertion that the following diagrams commute for all ${X \in \mathrm{ob} \left( \mathcal{T} \right)}$:

\begin{equation}
\begin{tikzcd}
Z^{\prime} \left( X \right) \oplus \varnothing \arrow[rr, "id_{Z^{\prime} \left( X \right)}^{\mathcal{B}} \oplus \varepsilon"] \arrow[swap, dd, "\rho_{Z^{\prime} \left( X \right)}^{\mathcal{B}}"] & & Z^{\prime} \left( X \right) \oplus Z^{\prime} \left( I \right) \arrow[dd, "\mu_{X, I}"]\\\\
Z^{\prime} \left( X \right) & & Z^{\prime} \left( X \otimes I \right) \arrow[ll, "Z^{\prime} \left( \rho_{X}^{\mathcal{T}} \right)"],
\end{tikzcd}
\end{equation}
i.e:

\begin{equation}
\forall X \in \mathrm{ob} \left( \mathcal{T} \right), \qquad Z^{\prime} \left( \rho_{X}^{\mathcal{T}} \right) \cup \left( \mu_{X, I} \cup id_{Z^{\prime} \left( X \right)}^{\mathcal{B}} \oplus \varepsilon \right) = \rho_{Z^{\prime} \left( X \right)}^{\mathcal{B}},
\end{equation}
and:

\begin{equation}
\begin{tikzcd}
\varnothing \oplus Z^{\prime} \left( X \right) \arrow[rr, "\epsilon \oplus id_{Z^{\prime} \left( X \right)}^{\mathcal{B}}"] \arrow[swap, dd, "\lambda_{Z^{\prime} \left( X \right)}^{\mathcal{B}}"] & & Z^{\prime} \left( I \right) \oplus Z^{\prime} \left( X \right) \arrow[dd, "\mu_{I, X}"]\\\\
Z^{\prime} \left( X \right) & & Z^{\prime} \left( I \otimes X \right) \arrow[ll, "Z^{\prime} \left( \lambda_{X}^{\mathcal{T}} \right)"],
\end{tikzcd}
\end{equation}
i.e:

\begin{equation}
\forall X \in \mathrm{ob} \left( \mathcal{T} \right), \qquad Z^{\prime} \left( \lambda_{X}^{\mathcal{T}} \right) \cup \left( \mu_{I, X} \cup \epsilon \oplus id_{Z^{\prime} \left( X \right)}^{\mathcal{B}} \right) = \lambda_{Z^{\prime} \left( X \right)}^{\mathcal{B}},
\end{equation}
respectively. Note that the definition presented here is for a \textit{lax} monoidal functor; if the coherence maps ${\varepsilon}$ and ${\mu_{X, Y}}$ were, additionally, either isomorphisms or identities for all ${X, Y \in \mathrm{ob} \left( \mathcal{T} \right)}$, then one would obtain a \textit{strong} or a \textit{strict} monoidal functor, respectively, instead. The monoidal functor ${Z^{\prime} : \left\langle \mathcal{T}, \otimes, I \right\rangle \to \left\langle \mathcal{B}, \oplus, \varnothing \right\rangle}$ is also symmetric, in the sense that the coherence maps ${\varepsilon}$ and ${\mu}$ also satisfy a further coherence condition with the braiding/symmetry isomorphisms ${\sigma^{\mathcal{T}}}$, ${\sigma^{\mathcal{B}}}$, represented by the assertion that the following diagram commutes for all ${X, Y \in \mathrm{ob} \left( \mathcal{T} \right)}$:

\begin{equation}
\begin{tikzcd}
Z^{\prime} \left( X \right) \oplus Z^{\prime} \left( Y \right) \arrow[rr, "\sigma_{Z^{\prime} \left( X \right), Z^{\prime} \left( Y \right)}^{\mathcal{B}}"] \arrow[swap, dd, "\mu_{X, Y}"] & & Z^{\prime} \left( X \right) \oplus Z^{\prime} \left( Y \right) \arrow[dd, "\mu_{Y,X}"]\\\\
Z^{\prime} \left( X \otimes Y \right) \arrow[swap, rr, "Z^{\prime} \left( \sigma_{X, Y}^{\mathcal{T}} \right)"] & & Z^{\prime} \left( Y \otimes X \right),
\end{tikzcd}
\end{equation}
i.e:

\begin{equation}
\forall X, Y \in \mathrm{ob} \left( \mathcal{T} \right), \qquad \mu_{Y, X} \cup \sigma_{Z^{\prime} \left( X \right), Z^{\prime} \left( Y \right)}^{\mathcal{B}}= Z^{\prime} \left( \sigma_{X, Y}^{\mathcal{T}} \right) \cup \mu_{X, Y}.
\end{equation}
If ${\left\langle \mathcal{T}, \otimes, I \right\rangle}$ and ${\left\langle \mathcal{B}, \oplus, \varnothing \right\rangle}$ were merely braided monoidal rather than symmetric monoidal categories, then a monoidal functor ${Z^{\prime}}$ satisfying the above coherence condition would instead be a \textit{braided} monoidal functor.

\begin{figure}[ht]
\centering
\includegraphics[width=0.795\textwidth]{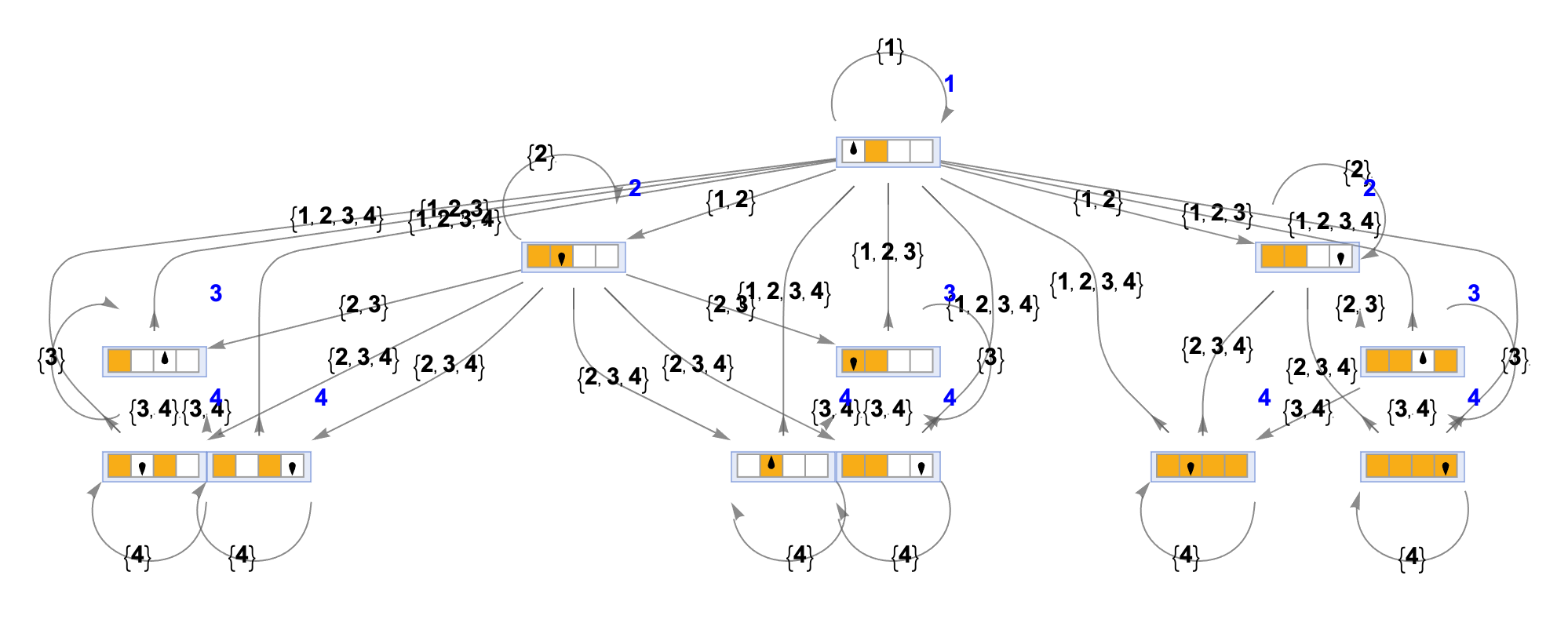}
\caption{A graph-theoretic representation of the category that is freely generated by the multiway evolution graph corresponding to the non-deterministic evolution of a 2-state, 2-color Turing machine constructed from the (parallel) composition of the Turing machine transition functions for rules 2506 and 3506 (considered as a quiver), with vertices/objects tagged with additional metadata corresponding to the step number on which they occur (shown in blue) and with edges/morphisms tagged with additional metadata corresponding to all intermediate step numbers traversed as part of the requisite computation.}
\label{fig:Figure11}
\end{figure}

Thus, if ${X_1}$, ${Y_1}$, ${X_2}$ and ${Y_2}$ represent Turing machine states reached at step numbers ${t_1}$, ${t_2}$, ${s_1}$ and ${s_2}$ respectively, and ${f : X_1 \to Y_1}$ and ${g : X_2 \to Y_2}$ represent the transitions between states ${X_1}$ and ${Y_1}$ and states ${X_2}$ and ${Y_2}$ respectively, then the (symmetric) monoidal functor ${Z^{\prime}}$:

\begin{equation}
\begin{tikzcd}
X_1 \arrow[swap, dd, "f"] & X_2 \arrow[swap, dd, "g"'{name=g}] & & & & t_1 \arrow[dd, "{\left[ t_1, t_2 \right] \cap \mathbb{N}}"'{name=f}] & s_1 \arrow[dd, "{\left[ s_1, s_2 \right] \cap \mathbb{N}}"]\\\\
Y_1 & Y_2 & & & & t_2 & s_2,
\arrow[mapsto, from=g, to=f, shorten=2.5em, "Z^{\prime}"]
\end{tikzcd}
\end{equation}
preserves the (symmetric) tensor product structure in the sense depicted by the following diagram:

\begin{equation}
\begin{tikzcd}
X_1 \otimes X_2 \arrow[dd, "f \otimes g"'{name=g}] & & & & t_1 \oplus s_1 \arrow[swap, dd, "{\left( \left[ t_1, t_2 \right] \cap \mathbb{N} \right) \oplus \left( \left[ s_1, s_2 \right] \cap \mathbb{N} \right)}"'{name=f}]\\\\
Y_1 \otimes Y_2 & & & & t_2 \oplus s_2
\arrow[mapsto, from=g, to=f, shorten=6em, "Z^{\prime}"].
\end{tikzcd}
\end{equation}
Hence, by equipping ${\mathcal{B}}$ with a tensor product structure, the resulting symmetric monoidal category ${\left\langle \mathcal{B}, \oplus, \varnothing \right\rangle}$ is actually a higher-dimensional cobordism category, in which the objects are potentially higher-dimensional manifolds (consisting of direct sums/disjoint unions of several 0-dimensional manifolds) and the morphisms are potentially higher-dimensional cobordisms (consisting of direct sums/disjoint unions of several 1-dimensional cobordisms), since there now exist two distinct directions in which cobordisms can be ``glued'' (either sequentially, via the standard union of contiguous intervals ${\cup}$, or in parallel, via the tensor product/direct sum of intervals ${\sqcup}$). For any given morphism in ${\mathcal{B}}$, the cardinalities of the various individual (1-dimensional) intervals of which it is composed represent, up to an additive constant, the time complexities of the corresponding deterministic/singleway computations in ${\mathcal{T}}$ (as in the singleway case analyzed in the previous section), whereas the number of distinct (1-dimensional) intervals appearing in the tensor product/direct sum within that morphism represents, up to an additive constant, the time complexity of the resulting composite non-deterministic/multiway computation in ${\mathcal{T}}$. As a consequence, we can conclude that, just as (singleway) computational irreducibility is reflected in the functoriality of ${Z^{\prime}}$, multicomputational irreducibility is reflected in the symmetric monoidal functoriality of ${Z^{\prime}}$, and just as computational reducibility corresponds precisely to a deformation of ${Z^{\prime}}$ away from a pure functor, multicomputational reducibility corresponds to a deformation of ${Z^{\prime}}$ from being symmetric monoidal. Another, perhaps cleaner, way to articulate this would be to say that computational reducibility corresponds to how much the map ${Z^{\prime}}$ distorts the sequential composition (${\circ}$) of computations in ${\mathcal{T}}$, while multicomputational reducibility corresponds to how much the map ${Z^{\prime}}$ distorts the parallel composition (${\otimes}$) of computations in ${\mathcal{T}}$. This construction is demonstrated in Figure \ref{fig:Figure11} for the case of the non-deterministic 2-state, 2-color Turing machine discussed previously, with each vertex/object tagged with its step number and each edge/morphism tagged with a list of step numbers traversed throughout the course of its corresponding computation. Note that the same algebraic axioms that describe how the time complexities attached to morphisms compose (namely identity, strict positivity and subadditivity/triangle inequality) in the sequential case under ${\circ}$ also hold in the parallel case under ${\otimes}$.

From this rather abstract description, it should be apparent that computational irreducibility and multicomputational irreducibility are essentially orthogonal concepts: the map ${Z^{\prime}}$ can distort sequential compositions to an arbitrary extent whilst keeping the tensor product structure entirely intact, or vice versa (or anything in between). This is a fairly intuitive consequence of the fact that computational irreducibility is a byproduct of the state \textit{evolution function} of a given multiway system (i.e. the function that specifies which states are obtainable in a single step from which other states, and which therefore provides the rules for constructing morphisms in ${\mathcal{T}}$), while multicomputational irreducibility is a byproduct of the state \textit{equivalence function} of a given multiway system (i.e. the function that specifies which pairs of states are to be considered equivalent, and which therefore provides the rules for merging/equating objects in ${\mathcal{T}}$), and the evolution and equivalence functions have entirely independent definitions: one can define a state evolution function with an arbitrarily high computational complexity whilst keeping the state equivalence function essentially trivial, or vice versa (or, once again, any reasonable intermediate). For instance, in the case of non-deterministic/multiway Turing machines considered thus far, the state equivalence function is elementary (two Turing machine states are considered equivalent if their tape states, head states and head positions are both identical, which is trivial to determine algorithmically), with all of the computational complexity originating from the state evolution function. On the other hand, in the case of (hyper)graph rewriting, as considered in the context of the Wolfram model\cite{wolfram3}\cite{gorard2}\cite{gorard3}, the state equivalence function is much more sophisticated, since it must account for (hyper)graph isomorphism, whose precise complexity class ${GI}$ remains unknown (i.e. it is not known whether ${GI}$ is $P$, ${NP}$-complete or ${NP}$-intermediate)\cite{booth}\cite{kobler}. In the analysis that follows, we shall be using a generalized version of the ``uniqueness tree'' isomorphism algorithm presented in \cite{gorard6}.

\begin{figure}[ht]
\centering
\includegraphics[width=0.395\textwidth]{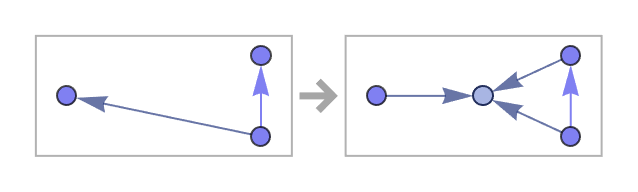}
\caption{A graphical representation of the (hyper)graph transformation rule corresponding to the set substitution system ${\left\lbrace \left\lbrace x, y \right\rbrace, \left\lbrace x, z \right\rbrace \right\rbrace \to \left\lbrace \left\lbrace x, z \right\rbrace, \left\lbrace x, w \right\rbrace, \left\lbrace y, w \right\rbrace, \left\lbrace z, w \right\rbrace \right\rbrace}$.}
\label{fig:Figure12}
\end{figure}

By defining a (directed/ordered) \textit{hypergraph} ${H = \left\langle V, E \right\rangle}$ in terms of a finite collection/multiset of ordered relations (hyperedges) between elements:

\begin{equation}
E \subseteq \mathcal{P} \left( V \right) \setminus \left\lbrace \varnothing \right\rbrace,
\end{equation}
where ${\mathcal{P}}$ denotes the power set, we can formalize the notion of hypergraph rewriting rules ${H_1 \to H_2}$ in terms of a \textit{span} of \textit{monomorphisms}\cite{ehrig}\cite{ehrig2} of the form:

\begin{equation}
\begin{tikzcd}
L & K \arrow[swap, l, "l"] \arrow[r, "r"] & R,
\end{tikzcd}
\end{equation}
in some category ${\mathcal{C}}$ (whose objects are hypergraphs, and whose morphisms represent subhypergraph inclusion maps), where $L$ is a hypergraph pattern designating the left-hand side of the rule, $R$ is a hypergraph pattern designating the right-hand side of the rule, and $K$ is a pattern designating the subhypergraph that remains invariant when the left-hand side is ``extracted'' and the right-hand side is ``injected''. (Note that \textit{hypergraph categories}, in the terminology of Kissinger\cite{kissinger4} and Fong\cite{fong}\cite{fong2}, namely symmetric monoidal categories equipped with a \textit{Frobenius algebra structure} which ensures that all string diagrams correspond to hypergraphs, constitute a particularly natural categorical setting in which to construct such a rewriting system). In the above, the morphisms ${l : K \to L}$ and ${r : K \to R}$ form a \textit{span} in the sense that they constitute a pair of morphisms with a common domain, and they are \textit{monomorphisms} in the sense that they are injective/left-cancellative: for any pairs of morphisms ${f_1 : X \to K}$ and ${f_2 : X \to K}$ that make either of the following diagrams commute:

\begin{equation}
\begin{tikzcd}
\forall X \arrow[r, bend left, "\forall f_1" {pos = 0.48}] \arrow[swap, r, bend right, "\forall f_2"] & K \arrow[r, "l"] & L
\end{tikzcd}, \qquad \text{ or } \qquad
\begin{tikzcd}
\forall X \arrow[r, bend left, "\forall f_1" {pos = 0.48}] \arrow[swap, r, bend right, "\forall f_2"] & K \arrow[r, "r"] & R,
\end{tikzcd}
\end{equation}
one necessarily has ${\left( f_1 : X \to K \right) = \left( f_2 : X \to K \right)}$, i.e:

\begin{multline}
\forall X \in \mathrm{ob} \left( \mathcal{C} \right), \qquad \forall \left( f_1 : X \to K \right), \left( f_2 : X \to K \right) \in \mathrm{hom} \left( \mathcal{C} \right),\\
\left( l \circ f_ 1 : X \to L \right) = \left( l \circ f_2 : X \to L \right) \qquad \implies \qquad \left( f_1 : X \to K \right) = \left( f_2 : X \to K \right),
\end{multline}
and:

\begin{multline}
\forall X \in \mathrm{ob} \left( \mathcal{C} \right), \qquad \forall \left( f_1 : X \to K \right), \left( f_2 : X \to K \right) \in \mathrm{hom} \left( \mathcal{C} \right),\\
\left( r \circ f_1 : X \to R \right) = \left( r \circ f_2 : X \to R \right) \qquad \implies \qquad \left( f_1 : X \to K \right) = \left( f_2 : X \to K \right).
\end{multline}
A graphical representation of such a hypergraph rewriting rule with relations/hyperedges of arity-2 (corresponding to the set substitution rule ${\left\lbrace \left\lbrace x, y \right\rbrace, \left\lbrace x, z \right\rbrace \right\rbrace \to \left\lbrace \left\lbrace x, z \right\rbrace, \left\lbrace x, w \right\rbrace, \left\lbrace y, w \right\rbrace, \left\lbrace z, w \right\rbrace \right\rbrace}$ - since any hypergraph rewriting rule of this form can always be reformulated as a symbolic set substitution rule acting on multisets of ordered relations between vertices) is shown in Figure \ref{fig:Figure12}.

\begin{figure}[ht]
\centering
\includegraphics[width=0.795\textwidth]{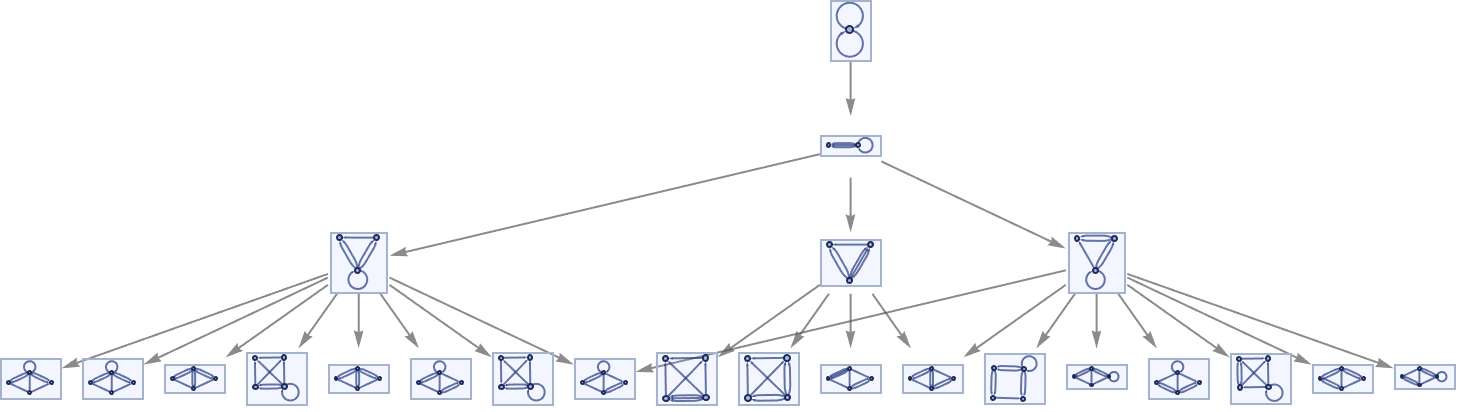}
\caption{A multiway evolution graph corresponding to the non-deterministic evolution of the set substitution system ${\left\lbrace \left\lbrace x, y \right\rbrace, \left\lbrace x, z \right\rbrace \right\rbrace \to \left\lbrace \left\lbrace x, z \right\rbrace, \left\lbrace x, w \right\rbrace, \left\lbrace y, w \right\rbrace, \left\lbrace z, w \right\rbrace \right\rbrace}$, starting from a ``double self-loop'' initial condition, for 3 steps; each edge represents a single application of the corresponding (hyper)graph rewriting rule.}
\label{fig:Figure13}
\end{figure}

\begin{figure}[ht]
\centering
\includegraphics[width=0.795\textwidth]{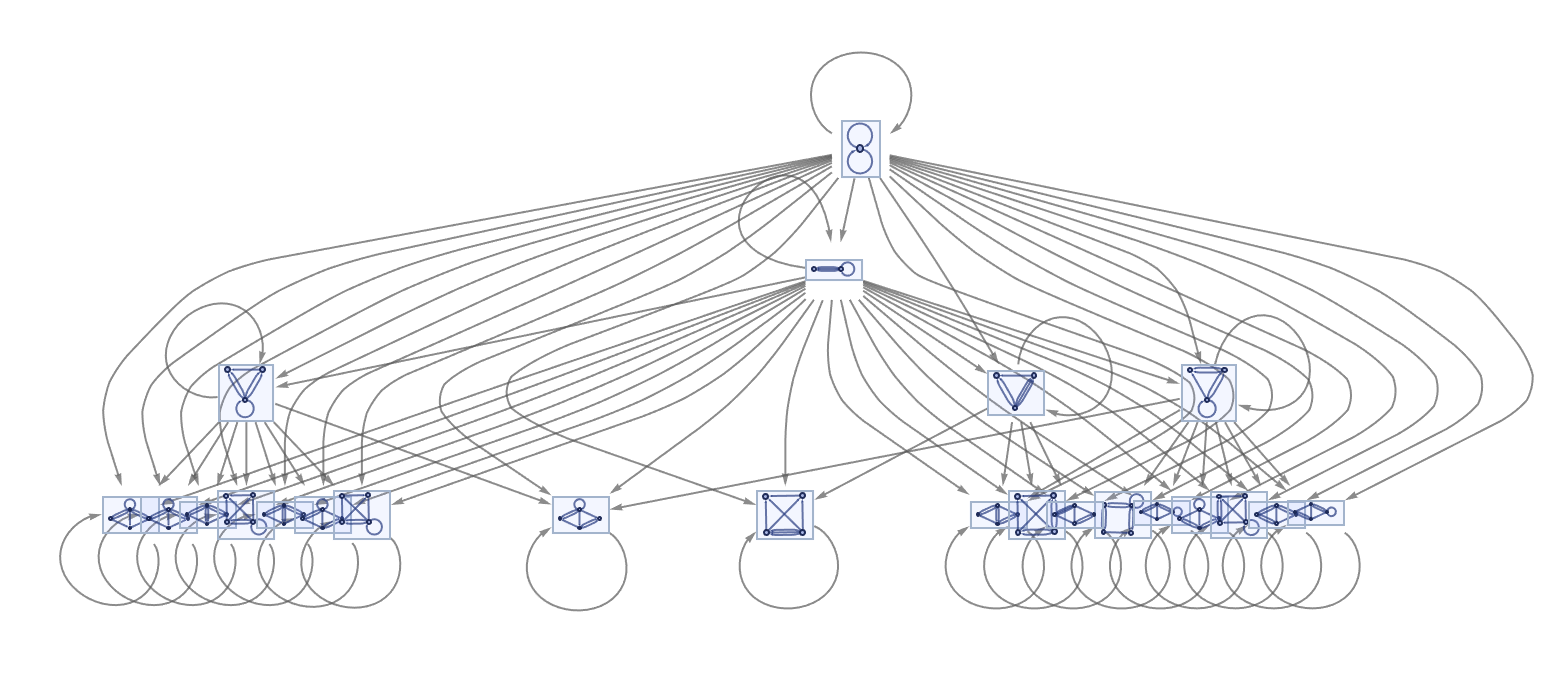}
\caption{A graph-theoretic representation of the category that is freely generated by the multiway evolution graph corresponding to the non-deterministic evolution of the set substitution system ${\left\lbrace \left\lbrace x, y \right\rbrace, \left\lbrace x, z \right\rbrace \right\rbrace \to \left\lbrace \left\lbrace x, z \right\rbrace, \left\lbrace x, w \right\rbrace, \left\lbrace y, w \right\rbrace, \left\lbrace z, w \right\rbrace \right\rbrace}$ (considered as a quiver), starting from a ``double self-loop'' initial condition, for 3 steps.}
\label{fig:Figure14}
\end{figure}

\begin{figure}[ht]
\centering
\includegraphics[width=0.795\textwidth]{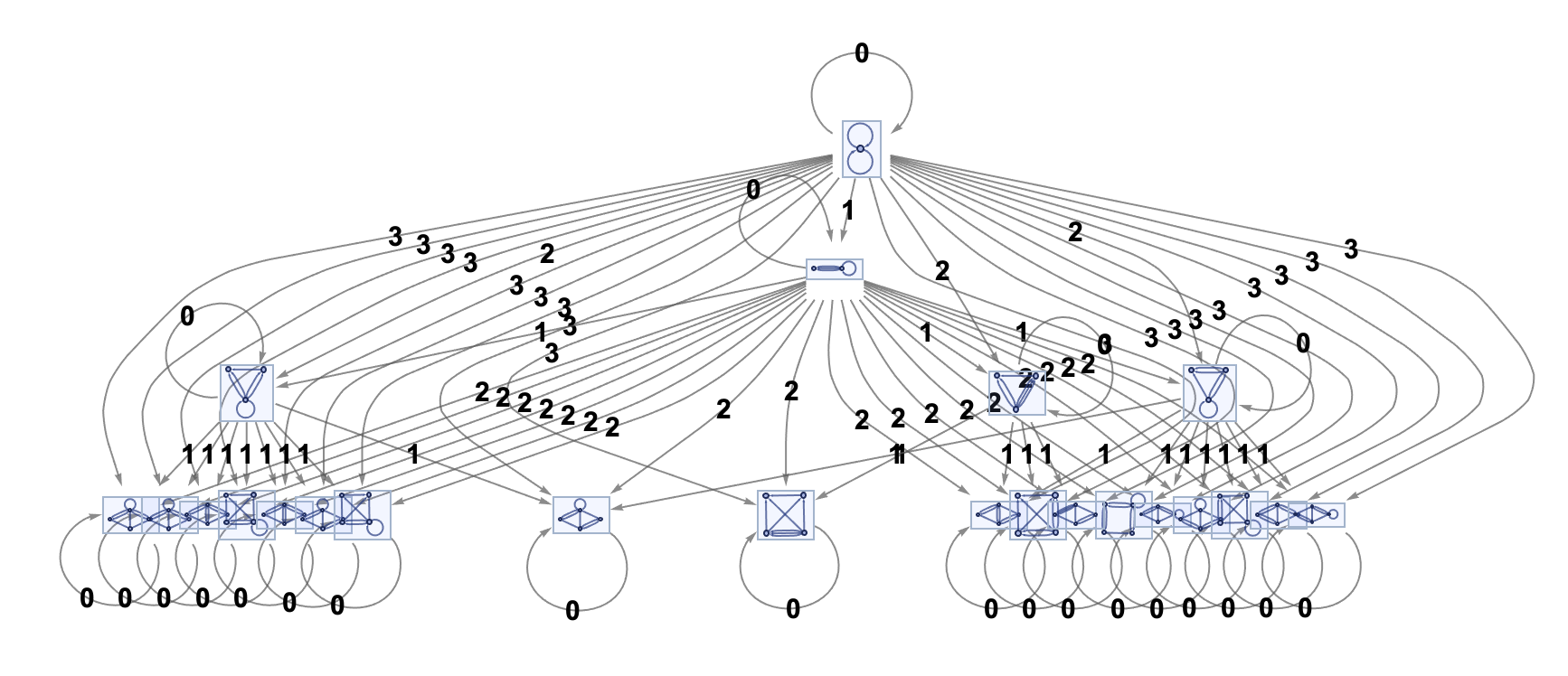}
\caption{A graph-theoretic representation of the category that is freely generated by the multiway evolution graph corresponding to the non-deterministic evolution of the set substitution system ${\left\lbrace \left\lbrace x, y \right\rbrace, \left\lbrace x, z \right\rbrace \right\rbrace \to \left\lbrace \left\lbrace x, z \right\rbrace, \left\lbrace x, w \right\rbrace, \left\lbrace y, w \right\rbrace, \left\lbrace z, w \right\rbrace \right\rbrace}$ (considered as a quiver), with edges/morphisms tagged with additional metadata corresponding to the number of ``steps'' (i.e. rewriting rule applications) required to perform the requisite computation.}
\label{fig:Figure15}
\end{figure}

\begin{figure}[ht]
\centering
\includegraphics[width=0.795\textwidth]{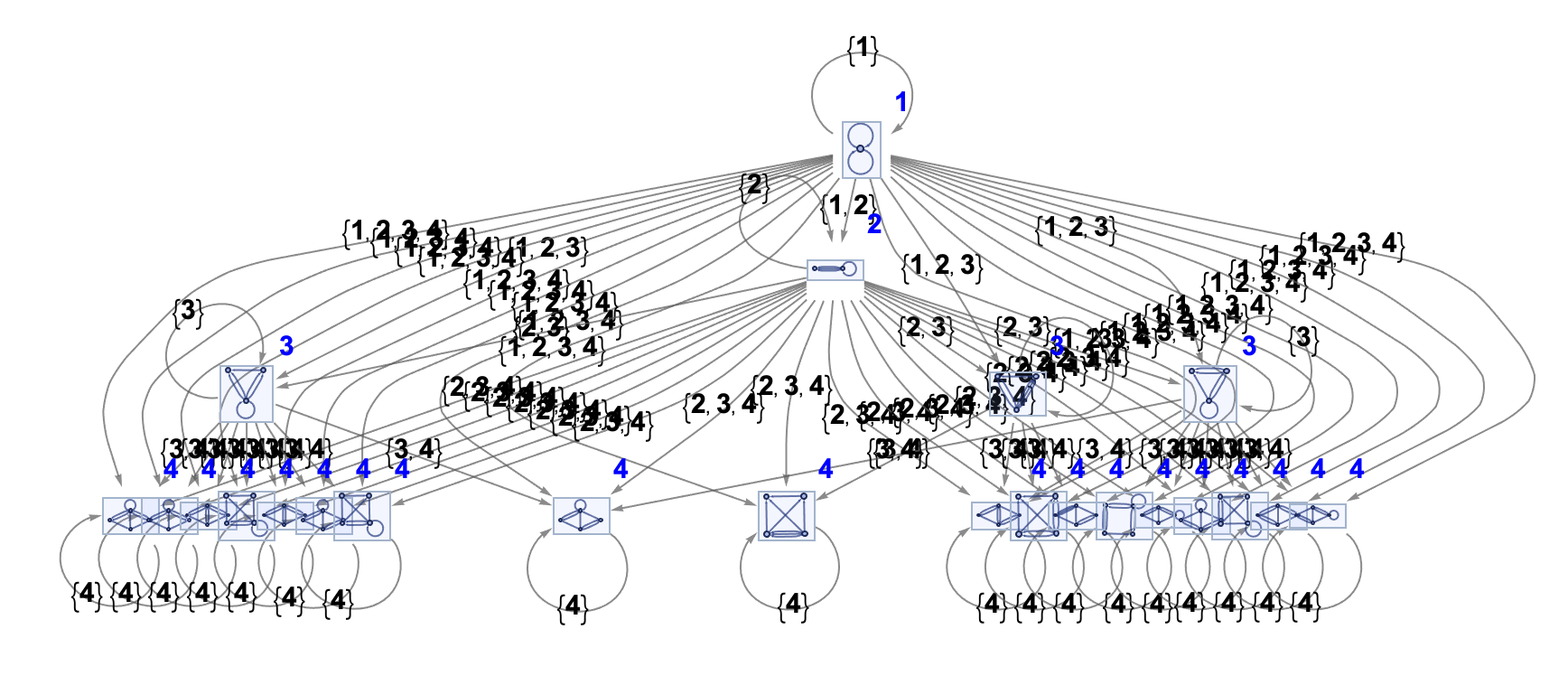}
\caption{A graph-theoretic representation of the category that is freely generated by the multiway evolution graph corresponding to the non-deterministic evolution of the set substitution system ${\left\lbrace \left\lbrace x, y \right\rbrace, \left\lbrace x, z \right\rbrace \right\rbrace \to \left\lbrace \left\lbrace x, z \right\rbrace, \left\lbrace x, w \right\rbrace, \left\lbrace y, w \right\rbrace, \left\lbrace z, w \right\rbrace \right\rbrace}$ (considered as a quiver), with vertices/objects tagged with additional metadata corresponding to the step number on which they occur (shown in blue) and with edges/morphisms tagged with additional metadata corresponding to all intermediate step numbers traversed as part of the requisite computation.}
\label{fig:Figure16}
\end{figure}

A hypergraph rewriting rule of the form presented above can then be said to \textit{match} a given hypergraph $G$ if there exists a morphism ${\left( m : L \to G \right) \in \mathrm{hom} \left( \mathcal{C} \right)}$, and the resulting hypergraph $H$ obtained by applying the rewriting rule at that match can be computed by means of the following \textit{double-pushout} diagram\cite{habel}:

\begin{equation}
\begin{tikzcd}
& L \arrow[swap, d, "m"] \arrow[swap, ddl, bend right, "\forall m^{*}"] & K \arrow[swap, l, "l"] \arrow[r, "r"] \arrow[d, "n"] & R \arrow[d, "p"] \arrow[ddr, bend left, "\forall p^{*}"] & \\
& G \arrow[dl, dashed, "\exists ! u_1"] & D \arrow[l, "g"] \arrow[swap, drr,  bend right, "\forall h^{*}"] \arrow[dll, bend left, "\forall g^{*}"] \arrow[swap, r, "h"] & H \arrow[swap, dr, dashed, "\exists ! u_2"] &\\
\forall G^{*} & & & & \forall H^{*}.
\end{tikzcd}
\end{equation}
More concretely, ${D \in \mathrm{ob} \left( \mathcal{C} \right)}$ is a hypergraph and ${\left( n : K \to D \right), \left( g : D \to G \right) \in \mathrm{hom} \left( \mathcal{C} \right)}$ are hypergraph inclusions such that the leftmost square commutes:

\begin{equation}
\left( g \circ n : K \to G \right) = \left( m \circ l : K \to G \right),
\end{equation}
and are universal in the sense that, for any hypergraph ${G^{*} \in \mathrm{ob} \left( \mathcal{C} \right)}$ equipped with inclusions:

\begin{equation}
\left( m^{*} : L \to G^{*} \right), \left( g^{*} : D \to G^{*} \right) \in \mathrm{hom} \left( \mathcal{C} \right),
\end{equation}
there exists a unique inclusion ${\left( u_1 : G \to G^{*} \right) \in \mathrm{hom} \left( \mathcal{C} \right)}$ such that:

\begin{equation}
\left( m^{*} : L \to G^{*} \right) = \left( u_1 \circ m : L \to G^{*} \right), \qquad \text{ and } \qquad \left( g^{*} : D \to G^{*} \right) = \left( u_1 \circ g : D \to G^{*} \right).
\end{equation}
This allows one to compute the ``residual'' hypergraph $D$ obtained by extracting out a subhypergraph isomorphic to $L$ from $G$ at the position defined by the rule match ${m : L \to G}$. Moreover, ${D, H \in \mathrm{ob} \left( \mathcal{C} \right)}$ are hypergraphs and ${\left( p : R \to H \right), \left( h : D \to H \right) \in \mathrm{hom} \left( \mathcal{C} \right)}$ are inclusions such that the rightmost square commutes:

\begin{equation}
\left( h \circ n : K \to H \right) = \left( p \circ r : K \to H \right),
\end{equation}
and are universal in the sense that, for any hypergraph ${H^{*} \in \mathrm{ob} \left( \mathcal{C} \right)}$ equipped with inclusions:

\begin{equation}
\left( p^{*} : R \to H^{*} \right), \left( h^{*} D \to H^{*} \right) \in \mathrm{hom} \left( \mathcal{C} \right),
\end{equation}
there exists a unique inclusion ${\left( u_2 : H \to H^{*} \right) \in \mathrm{hom} \left( \mathcal{C} \right)}$ such that:

\begin{equation}
\left( p^{*} : R \to H^{*} \right) = \left( u_2 \circ p : R \to H^{*} \right), \qquad \text{ and } \qquad \left( h^{*} : D \to H^{*} \right) = \left( u_2 \circ h : D \to H^{*} \right).
\end{equation}
This, in turn, allows one to compute the resulting hypergraph $H$ obtained by gluing a subhypergraph isomorphic to $R$ into the residual hypergraph $D$ at the position defined by the injection map ${m : K \to D}$ (from the invariant subhypergraph to the residual hypergraph). This double-pushout construction yields the class of possible hypergraph transitions from which we are able to construct a multiway evolution graph (whose vertices represent hypergraphs and whose edges represent hypergraph rewrites), by a process that is directly analogous to the Turing machine case considered previously: an explicit example for the hypergraph rewriting rule presented above is shown in Figure \ref{fig:Figure13}, and the category that is freely generated by this multiway evolution graph (considered as a quiver) is shown in Figure \ref{fig:Figure14}. Figure \ref{fig:Figure15} shows each morphism/edge tagged with the number of hypergraph transitions necessary to perform the corresponding computation, illustrating that the composition ${\circ}$ and tensor product ${\otimes}$ operations are not purely additive (although they are both somewhat close), thus indicating that the system is largely (multi)computationally irreducible, but not entirely so. The construction with each vertex/object tagged with its step number and each edge/morphism tagged with a list of step numbers traversed throughout the course of its corresponding computation is shown in Figure \ref{fig:Figure16}. As expected, we see that the hypergraph rewriting system is more computationally reducible than multicomputationally so (i.e. the composition operation ${\circ}$ is distorted more by the map ${Z^{\prime}}$ than the tensor product operation ${\otimes}$): a consequence of the fact that the state equivalence function (based on hypergraph isomorphism) is more computationally complex than it was for the non-deterministic Turing machine case previously.

The usual setting in which double-pushout rewriting takes place is within an \textit{adhesive category}\cite{lack}, and thus adhesivity is the usual condition that one would impose on the category ${\mathcal{C}}$ described above; loosely speaking, adhesivity provides sufficient conditions for pushouts to be ``glued'' along monomorphisms in the necessary manner for double-pushout diagrams to be constructed. More formally, a category is adhesive if it has \textit{pullbacks}, and all pushouts along monomorphisms satisfy the \textit{van-Kampen square} condition. Having pullbacks simply entails that, for some \textit{cospan} (i.e. a pair of morphisms with a common codomain):

\begin{equation}
\begin{tikzcd}
Y \arrow[r, "f"] & X & Z \arrow[swap, l, "g"]
\end{tikzcd}
\end{equation}
in ${\mathcal{C}}$, there exists an object ${P \in \mathrm{ob} \left( \mathcal{C} \right)}$ and a pair of morphisms ${\left( f^{\prime} : P \to Z \right), \left( g^{\prime} : P \to Y \right) \in \mathrm{hom} \left( \mathcal{C} \right)}$ such that the following square commutes:

\begin{equation}
\begin{tikzcd}
P \arrow[r, "f^{\prime}"] \arrow[swap, d, "g^{\prime}"] & Z \arrow[d, "g"]\\
Y \arrow[swap, r, "f"] & X,
\end{tikzcd}
\end{equation}
i.e:

\begin{equation}
\left( f \circ g^{\prime} : P \to X \right) = \left( g \circ f^{\prime} : P \to X \right),
\end{equation}
and that are universal in the sense that, for any object ${P^{*} \in \mathrm{ob} \left( \mathcal{C} \right)}$ equipped with morphisms:

\begin{equation}
\left( f^{*} : P^{*} \to Z \right), \left( g^{*} : P^{*} \to Y \right) \in \mathrm{hom} \left( \mathcal{C} \right),
\end{equation}
there exists a unique morphism ${\left( u : P^{*} \to P \right) \in \mathrm{hom} \left( \mathcal{C} \right)}$ such that:

\begin{equation}
\left( f^{*} : P^{*} \to Z \right) = \left( f^{\prime} \circ u : P^{*} \to Z \right), \qquad \text{ and } \qquad \left( g^{*} : P^{*} \to Y \right) = \left( g^{\prime} \circ u : P^{*} \to Y \right),
\end{equation}
i.e. the following diagram commutes:

\begin{equation}
\begin{tikzcd}
\forall P^{*} \arrow[drr, bend left, "\forall f^{*}"] \arrow[swap, ddr, bend right, "\forall g^{*}"] \arrow[dr, dashed, "\exists ! u"] & &\\
& P \arrow[r, "f^{\prime}"] \arrow[swap, d, "g^{\prime}"] & Z \arrow[d, "g"]\\
& Y \arrow[swap, r, "f"] & X.
\end{tikzcd}
\end{equation}
Though the condition of having pullbacks along cospans is relatively straightforward to state and understand, the van-Kampen square condition on pushouts along monomorphisms is, on the other hand, far more opaque and technical. It asserts that, if the object ${W \in \mathrm{ob} \left( \mathcal{C} \right)}$ and the morphisms ${\left( f^{\prime} : Y \to W \right), \left( g^{\prime} : Z \to W \right) \in \mathrm{hom} \left( \mathcal{C} \right)}$ in the following diagram:

\begin{equation}
\begin{tikzcd}
X \arrow[r, "f"] \arrow[swap, d, "g"] & Z \arrow[d, "g^{\prime}"] \arrow[ddr, bend left, "\forall g^{*}"] &\\
Y \arrow[swap, r, "f^{\prime}"] \arrow[swap, drr, bend right, "\forall f^{*}"] & W \arrow[dr, dashed, "\exists ! u"] &\\
& & \forall W^{*},
\end{tikzcd}
\end{equation}
constitute a pushout of the span ${\left( f : X \to Z \right), \left( g : X \to Y \right) \in \mathrm{hom} \left( \mathcal{C} \right)}$, i.e:

\begin{equation}
\left( f^{\prime} \circ g : X \to W \right) = \left( g^{\prime} \circ f : X \to W \right),
\end{equation}
and:

\begin{multline}
\forall W^{*} \in \mathrm{ob} \left( \mathcal{C} \right), \qquad \forall \left( f^{*} : Y \to W^{*} \right), \left( g^{*} : Z \to W^{*} \right) \in \mathrm{hom} \left( \mathcal{C} \right),\\
\text{ such that } \qquad \left( f^{*} \circ g : X \to W^{*} \right) = \left( g^{*} \circ f : X \to W^{*} \right), \qquad \exists ! \left( u : W \to W^{*} \right) \in \mathrm{hom} \left( \mathcal{C} \right),
\end{multline}
such that:

\begin{equation}
\left( f^{*} : Y \to W^{*} \right) = \left( u \circ f^{\prime} : Y \to W^{*} \right), \qquad \text{ and } \qquad \left( g^{*} : Z \to W^{*} \right) = \left( u \circ g^{\prime} : Z \to W^{*} \right),
\end{equation}
then that pushout is a van-Kampen square if and only if, for every commutative cube of the form:

\begin{equation}
\begin{tikzcd}
X^{\prime} \arrow[rrr, "f_h"] \arrow[swap, ddd, "g_h"] \arrow[dr, "h_x"] & & & Z^{\prime} \arrow[swap, dl, "h_Z"] \arrow[ddd, "g_{h}^{\prime}"]\\
& X \arrow[r, "f"] \arrow[swap, d, "g"] & Z \arrow[d, "g^{\prime}"] &\\
& Y \arrow[swap, r, "f^{\prime}"] & W &\\
Y^{\prime} \arrow[ur, "h_Y"] \arrow[swap, rrr, "f_{h}^{\prime}"] & & & W^{\prime} \arrow[swap, ul, "h_W"],
\end{tikzcd}
\end{equation}
for which the top and left faces are pullback squares, in other words if and only if the object ${X^{\prime} \in \mathrm{ob} \left( \mathcal{C} \right)}$ together with the morphisms ${\left( h_X : X^{\prime} \to X \right), \left( g_h : X^{\prime} \to Y{^\prime} \right) \in \mathrm{hom} \left( \mathcal{C} \right)}$, and the object ${X^{\prime} \in \mathrm{ob} \left( \mathcal{C} \right)}$ together with the morphisms ${\left( f_h : X^{\prime} \to Z^{\prime} \right), \left( h_X : X^{\prime} \to X \right) \in \mathrm{hom} \left( \mathcal{C} \right)}$ in the following pair of diagrams:

\begin{equation}
\begin{tikzcd}
& & \forall X^{*} \arrow[swap, dll, bend right, "\forall g_{h}^{*}"] \arrow[ddl, bend left, "\forall h_{X}^{*}"] \arrow[swap, dl, dashed, "\exists ! u"]\\
Y^{\prime} \arrow[swap, d, "h_Y"] & X^{\prime} \arrow[swap, l, "g_h"] \arrow[d, "h_X"] &\\
Y & X \arrow[l, "g"] &,
\end{tikzcd} \qquad \text{ and } \qquad
\begin{tikzcd}
\forall X^{*} \arrow[drr, bend left, "\forall f_{h}^{*}"] \arrow[swap, ddr, bend right, "\forall h_{X}^{*}"] \arrow[dr, dashed, "\exists ! u"] & &\\
& X^{\prime} \arrow[r, "f_h"] \arrow[swap, d, "h_X"] & Z^{\prime} \arrow[d, "h_Z"]\\
& X \arrow[swap, r, "f"] & Z,
\end{tikzcd}
\end{equation}
constitute pullbacks of the cospans ${\left( g : X \to Y \right), \left( h_Y : Y^{\prime} \to Y \right) \in \mathrm{hom} \left( \mathcal{C} \right)}$ and ${\left( f : X \to Z \right), \left( h_Z : Z^{\prime} \to Z \right) \in \mathrm{hom} \left( \mathcal{C} \right)}$, respectively, i.e:

\begin{equation}
\left( g \circ h_X : X^{\prime} \to Y \right) = \left( h_Y \circ g_h : X^{\prime} \to Y \right), \qquad \text{ and } \qquad \left( f \circ h_X : X^{\prime} \to Z \right) = \left( h_Z \circ f_h : X^{\prime} \to Z \right),
\end{equation}
and, moreover:

\begin{multline}
\forall X^{\prime} \in \mathrm{ob} \left( \mathcal{C} \right), \qquad \forall \left( g_{h}^{*} : X^{*} \to Y^{\prime} \right), \left( h_{X}^{*} : X^{*} \to X \right) \in \mathrm{hom} \left( \mathcal{C} \right),\\
\text{ such that } \qquad \left( g \circ h_{X}^{*} : X^{*} \to Y \right) = \left( h_Y \circ g_{h}^{*} : X^{*} \to Y \right), \qquad \exists ! \left( u : X^{*} \to X^{\prime} \right) \in \mathrm{hom} \left( \mathcal{C} \right),
\end{multline}
such that:

\begin{equation}
\left( h_{X}^{*} : X^{*} \to X \right) = \left( h_X \circ u : X^{*} \to X \right), \qquad \text{ and } \qquad \left( g_{h}^{*} : X^{*} \to Y^{\prime} \right) = \left( g_h \circ u : X^{*} \to Y^{\prime} \right),
\end{equation}
for the case of the first (leftmost) diagram, and:

\begin{multline}
\forall X^{\prime} \in \mathrm{ob} \left( \mathcal{C} \right), \qquad \forall \left( f_{h}^{*} : X^{*} \to Z^{\prime} \right), \left( h_{X}^{*} : X^{*} \to X \right) \in \mathrm{hom} \left( \mathcal{C} \right),\\
\text{ such that } \qquad \left( f \circ h_{X}^{*} : X^{*} \to Z \right) = \left( h_Z \circ f_{h}^{*} : X^{*} \to Z \right), \qquad \exists ! \left( u : X^{*} \to X^{\prime} \right) \in \mathrm{hom} \left( \mathcal{C} \right),
\end{multline}
such that:

\begin{equation}
\left( h_{X}^{*} : X^{*} \to X \right) = \left( h_X \circ u : X^{*} \to X \right), \qquad \text{ and } \qquad \left( f_{h}^{*} : X^{*} \to Z^{\prime} \right) = \left( f_h \circ u : X^{*} \to Z^{\prime} \right),
\end{equation}
for the case of the second (rightmost) diagram, then a certain compatibility condition is satisfied between the pushout and pullback squares. In detail, this compatibility condition states that the rear face is a pushout square, in other words the object ${W^{\prime} \in \mathrm{ob} \left( \mathcal{C} \right)}$ and the morphisms ${\left( f_{h}^{\prime} : Y^{\prime} \to W^{\prime} \right), \left( g_{h}^{\prime} : Z^{\prime} \to W^{\prime} \right) \in \mathrm{hom} \left( \mathcal{C} \right)}$ in the following diagram:

\begin{equation}
\begin{tikzcd}
X^{\prime} \arrow[r, "f_h"] \arrow[swap, d, "g_h"] & Z^{\prime} \arrow[d, "g_{h}^{\prime}"] \arrow[ddr, bend left, "\forall g_{h}^{*}"]&\\
Y^{\prime} \arrow[swap, r, "f_{h}^{\prime}"] \arrow[swap, drr, bend right, "\forall f_{h}^{*}"] & W^{\prime} \arrow[dr, dashed, "\exists ! u"] &\\
& & \forall W^{*},
\end{tikzcd}
\end{equation}
constitute a pushout of the span ${\left( f_h : X^{\prime} \to Z^{\prime} \right), \left( g_h : X^{\prime} \to Y^{\prime} \right) \in \mathrm{hom} \left( \mathcal{C} \right)}$, i.e:

\begin{equation}
\left( f_{h}^{\prime} \circ g_h : X^{\prime} \to W^{\prime} \right) = \left( g_{h}^{\prime} \circ f_h : X^{\prime} \to W^{\prime} \right),
\end{equation}
and:

\begin{multline}
\forall W^{*} \in \mathrm{ob} \left( \mathcal{C} \right), \qquad \forall \left( f_{h}^{*} : Y^{\prime} \to W^{*} \right), \left( g_{h}^{*} : Z^{\prime} \to W^{*} \right) \in \mathrm{hom} \left( \mathcal{C} \right),\\
\text{ such that } \qquad \left( f_{h}^{*} \circ g_h : X^{\prime} \to W^{*} \right) = \left( g_{h}^{*} \circ f_h : X^{\prime} \to W^{*} \right), \qquad \exists ! \left( u : W^{\prime} \to W^{*} \right),
\end{multline}
such that:

\begin{equation}
\left( f_{h}^{*} : Y^{\prime} \to W^{*} \right) = \left( u \circ f_{h}^{\prime} : ^{\prime} \to W^{*} \right), \qquad \text{ and } \qquad \left( g_{h}^{*} : Z^{\prime} \to W^{*} \right) = \left( u \circ g_{h}^{\prime} : Z^{\prime} \to W^{*} \right),
\end{equation}
if and only if the bottom and right faces are pullback squares, in other words if and only if the object ${Y^{\prime} \in \mathrm{ob} \left( \mathcal{C} \right)}$ together with the morphisms ${\left( f_{h}^{\prime} : Y^{\prime} \to W^{\prime} \right), \left( h_Y : Y^{\prime} \to Y \right) \in \mathrm{hom} \left( \mathcal{C} \right)}$, and the object ${Z^{\prime} \in \mathrm{ob} \left( \mathcal{C} \right)}$ together with the morphisms ${\left( g_{h}^{\prime} : Z^{\prime} \to W^{\prime} \right), \left( h_Z : Z^{\prime} \to Z \right) \in \mathrm{hom} \left( \mathcal{C} \right)}$ in the following pair of diagrams:

\begin{equation}
\begin{tikzcd}
& Y \arrow[r, "f^{\prime}"] & W\\
& Y^{\prime} \arrow[u, "h_Y"] \arrow[swap, r, "f_{h}^{\prime}"] & W^{\prime} \arrow[swap, u, "h_W"]\\
\forall Y^{*} \arrow[uur, bend left, "\forall h_{Y}^{*}"] \arrow[swap, urr, bend right, "\forall f_{h}^{*}"] \arrow[ur, dashed, "\exists ! u"]
\end{tikzcd} \qquad \text{ and } \qquad
\begin{tikzcd}
W & Z \arrow[swap, l, "g^{\prime}"] &\\
W^{\prime} \arrow[u, "h_W"] & Z^{\prime} \arrow[swap, u, "h_Z"] \arrow[l, "g_{h}^{\prime}"] &\\
& & \forall Z^{*} \arrow[ull, bend left, "\forall g_{h}^{*}"] \arrow[swap, uul, bend right, "\forall h_{Z}^{*}"] \arrow[swap, ul, dashed, "\exists ! u"],
\end{tikzcd}
\end{equation}
constitute pullbacks of the cospans ${\left( f^{\prime} : Y \to W \right), \left( h_W : W^{\prime} \to W \right) \in \mathrm{hom} \left( \mathcal{C} \right)}$ and ${\left( g^{\prime} : Z \to W \right), \left( h_W : W^{\prime} \to W \right) \in \mathrm{hom} \left( \mathcal{C} \right)}$, respectively, i.e:

\begin{equation}
\left( h_W \circ f_{h}^{\prime} : Y^{\prime} \to W \right) = \left( f^{\prime} \circ h_Y : Y^{\prime} \to W \right), \qquad \text{ and } \qquad \left( h_W \circ g_{h}^{\prime} : Z^{\prime} \to W \right) = \left( g^{\prime} \circ h_Z : Z^{\prime} \to W \right),
\end{equation}
and, moreover:

\begin{multline}
\forall Y^{*} \in \mathrm{ob} \left( \mathcal{C} \right), \qquad \forall \left( f_{h}^{*} : Y^{*} \to W^{\prime} \right), \left( h_{Y}^{*} : Y^{*} \to Y \right) \in \mathrm{hom} \left( \mathcal{C} \right),\\
\text{ such that } \qquad \left( h_W \circ f_{h}^{*} : Y^{*} \to W \right) = \left( f^{\prime} \circ h_{Y}^{*} : Y^{*} \to W \right), \qquad \exists ! \left( u : Y^{*} \to Y^{\prime} \right) \in \mathrm{hom} \left( \mathcal{C} \right),
\end{multline}
such that:

\begin{equation}
\left( f_{h}^{*} : Y^{*} \to W^{\prime} \right) = \left( f_{h}^{\prime} \circ u : Y^{*} \to W^{\prime} \right), \qquad \text{ and } \qquad \left( h_{Y}^{*} : Y^{*} \to Y \right) = \left( h_Y \circ u : Y^{*} \to Y \right),
\end{equation}
for the case of the first (leftmost) diagram, and:

\begin{multline}
\forall Z^{*} \in \mathrm{ob} \left( \mathcal{C} \right), \qquad \forall \left( g_{h}^{*} : Z^{*} \to W^{\prime} \right), \left( h_{Z}^{*} : Z^{*} \to Z \right) \in \mathrm{hom} \left( \mathcal{C} \right),\\
\text{ such that } \qquad \left( h_W \circ g_{h}^{*} : Z^{*} \to W \right) = \left( g^{\prime} \circ h_{Z}^{*} : Z^{*} \to W \right), \qquad \exists ! \left( u : Z^{*} \to Z^{\prime} \right) \in \mathrm{hom} \left( \mathcal{C} \right),
\end{multline}
such that:

\begin{equation}
\left( g_{h}^{*} : Z^{*} \to W^{\prime} \right) = \left( g_{h}^{\prime} \circ u : Z^{*} \to W^{\prime} \right), \qquad \text{ and } \qquad \left( h_{Z}^{*} : Z^{*} \to Z \right) = \left( h_Z \circ u : Z^{*} \to Z \right).
\end{equation}
for the case of the second (rightmost) diagram. Although the category of hypergraphs and subhypergraph inclusion maps considered in the case of Wolfram model evolution is not strictly an adhesive category (due to the arbitrary connectivity of vertices within each hyperedge, implying that not all pushouts along monomorphisms are guaranteed to exist), as noted by Kissinger\cite{kissinger}\cite{kissinger2}\cite{kissinger3} it constitutes a full subcategory of an adhesive category, and can be ``embedded'' in the ambient adhesive category in such a way as to inherit sufficient ``adhesivity'' to allow double-pushout rewriting to be performed (through the mechanisms of either \textit{selective adhesivity} or \textit{partial adhesivity} - essentially by allowing the functor that embeds the subcategory into the adhesive category to preserve monomorphisms).

As detailed in \cite{gorard7}, one can construct both the ordinary composition structure and the symmetric monoidal structure of the resulting category ${\mathcal{T}}$ of hypergraphs and hypergraph rewritings in a very explicit way using a combination of the \textit{concurrency} and \textit{parallelism} theorems from algebraic graph transformation theory. Specifically, if one has a pair of \textit{hypergraph productions} ${p_1}$ and ${p_2}$ (i.e. two spans of monomorphisms corresponding to two hypergraph rewrites) rewriting hypergraph $G$ to $H$ and hypergraph $H$ to ${G^{\prime}}$ respectively (known as an \textit{$E$-related transformation sequence}), then the concurrency theorem allows one to compose the productions to obtain an \textit{$E$-concurrent} hypergraph production ${p_1 *_{E} p_2}$ of the form:

\begin{equation}
\begin{tikzcd}
& H \arrow[dr, Rightarrow, "p_2"] &\\
G \arrow[ur, Rightarrow, "p_1"] \arrow[swap, rr, Rightarrow, "p_1 *_{E} p_2"] & & G^{\prime},
\end{tikzcd}
\end{equation}
thus giving rise to the ordinary (sequential) composition of morphisms ${\circ}$ in ${\mathcal{T}}$. On the other hand, if one has a pair of hypergraph productions ${p_1}$ and ${p_2}$ yielding two different, \textit{sequentially-independent transformation sequences} $G$ to ${H_1}$ to ${G^{\prime}}$ and $G$ to ${H_2}$ to ${G^{\prime}}$ (obtained by applying ${p_1}$ then ${p_2}$ vs. ${p_2}$ then ${p_1}$), then the parallelism theorem allows one to compose the productions to obtain a \textit{parallel hypergraph production} ${p_1 + p_2}$ of the form:

\begin{equation}
\begin{tikzcd}
& G \arrow[swap, dl, Rightarrow, "p_1"] \arrow[dr, Rightarrow, "p_2"] \arrow[dd, Rightarrow, "p_1 + p_2"] &\\
H_1 \arrow[swap, dr,Rightarrow, "p_2"] & & H_2 \arrow[dl, Rightarrow, "p_1"]\\
& G^{\prime} &,
\end{tikzcd}
\end{equation}
thus giving rise to the tensor product (parallel) composition of morphisms ${\otimes}$ in ${\mathcal{T}}$, as required.

\section{Correspondence with Categorical Quantum Mechanics and Functorial Quantum Field Theory}
\label{sec:Section3}

In conventional (non-relativistic) quantum mechanics, one typically associates to each moment of time ${t \in \mathbb{R}}$ a corresponding (Hilbert) space of states ${V_t}$ for the system, and to every interval of time ${\left[ t_1, t_2 \right]}$, where ${t_1, t_2 \in \mathbb{R}}$ such that ${t_1 \leq t_2}$, a corresponding linear (unitary) time evolution operator ${\hat{U} \left( t_1, t_2 \right) : V_{t_1} \to V_{t_2}}$. If the Hamiltonian ${\hat{H} \left( t \right)}$ is a Hermitian/self-adjoint operator that depends smoothly on ${t \in \mathbb{R}}$, then we can express ${\hat{U} \left( t_1, t_2 \right)}$ explicitly in terms of the following Dyson formula (otherwise known more generally as an iterated integral expansion for parallel transport) for the time-dependent Schr\"odinger equation:

\begin{equation}
\hat{U} \left( t_1, t_2 \right) = P \exp \left( \frac{i}{\hbar} \int_{t_1}^{t_2} \hat{H} \left( t \right) d t \right),
\end{equation}
where ${P \exp}$ denotes the path-ordered exponential operator for non-commutative algebras; for the time-independent case in which the Hamiltonian ${\hat{H}}$ is fixed, this simplifies to just an ordinary exponential:

\begin{equation}
\forall t \in \mathbb{R}, \qquad \hat{H} \left( t \right) = \hat{H}, \qquad \implies \qquad \hat{U} \left( t_1, t_2 \right) = \exp \left( \frac{i}{\hbar} \left( t_2 - t_1 \right) \hat{H} \right).
\end{equation}
This explicit representation of ${\hat{U} \left( t_1, t_2 \right)}$ makes manifest one of the fundamental features of quantum mechanical time evolution: that it is necessarily \textit{local} in time. In other words, due to the linearity of integration, the ``global'' time interval ${\left[ t_1, t_2 \right]}$ can always be subdivided into many ``local'' subintervals, in such a way that the single global time evolution is obtained by integrating up the effects of the many local time evolutions. More formally, we have:

\begin{equation}
\forall t_1, t_2, t_3 \in \mathbb{R}, \qquad \text{ such that } \qquad t_1 \leq t_2 \leq t_3, \qquad \hat{U} \left( t_1, t_3 \right) = \hat{U} \left( t_2, t_3 \right) \circ \hat{U} \left( t_1, t_2 \right),
\end{equation}
in other words, time evolution has a natural composition structure ${\circ}$ such that, when combined with the (mostly, though not entirely, innocent) condition that:

\begin{equation}
\forall t \in \mathbb{R}, \qquad \hat{U} \left( t, t \right) = id_{V_t},
\end{equation}
we see that time evolution in quantum mechanics satisfies the requisite axioms of a category (with associativity inherited from the associativity of products of linear operators); this was ultimately the insight underlying Abramsky and Coecke's formulation of \textit{categorical quantum mechanics}\cite{abramsky}\cite{abramsky2}. Thus, if ${\mathbf{Bord}_{1}^{Riem}}$ refers to the 1-dimensional category of (Riemannian) manifolds and their cobordisms, and ${\mathbf{Vect}}$ refers to the category of vector spaces and their linear isomorphisms, then the statement of locality of time evolution in quantum mechanics simply becomes a statement that the map:

\begin{equation}
Z : \mathbf{Bord}_{1}^{Riem} \to \mathbf{Vect},
\end{equation}
is a functor, i.e. (at least in this context) locality \textit{is} functoriality. This functoriality between the categories ${\mathbf{Bord}_{1}^{Riem}}$ and ${\mathbf{Vect}}$ can be illustrated diagrammatically as follows:

\begin{equation}
\begin{tikzcd}
& & t_2 \arrow[swap, ddrr, "{\left[ t_2, t_3 \right]}"'{name=g}] & & & & & & & V_{t_2} \arrow[ddrr, "{\hat{U} \left( t_2, t_3 \right)}"] & &\\\\
t_1 \arrow[uurr, "{\left[ t_1, t_2 \right]}"] \arrow[swap, rrrr, "\substack{\left[ t_2, t_3 \right] \cup \left[ t_1, t_2 \right] \\ = \left[ t_1, t_3 \right]}"] & & & & t_3 & & & V_{t_1} \arrow[swap, uurr, "{\hat{U} \left( t_1, t_2 \right)}"'{name=f}] \arrow[swap, rrrr, "\substack{\hat{U} \left( t_2, t_3 \right) \circ \hat{U} \left( t_1, t_2 \right) \\ = \hat{U} \left( t_1, t_3 \right)}"] & & & & V_{t_3}.
\arrow[mapsto, from=g, to=f, shorten=4em, "Z"].
\end{tikzcd}
\end{equation}
If we think of the (Hilbert) spaces of states ${V_t}$ as being data structures, and the unitary time evolution operators ${\hat{U} \left( t_1, t_2 \right)}$ as being elementary computations (as is the case in, for instance, quantum information theory, where they represent the actions of compositions of quantum gates), then this looks structurally very similar to the continuous case of the functor ${Z^{\prime} : \mathcal{T} \to \mathcal{B}}$ from a category of data structures and computations to a category of manifolds and cobordisms, namely:

\begin{equation}
\begin{tikzcd}
& & Y \arrow[swap, ddrr, "g"'{name=g}] & & & & & & & t_2 \arrow[ddrr, "{\left[ t_2, t_3 \right]}"] & &\\\\
X \arrow[uurr, "f"] \arrow[swap, rrrr, "g \circ f"] & & & & Z & & & t_1 \arrow[swap, uurr, "{\left[ t_1, t_2 \right]}"'{name=f}] \arrow[swap, rrrr, "\substack{\left[ t_2, t_3 \right] \cup \left[ t_1, t_2 \right] \\ = \left[ t_1, t_3 \right]}"] & & & & t_3,
\arrow[mapsto, from=g, to=f, shorten=5em, "Z^{\prime}"]
\end{tikzcd}
\end{equation}
considered in the preceding sections, albeit with the domain and the codomain categories swapped around.

In the most general case of a functorial quantum mechanics theory, defined by some functor from cobordisms to computations ${Z: \mathcal{B} \to \mathcal{T}}$, although it is not necessarily the case that $Z$ will always be a strict inverse of ${Z^{\prime} : \mathcal{T} \to \mathcal{B}}$, the two functors will at least be \textit{adjoint} to one another.  Formally, the statement that ${Z^{\prime} : \mathcal{T} \to \mathcal{B}}$ is \textit{left adjoint} to ${Z : \mathcal{B} \to \mathcal{T}}$ corresponds to the assertion that, for every object (manifold) ${X \in \mathrm{ob} \left( \mathcal{B} \right)}$, there exists a \textit{universal morphism} (cobordism) ${\left( \varepsilon_X : Z^{\prime} \left( Z \left( X \right) \right) \to X \right) \in \mathrm{hom} \left( \mathcal{B} \right)}$ from ${Z^{\prime}}$ to $X$ for some object (data structure) ${Z \left( X \right) \in \mathrm{ob} \left( \mathcal{T} \right)}$, where the universality property necessitates that, for any object (data structure) ${Y \in \mathrm{ob} \left( \mathcal{T} \right)}$ and any morphism (cobordism) ${\left( f : Z^{\prime} \left( Y \right) \to X \right) \in \mathrm{hom} \left( \mathcal{B} \right)}$, there exists a unique morphism (computation) ${\left( g : Y \to Z \left( X \right) \right) \in \mathrm{hom} \left( \mathcal{T} \right)}$ such that:

\begin{equation}
\left( \varepsilon_X \circ Z^{\prime} \left( g \right) : Z^{\prime} \left( Y \right) \to X \right) = \left( f : Z^{\prime} \left( Y \right) \to X \right).
\end{equation}
This condition may be restated succinctly via the following commutative diagram:

\begin{equation}
\begin{tikzcd}
Z^{\prime} \left( \forall Y \right) \arrow[swap, dd, dashed, "Z^{\prime} \left( \exists ! g \right)"] \arrow[ddrr, "\forall f"] & &\\\\
Z^{\prime} \left( \exists Z \left( X \right) \right) \arrow[swap, rr, "\exists \varepsilon_X"] & & \forall X.
\end{tikzcd}
\end{equation}
On the other hand, the statement that ${Z : \mathcal{B} \to \mathcal{T}}$ is \textit{right adjoint} to ${Z^{\prime} : \mathcal{T} \to \mathcal{B}}$ corresponds to the assertion that, for every object (data structure) ${Y \in \mathrm{ob} \left( \mathcal{T} \right)}$, there exists a \textit{universal morphism} (computation) ${\left( \eta_Y : Y \to Z \left( Z^{\prime} \left( Y \right) \right) \right) \in \mathrm{hom} \left( \mathcal{T} \right)}$ from $Y$ to $Z$ for some object (manifold) ${Z^{\prime} \left( Y \right) \in \mathrm{ob} \left( \mathcal{B} \right)}$, where the universality property necessitates that, for any object (manifold) ${X \in \mathrm{ob} \left( \mathcal{B} \right)}$ and any morphism (computation) ${\left( g : Y \to Z \left( X \right) \right) \in \mathrm{hom} \left( \mathcal{T} \right)}$, there exists a unique morphism (cobordism) ${\left( f : Z^{\prime} \left( Y \right) \to X \right) \in \mathrm{hom} \left( \mathcal{B} \right)}$ such that:

\begin{equation}
\left( Z \left( f \right) \circ \eta_Y : Y \to Z \left( X \right) \right) = \left( g : Y \to Z \left( X \right) \right).
\end{equation}
This condition may be restated succinctly via the following commutative diagram:

\begin{equation}
\begin{tikzcd}
\forall Y \arrow[rr, "\exists \eta_Y"] \arrow[ddrr, "\forall g"] & & Z \left( \exists Z^{\prime} \left( Y \right) \right) \arrow[dd, dashed, "Z \left( \exists ! f \right)"]\\\\
& & Z \left( \forall X \right).
\end{tikzcd}
\end{equation}
It is in this rather precise sense that we are able to claim that the irreducibility of computations in computational complexity theory is dual/adjoint to the locality of time evolution in categorical quantum mechanics: for any functor ${Z^{\prime} : \mathcal{T} \to \mathcal{B}}$ describing an irreducible computation, we can uniquely construct a corresponding functor ${Z : \mathcal{B} \to \mathcal{T}}$ describing a local quantum time evolution such that:

\begin{multline}
\forall X, X^{\prime} \in \mathrm{ob} \left( \mathcal{B} \right), \qquad \forall \left( f: X^{\prime} \to X \right) \in \mathrm{hom} \left( \mathcal{B} \right),\\
\left( \varepsilon_X \circ Z^{\prime} \left( Z \left( f \right) \right) : Z^{\prime} \left( Z \left( X^{\prime} \right) \right) \to X \right) = \left( f \circ \varepsilon_{X^{\prime}} : Z^{\prime} \left( Z \left( X^{\prime} \right) \right) \to X \right),
\end{multline}
and, conversely, for any functor ${Z : \mathcal{B} \to \mathcal{T}}$ describing a local quantum time evolution, we can uniquely construct a corresponding functor ${Z^{\prime} : \mathcal{T} \to \mathcal{B}}$ describing an irreducible computation such that:

\begin{multline}
\forall Y, Y^{\prime} \in \mathrm{ob} \left( \mathcal{T} \right), \qquad \forall \left( g : Y \to Y^{\prime} \right) \in \mathrm{hom} \left( \mathcal{T} \right),\\
\left( Z \left( Z^{\prime} \left( g \right) \right) \circ \eta_Y : Y \to Z \left( Z^{\prime} \left( Y^{\prime} \right) \right) \right) = \left( \eta_{Y^{\prime}} \circ g : Y \to Z \left( Z^{\prime} \left( Y^{\prime} \right) \right) \right),
\end{multline}
as described by the following pair of commutative diagrams:

\begin{equation}
\begin{tikzcd}
Z^{\prime} \left( Z \left( X^{\prime} \right) \right) \arrow[rr, "Z^{\prime} \left( Z \left( f \right) \right)"] \arrow[swap, dd, "\varepsilon_{X^{\prime}}"] & & Z^{\prime} \left( Z \left( X \right) \right) \arrow[dd, "\varepsilon_X"]\\\\
X^{\prime} \arrow[swap, rr, "f"] & & X,
\end{tikzcd} \qquad \text{ and } \qquad
\begin{tikzcd}
Y \arrow[rr, "\eta_Y"] \arrow[swap, dd, "g"] & & Z \left( Z^{\prime} \left( Y \right) \right) \arrow[dd, "Z \left( Z^{\prime} \left( g \right) \right)"]\\\\
Y^{\prime} \arrow[swap, rr, "\eta_{Y^{\prime}}"] & & Z \left( Z^{\prime} \left( Y^{\prime} \right) \right).
\end{tikzcd}
\end{equation}
Somewhat more cryptically, this enables us to make the claim that, in this very restricted sense, computational complexity theory is dual/adjoint to (non-relativistic) quantum mechanics.

In the non-deterministic/multicomputational case, in which categories ${\mathcal{T}}$ and ${\mathcal{B}}$ are both equipped with a (symmetric) monoidal structure, and thus in which the adjoint functors ${Z^{\prime}}$ and $Z$ are now (symmetric) monoidal functors:

\begin{equation}
Z^{\prime} : \left\langle T, \otimes, I \right\rangle \to \left\langle \mathcal{B}, \oplus, \varnothing \right\rangle, \qquad \text{ and } \qquad Z : \left\langle \mathcal{B}, \oplus, \varnothing \right\rangle \to \left\langle \mathcal{T}, \otimes, I \right\rangle,
\end{equation}
we obtain a higher-dimensional analog of the time evolution functor for non-relativistic quantum mechanics on the right-hand side of the adjunction. For the example considered initially, in which the category ${\mathcal{T}}$ of data structures and computations is abstractly represented as a category of vector spaces and linear isomorphisms (with the tensor product operation given by the usual tensor product of vector spaces), this functor consequently takes the form:

\begin{equation}
Z : \mathbf{Bord}_{d}^{Riem} \to \mathbf{Vect},
\end{equation}
where ${d > 1}$. In the context of functorial approaches to quantum field theory (e.g. in topological field theories or 2-dimensional conformal field theories)\cite{schreiber}\cite{baez3}, such a functor plays the role of a \textit{propagator}, i.e. the Lorentz-invariant analog of the time evolution functor from non-relativistic quantum mechanics: to every codimension-1 spacelike hypersurface ${M_{d - 1} \in \mathrm{ob} \left( \mathbf{Bord}_{d}^{Riem} \right)}$, this functor assigns a corresponding vector space ${Z \left( M_{d - 1} \right) \in \mathrm{ob} \left( \mathbf{Vect} \right)}$ designating the space of states over that hypersurface, and to every cobordism/spacetime/worldvolume ${M \in \mathrm{hom} \left( \mathbf{Bord}_{d}^{Riem} \right)}$ with boundaries ${\partial \left( M \right)}$, i.e.:

\begin{equation}
\left( M : \partial_{in} \left( M \right) \to \partial_{out} \left( M \right) \right) \in \mathrm{hom} \left( \mathbf{Bord}_{d}^{Riem} \right),
\end{equation}
this functor assigns a corresponding linear isomorphism:

\begin{equation}
\left( Z \left( M \right) : Z \left( \partial_{in} \left( M \right) \right) \to Z \left(  \partial_{out} \left( M \right) \right) \right) \in \mathrm{hom} \left( \mathbf{Vect} \right),
\end{equation}
designating the propagator/\textit{scattering amplitude}/\textit{S-matrix} for a process of shape $M$ mapping from hypersurface ${\partial_{in} \left( M \right)}$ to hypersurface ${\partial_{out} \left( M \right)}$.  Hence, just as computational irreducibility may be said to be formally dual/adjoint to locality of time evolution in quantum mechanics, multicomputational irreducibility may be said to be formally dual/adjoint to adherence to the Atiyah-Segal \textit{sewing laws}\cite{atiyah}\cite{atiyah2}\cite{segal} in functorial quantum field theory, under which the path integral over a domain ${\Sigma}$ which can be decomposed into subdomains ${\Sigma_1}$ and ${\Sigma_2}$ (i.e. ${\Sigma = \Sigma_1 \sqcup \Sigma_2}$) must be equal to the path integral over subdomain ${\Sigma_1}$ composed with the path integral over subdomain ${\Sigma_2}$.

Conventionally, the symmetric monoidal functors ${Z : \mathbf{Bord}_{1}^{Riem} \to \mathbf{Vect}}$ and ${Z : \mathbf{Bord}_{d}^{Riem} \to \mathbf{Vect}}$ that define time evolution in category quantum mechanics and functorial quantum field theory are assumed to be \textit{strong} (in the sense described previously that the coherence maps ${\epsilon}$ and ${\mu_{X, Y}}$ are isomorphisms for all ${X, Y \in \mathrm{ob} \left( \mathbf{Vect} \right)}$). However, simply preserving the tensor product structure of cobordisms/vector spaces is not sufficient to define a truly physical theory of quantum mechanics or quantum fields: at the very least, one must also preserve the \textit{dagger structure} of time evolution (which generalizes the Hermitian adjoint operation on linear transformations, otherwise known as the conjugate transpose in the finite-dimensional case), as well as the \textit{compact structure} of vector spaces (which generalizes the operation of taking duals of a finite-dimensional vector space). In this setting, we say that ${\mathbf{Vect}}$ is a \textit{dagger category}\cite{burgin}\cite{lambek} to mean that it is equipped with an \textit{involutive contravariant endofunctor} ${\dagger : \mathbf{Vect} \to \mathbf{Vect}}$, i.e. a functor from ${\mathbf{Vect}}$ to itself which has the effect of swapping the sources and targets of each morphism:

\begin{equation}
\forall \left( f : X \to Y \right) \in \mathrm{hom} \left( \mathbf{Vect} \right), \qquad \exists \left( f^{\dagger} : Y \to X \right) \in \mathrm{hom} \left( \mathbf{Vect} \right),
\end{equation}
and reversing the direction of composition:

\begin{equation}
\forall \left( f : X \to Y \right), \left( g : Y \to Z \right) \in \mathrm{hom} \left( \mathbf{Vect} \right), \qquad \left( \left( g \circ f \right)^{\dagger} : Z \to X \right) = \left( f^{\dagger} \circ g^{\dagger} : Z \to X \right),
\end{equation}
which acts as the identity on objects:

\begin{equation}
\forall X \in \mathrm{ob} \left( \mathbf{Vect} \right), \qquad X^{\dagger} = X, \qquad \text{ and } \qquad \left( id_{X}^{\dagger} : X \to X \right) = \left( id_{X} : X \to X \right),
\end{equation}
and which is involutive in the sense that it acts as its own inverse functor:

\begin{equation}
\forall \left( f : X \to Y \right) \in \mathrm{hom} \left( \mathbf{Vect} \right), \qquad \left( \left( f^{\dagger} \right)^{\dagger} : X \to Y \right) = \left( f : X \to Y \right).
\end{equation}
Since ${\mathbf{Vect}}$ is also a symmetric monoidal category, we would ideally like for the dagger structure ${\dagger}$ to be compatible with the tensor product structure ${\otimes}$, meaning that:

\begin{multline}
\forall \left( f: X \to Y \right), \left( g : Z \to W \right) \in \mathrm{hom} \left( \mathbf{Vect} \right),\\
\left( \left( f \otimes g \right)^{\dagger} : Y \otimes W \to X \otimes Z \right) = \left( f^{\dagger} \otimes g^{\dagger} : Y \otimes W \to X \otimes Z \right),
\end{multline}
in such a way that one maintains compatibility with the defining natural isomorphisms of the symmetric monoidal structure, namely the associator isomorphism ${\alpha}$:

\begin{multline}
\forall X, Y, Z \in \mathrm{ob} \left( \mathbf{Vect} \right),\\
\left( \alpha_{X, Y, Z}^{\dagger} : \left( X \otimes Y \right) \otimes Z \to X \otimes \left( Y \otimes Z \right) \right) = \left( \alpha_{X, Y, Z}^{-1} : \left( X \otimes Y \right) \otimes Z \to X \otimes \left( Y \otimes Z \right) \right),
\end{multline}
the left and right unitor isomorphisms ${\lambda}$:

\begin{equation}
\forall X \in \mathrm{ ob} \left( \mathbf{Vect} \right), \qquad \left( \lambda_{X}^{\dagger} : X \to I \otimes X \right) = \left( \lambda_{X}^{-1} : X \to I \otimes X \right),
\end{equation}
and ${\rho}$:

\begin{equation}
\forall X \in \mathrm{ob} \left( \mathbf{Vect} \right), \qquad \left( \rho_{X}^{\dagger} : X \to X \otimes I \right) = \left( \rho_{X}^{-1} : X \to X \otimes I \right),
\end{equation}
and the symmetry/braiding isomorphism ${\sigma}$:

\begin{equation}
\forall X, Y \in \mathrm{ob} \left( \mathbf{Vect} \right), \qquad \left( \sigma_{X, Y}^{\dagger} : Y \otimes X \to X \otimes Y \right) = \left( \sigma_{X, Y}^{-1} : Y \otimes X \to X \otimes Y \right).
\end{equation}
For the case of vector spaces (respectively finite-dimensional vector spaces), the Hermitian adjoint operation (respectively the conjugate transpose operation) clearly furnishes ${\mathbf{Vect}}$/${\mathbf{FdVect}}$ with a canonical dagger structure. For the case of the cobordism categories ${\mathbf{Bord}_{1}^{Riem}}$ and ${\mathbf{Bord}_{d}^{Riem}}$, the operation of ``time reversal'' (i.e. inversion of the orientation of cobordisms) yields a compatible dagger structure. Likewise, for the case of the category ${\mathcal{T}}$ of data structures and computations, obvious dagger structures exist for the cases of both Turing machine evolution and hypergraph rewriting considered previously (since for any given Turing machine rule, it is always possible to find a Turing machine rule of the same signature that reverses its evolution, and for any given hypergraph rewriting system, one can always swap the $L$ and $R$ objects within the span of monomorphisms that defines the rewriting rule in order to obtain a time-reversed version of the same evolution).

On the other hand, we also say that ${\mathbf{FdVect}}$ (the category of finite-dimensional vector spaces) is a \textit{compact closed}\cite{kelly3}\cite{doplicher} symmetric monoidal category as a shorthand for saying that every object ${X \in \mathrm{ob} \left( \mathbf{FdVect} \right)}$ has a corresponding dual object ${X^{*} \in \mathrm{ob} \left( \mathbf{FdVect} \right)}$ that is unique up to canonical isomorphism, which is equipped with a pair of morphisms ${\eta_X}$ and ${\varepsilon_X}$ of the form:

\begin{equation}
\left( \eta_X : I \to A^{*} \otimes A \right), \left( \varepsilon_A : A \otimes A^{*} \to I \right) \in \mathrm{hom} \left( \mathbf{FdVect} \right),
\end{equation}
known as the \textit{unit} and \textit{counit} morphisms, respectively, satisfying the following pair of coherence conditions (sometimes known as the \textit{yanking conditions}):

\begin{multline}
\forall X \in \mathrm{ob} \left( \mathbf{FdVect} \right),\\
\left( \lambda_X \circ \left( \left( \varepsilon_X \otimes id_X \right) \circ \left( \alpha_{X, X^{*}, X} \circ \left( \left( id_{X} \otimes \eta_X \right) \circ \rho_{X}^{-1} \right) \right) \right) : X \to X \right) = \left( id_X : X \to X \right),
\end{multline}
and:

\begin{multline}
\forall X \in \mathrm{ob} \left( \mathbf{FdVect} \right),\\
\left( \rho_{X^{*}} \circ \left( \left( id_{X^{*}} \otimes \varepsilon_{X} \right) \circ \left( \alpha_{X^{*}, X, X^{*}}^{-1} \circ \left( \left( \eta_X \otimes id_{X^{*}} \right) \circ \lambda_{X^{*}}^{-1} \right) \right) \right) : X^{*} \to X^{*} \right) = \left( id_{X^{*}} : X^{*} \to X^{*} \right).
\end{multline}
We can reformulate these two yanking conditions diagrammatically as the statement that the following pair of diagrams commute for all ${X \in \mathrm{ob} \left( \mathbf{FdVect} \right)}$:

\begin{equation}
\begin{tikzcd}
X \arrow[rr, "\rho_{X}^{-1}"] \arrow[swap, ddddrrrrrr, equal, "id_X"] & & X \otimes I \arrow[rr, "id_{X} \otimes \eta_{X}"] & & X \otimes \left( X^{*} \otimes X \right) \arrow[rr, "\alpha_{X, X^{*}, X}"] & & \left( X \otimes X^{*} \right) \otimes X \arrow[dd, "\varepsilon_{X} \otimes id_X"]\\\\
& & & & & & I \otimes X \arrow[dd,"\lambda_X"]\\\\
& & & & & & X,
\end{tikzcd}
\end{equation}
and:

\begin{equation}
\begin{tikzcd}
X^{*} \arrow[rr, "\lambda_{X^{*}}^{-1}"] \arrow[swap, ddddrrrrrr, equal, "id_{X^{*}}"] & & I \otimes X^{*} \arrow[rr, "\eta_X \otimes id_{X^{*}}"] & & \left( X^{*} \otimes X \right) \otimes X^{*} \arrow[rr, "\alpha_{X^{*}, X, X^{*}}^{-1}"] & & X^{*} \otimes \left( X \otimes X^{*} \right) \arrow[dd, "id_{X^{*}} \otimes \varepsilon_X"]\\\\
& & & & & & X^{*} \otimes I \arrow[dd, "\rho_{X^{*}}"]\\\\
& & & & & & X^{*}.
\end{tikzcd}
\end{equation}
If, moreover, there exists a dagger structure ${\dagger}$ that is compatible with the compact structure in such a way that the unit and counit morphisms ${\eta}$ and ${\varepsilon}$ can be related by means of the dagger operation (as indeed is the case for the category of finite-dimensional vector spaces and their linear isomorphisms), in other words if the following diagram commutes for all objects ${X \in \mathrm{ob} \left( \mathbf{FdVect} \right)}$:

\begin{equation}
\begin{tikzcd}
I \arrow[r, "\varepsilon_{X}^{\dagger}"] \arrow[swap, dr, "\eta_X"] & X \otimes X^{*} \arrow[d, "\sigma_{X, X^{*}}"]\\
& X^{*} \otimes X,
\end{tikzcd}
\end{equation}
i.e:

\begin{equation}
\forall X \in \mathrm{ob} \left( \mathbf{FdVect} \right), \qquad \left( \sigma_{X, X^{*}} \circ \varepsilon_{X}^{\dagger} : I \to X^{*} \otimes X \right) = \left( \eta_X : I \to X^{*} \otimes X \right),
\end{equation}
then we describe the symmetric monoidal category as being \textit{dagger compact}. For the case of finite-dimensional vector spaces, the passage to the dual vector space (i.e. the of vector space of linear forms under pointwise addition and scalar multiplication) furnishes ${\mathbf{FdVect}}$ with a canonical compact structure\cite{selinger}\cite{hasegawa}; for the case of the infinite-dimensional vector spaces in ${\mathbf{Vect}}$, dual spaces also exist, although they are inherently less ``well-behaved'' (since the axiom of choice here implies that the dual space is always of strictly larger dimension than the original space) and satisfy fewer compatibility conditions. In topological quantum field theories, in which the number of degrees of freedom (and therefore the dimensionality of the spaces of states) is always finite,  the propagator for processes of shape $M$ from hypersurface ${\partial_{in} \left( M \right)}$ to hypersurface ${\partial_{out} \left( M \right)}$, namely:

\begin{equation}
Z \left( M \right) : Z \left( \partial_{in} \left( M \right) \right) \to Z \left( \partial_{out} \left( M \right) \right),
\end{equation}
generated by the functor ${Z : \mathbf{Bord}_{d}^{Riem} \to \mathbf{Vect}}$ can be formulated in terms of dual spaces as:

\begin{equation}
Z \left( M \right) : \mathbb{C} \to Z \left( \partial \left( M \right) \right) = Z \left( \partial_{out} \left( M \right) \right) \otimes Z \left( \partial_{in} \left( M \right) \right)^{*},
\end{equation}
i.e. in the form of a \textit{correlator} or \textit{$n$-point function}. For the category of hypergraphs and subhypergraph inclusion maps, every hypergraph has a dual (that is trivially compatible with the dagger structure) given by the interchange of its vertex set $V$ and its hyperedge set $E$, i.e if:

\begin{equation}
H = \left\langle V = \left\lbrace v_i \left\vert i \in I_v \right. \right\rbrace, E = \left\lbrace e_i \left\vert i \in I_e, e_i \subseteq V, e_i \neq \emptyset \right. \right\rbrace \right\rangle,
\end{equation}
then:

\begin{equation}
H^{*} = \left\langle V^{*} = E, E^{*} = \left\lbrace \left\lbrace e_i \left\vert v_m \in e_i \right. \right\rbrace \left\vert m \in I_v \right. \right\rbrace \right\rangle,
\end{equation}
where ${I_v}$ and ${I_e}$ are index sets for the vertices and for the hyperedges, respectively. Corresponding dual structures may well exist for other categories of data structures and computations too (such as the category of Turing machine states and transitions), but, if they do, then their explicit forms remain unknown at present.

Just as we have shown that deformation of the sequential composition structure ${\circ}$ in an ordinary category can be used to characterize and quantify computational reducibility, and that deformation of the parallel composition structure/tensor product ${\otimes}$ in a symmetric monoidal category can be used to characterize and quantify multicomputational reducibility, it is entirely conceivable that, as a consequence of this formal duality/adjunction relating computational complexity theory to quantum mechanics and multicomputational complexity theory to quantum field theory, these various other algebraic properties and structures inherent to the physical theories will end up having rather natural, and purely complexity-theoretic, analogs in the abstract theory of computation. For instance, it is plausible that deformation of the dagger structure ${\dagger}$ in a dagger symmetric monoidal category may provide some means of characterizing the irreversibility (meaning the pragmatic computational difficulty of reversing a computation that is, in principle, fully reversible) of certain classes of (multi)computations - a concept which is, of course, deeply conceptually related to (multi)computational irreducibility itself - while deformation of the compact structure ${\eta}$, ${\varepsilon}$ in a dagger compact category may be useful for quantifying the computational difficulty involved in exchanging outputs/values with inputs/arguments in a multicomputation consisting of one or more multi-argument, multi-valued functions (e.g. as represented by a tensor network or a monoidal string diagram). These exciting possibilities remain topics for future investigation, as further detailed within the concluding remarks below.

\section{Concluding Remarks}
\label{sec:Section4}

Throughout the course of this article, we have sought to develop a systematic procedure by which the abstract \textit{syntax} of a category may be endowed with a concrete computational \textit{semantics}, in which all objects are interpreted as data structures and all morphisms are interpreted as computations, in such a way that each morphism carries with it certain metadata corresponding to the time complexity of its underlying computation, along with an algebra that defines how these time complexities behave under composition. We have shown that this can be achieved by defining a map from the category of data structures and computations to a discrete cobordism category (of 0-dimensional manifolds and discrete cobordisms/intervals), in such a way that the irreducibility of a given computation is characterized by the extent to which this map preserves additivity of time complexities under sequential composition (i.e. its functoriality), and such that the multicomputational irreducibility of a given multicomputation (described by the case of higher-dimensional manifolds and discrete cobordisms, in which both categories carry an additional symmetric monoidal structure) is characterized by the extent to which this map preserves additivity of time complexities under parallel/tensor product composition (i.e. its symmetric monoidal functoriality). This latter extension is achieved by exploiting the fact that symmetric monoidal categories provide a convenient compositional semantics for describing multiway systems, branchial graphs and the general algebraic structure of non-deterministic computations. In so doing, we have effectively defined the outlines of a potential novel extension to the standard techniques of category theory and monoidal category theory, in which the compositions of morphisms (both in sequence and in parallel) carry with them additional algebraic structure resulting from the constraints of computational complexity theory. This new formalism brings with it many compelling directions for future research and exploration, a few of which we shall indicate below.

Firstly, although traditional computational complexity theory has tended to restrict itself to the investigation of complexity classes with very simply and well-defined algebraic constraints (e.g. $P$, ${EXP}$, ${NP}$, etc.), this category-theoretic formalism, along with the explicit computational tools developed for the purposes of the present article, potentially enables one to conduct a systematic and empirical investigation of computational complexity classes exhibiting far less a priori algebraic structure. Further, by considering the extension to monoidal categories, it also enables the rigorous investigation of \textit{multicomputational} complexity theory, which differs fundamentally from ordinary non-deterministic complexity theory in that, within a standard non-deterministic complexity class (such as ${NP}$), although the non-deterministic nature of the computation implies that one is inevitably forced to consider a multiway system of different singleway/deterministic computations, the complexity class itself is ultimately concerned only with the time complexity along a single branch of that system (albeit a branch whose identity may not necessarily be known in advance). Multicomputational complexity theory, on the other hand, investigates the computational complexity of the multiway system itself, and specifically of the tensor product structure of its branchial graphs (encoding, as it does, the complex branching and merging behavior of the entire multiway system), featuring, as discussed previously, contributions from the complexity-theoretic properties of \textit{both} the state \textit{evolution function} and the state \textit{equivalence function} of the multiway system, and, most crucially, the properties of the interactions between the two. To the best of the author's knowledge, the space of possible multicomputational complexity classes remains essentially unexplored, and constitutes a major topic for planned future examination.

\begin{figure}[ht]
\centering
\includegraphics[width=0.595\textwidth]{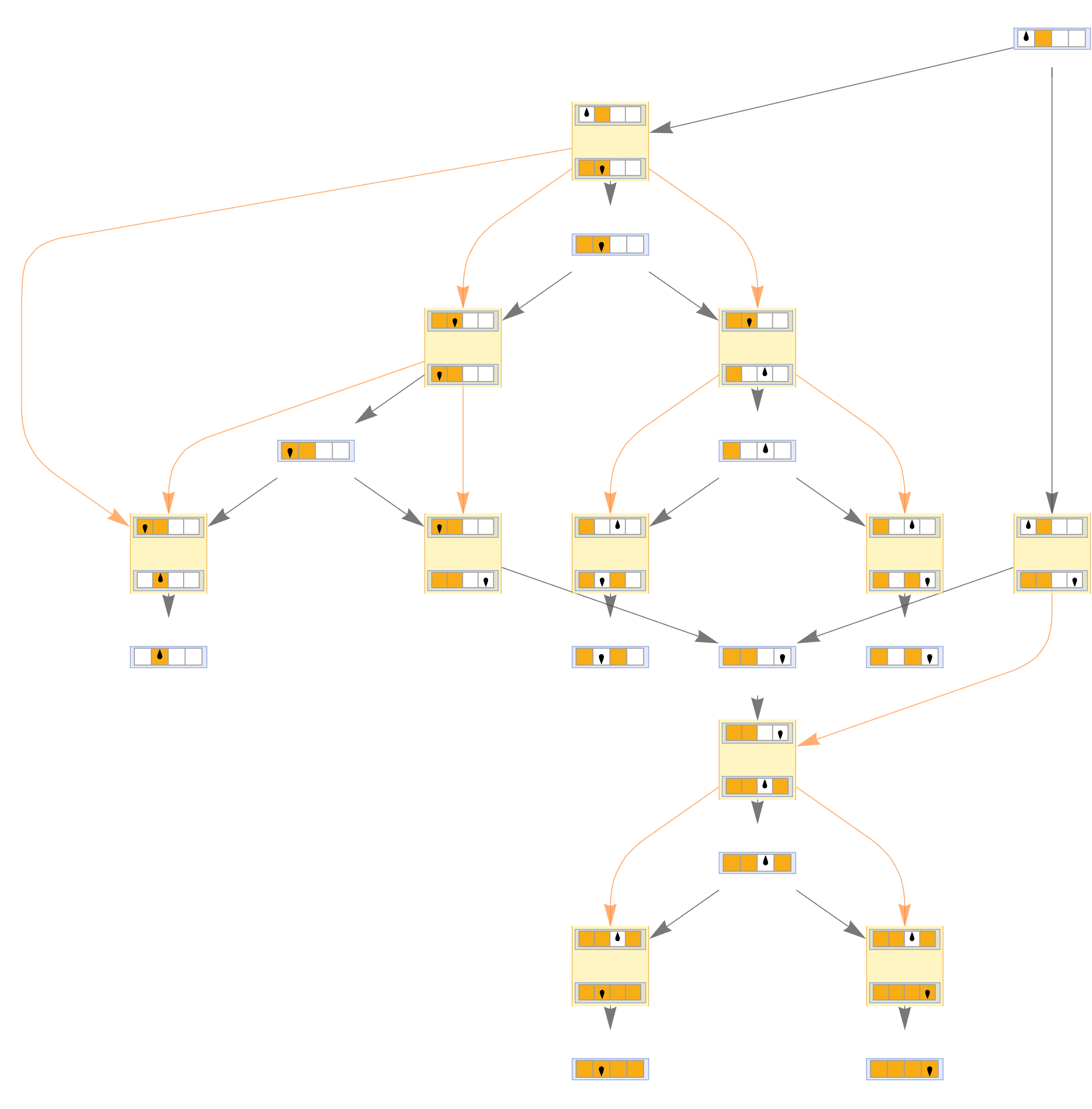}
\caption{A multiway evolution causal graph corresponding to the non-deterministic evolution of a 2-state, 2-color Turing machine constructed from the (parallel) composition of the Turing machine transition functions for rules 2506 and 3506, starting from the single tape state ${\left\lbrace 0, 1, 0, 0 \right\rbrace}$, for 3 steps; the gray edges are the usual evolution edges, while each orange edge represents a causal relationship between two transitions.}
\label{fig:Figure17}
\end{figure}

Next, all of the analysis presented within this article has focused exclusively on singleway/multiway evolution structure, and has neglected any consideration of causality; however, a natural causal semantics does exist for the case of hypergraph rewriting\cite{gorard8}\cite{gorard9} (and, indeed, for non-deterministic Turing machine evolution, as depicted as part of the \textit{multiway evolution causal graph} shown in Figure \ref{fig:Figure17} for the non-deterministic 2-state, 2-color Turing machine analyzed previously), allowing one to encode formally, by means of an explicit partial order, the notion of one transition/rewriting event only being applicable if another transition/rewriting event had previously occurred. Just as the multiway evolution graph can be interpreted as freely generating a symmetric monoidal category, the multiway evolution causal graph (featuring a combination of both evolution edges, indicating transitions/rewriting events, and causal edges, indicating causal relationships between those transitions/rewriting events) can be interpreted as freely generating a weak 2-category\cite{benabou}, with the 2-cells representing causal relationships. These 2-cells are also equipped with their own tensor product operation, usually denoted ${\otimes_{C}}$, that satisfies the axioms of a \textit{partial} monoidal structure in the sense defined by Coecke and Lal in the context of causal categories\cite{coecke}, and allows one to compose causal relationships between \textit{spacelike-separated} (causally-independent) events in parallel, but not between \textit{timelike-separated} (causally-dependent) events, which may only be composed in sequence. This leads to the intriguing possibility that one may be able to equip the discrete cobordism category ${\mathcal{B}}$ with a second (partial) monoidal structure and, with it, to examine the 2-functoriality of the resulting map ${Z^{\prime} : \mathcal{T} \to \mathcal{B}}$ as a proxy for extending the definition of multicomputational irreducibility so as also to incorporate some kind of inherent \textit{complexity of causality}, and not simply of evolution (essentially by examining the extent to which ${Z^{\prime}}$ also distorts the sequential and parallel composition of the 2-cells representing the causal structure of the system). The aim of such an undertaking would be to formalize a notion of \textit{causal irreducibility}, in which the complete causal structure of a causally irreducible multiway system cannot be preempted without effectively tracing the explicit causal relationships between all transitions/rewriting events, spanning across all branches of the multiway evolution graph.

As discussed within the preceding section, there are also various features of the formal duality/adjunction relationship between (multi)computational complexity theory and categorical quantum mechanics/functorial quantum field theory that might potentially yield additional, purely complexity-theoretic, extensions to this formalism. For instance, it has often been assumed that the practical irreversibility of computations which are in principle reversible (e.g. in the case of one-way functions in cryptography) is merely a byproduct of computational irreducibility, but this assumption tacitly presupposes a compatibility condition between the irreducibilities of forward and backward evolution that may or may not hold for certain classes of systems. By equipping our category ${\mathcal{T}}$ with an involutive dagger structure ${\dag}$ and examining the effects of the map ${Z^{\prime}}$ on that structure using methods from categorical quantum mechanics, it is conceivable that we may be able to disentangle the irreducibility of evolution from the irreducibility of reversal in a relatively fine-grained way. Deeply related to this is the irreducibility of swapping arguments/inputs with values/outputs in a composition of several multi-argument, multi-valued functions; in the case of a tensor network or monoidal string diagram (where this swapping operation is encoded as the raising and lowering of contravariant/covariant indices), such operations are usually assumed to be computationally trivial, but for the case of computations whose reversal operation is irreducible this need not be the case. Equipping the category ${\mathcal{T}}$ with a compact structure, with unit/counit ${\eta}$/${\varepsilon}$, and observing how the compact structure gets distorted under the action of ${Z^{\prime}}$ may permit one to quantify the complexities of these operations in a more meaningful way.

\begin{figure}[ht]
\centering
\includegraphics[width=0.695\textwidth]{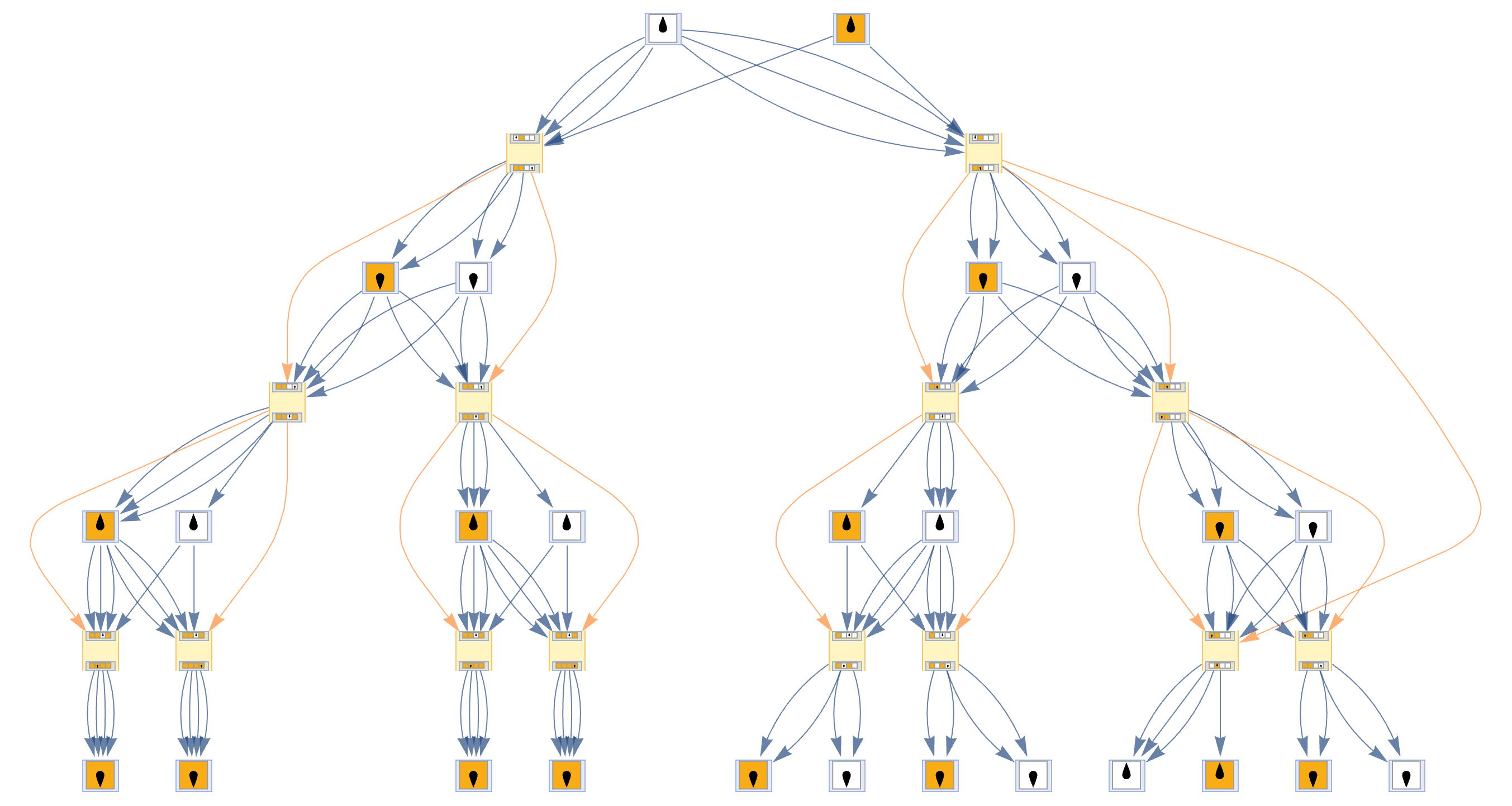}
\caption{A ``glocal'' multiway evolution causal graph corresponding to the non-deterministic evolution of a 2-state, 2-color Turing machine constructed from the (parallel) composition of the Turing machine transition functions for rules 2506 and 3506, starting from the single tape state ${\left\lbrace 0, 1, 0, 0 \right\rbrace}$, for 3 steps; the orange edges represent causal relationships between transitions, while each gray edge represents either the ``ingestion'' or the ``egestion'' of a single ``token'' (tape position) into or out of a single transition.}
\label{fig:Figure18}
\end{figure}

\begin{figure}[ht]
\centering
\includegraphics[width=0.495\textwidth]{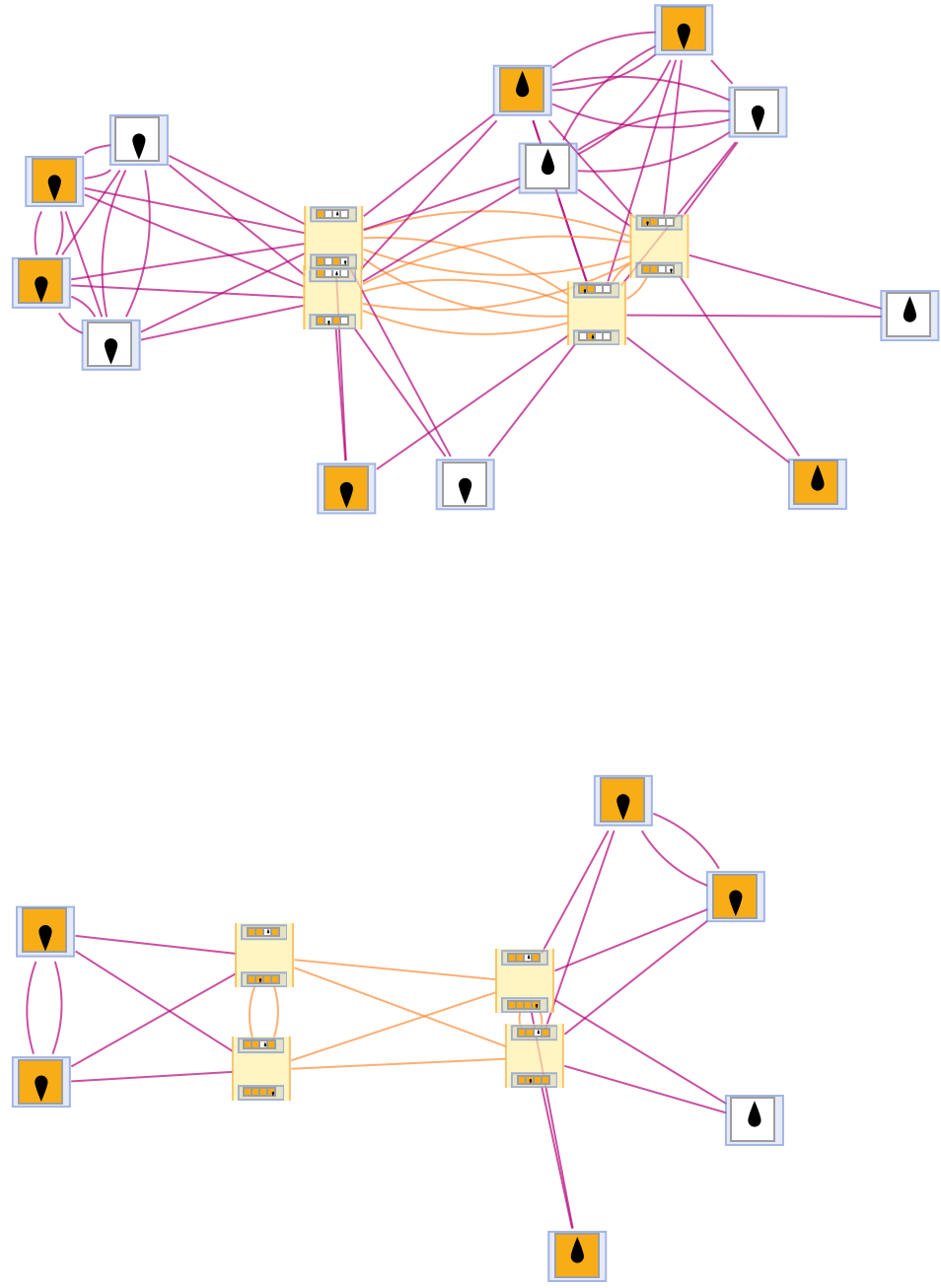}
\caption{The corresponding ``glocal'' branchial graph associated to the default ``foliation'' of the glocal multiway evolution causal graph for the non-deterministic evolution of the 2-state, 2-color Turing machine constructed from parallel composition of the transition functions for rules 2506 and 3506, after 3 steps, showing a mixture of both ``spatial'' and ``branchial'' tensor product structure.}
\label{fig:Figure19}
\end{figure}

However, the conventional multiway system formalism does not make full use of the compact structure that is available within dagger compact categories, since transitions/events in a multiway system are traditionally single-argument but potentially multi-valued (that is, a state evolution function conventionally takes in a single state as input and produces a list of possible successor states as output). On the other hand, a \textit{glocal} multiway system promotes the transitions/events to being true multi-argument functions (and, hence, to true symbolic tensors in a tensor network/string diagram representation of the multiway system) by effectively ``shattering'' the states into their constituent ``tokens'' (for instance, into individual hyperedges for the case of hypergraph rewriting systems, or individual tape positions for the case of Turing machines) and then reassembling them on-demand for the application of particular transitions/events; the term ``glocal'' refers here to the fact that the tokens are local but the events are global, and so in particular the multiway equivalence function acts at the level of events rather than at the level of states. An example of a glocal multiway evolution causal graph, for the same non-deterministic 2-state, 2-color Turing machine as before, is shown in Figure \ref{fig:Figure18}, with its associated glocal branchial graph shown in Figure \ref{fig:Figure19}. Note that, on a given glocal branchial graph, some of the tokens are separated spatially (i.e. they correspond to different regions of the tape), whilst some of the tokens are separated ``branchially'' (i.e. they correspond to the same region of the tape, but on two different branches of the multiway system), and so, unlike the branchial graphs considered previously, glocal branchial graphs actually encode two different tensor product structures: the standard symmetric monoidal structure inherited from the multiway system, and a ``spatial'' tensor product structure (which is really identical to the causal tensor product structure ${\otimes_C}$ described above). The compatibility conditions between these two tensor product structures are not known (for instance, it is unknown whether they form a \textit{rig} category in special cases, although this possibility is unlikely) and remain a topic for future research. Note that the question of whether the ordering of tokens matters (as for Turing machine tape states) or does not (as for hyperedges in a hypergraph) corresponds to the question of whether the ``spatial'' part of the category is symmetric monoidal or simply monoidal. Rather excitingly, just as the multiway tensor product structure enabled us to quantify multicomputational complexity, it is conceivable that the spatial tensor product structure (once it is better understood) may enable us to quantify space complexity, and thus to examine questions surrounding (e.g.) the trade-offs between space complexity and time complexity in arbitrary (multi)computations. A meticulous treatment of the relationship between multiway systems, the notion of ``spatiality'', higher categories and type theory was previously conducted by Arsiwalla\cite{arsiwalla}\cite{arsiwalla2} within the context of Shulman's cohesive homotopy type theory\cite{shulman}.

Finally, there are several conceptual consequences of the realization of the \textit{orthogonality} (independence) that exists between the complexity of state evolution and state equivalence (and thus between computational and multicomputational irreducibility). For instance, the computational interpretation of the second law of thermodynamics advocated by Wolfram\cite{wolfram2} implies that entropy increase is a consequence of computational irreducibility, wherein the progression of a reversible computation can have the effect of ``encrypting'' the details of its initial conditions (such that, even if the computation is in principle reversible, in practice it can represent an arbitrarily hard problem of cryptanalysis to enact that reversal). However, synthesizing this idea with the ``orthogonality'' principle indicates that there should exist at least two distinct concepts of entropy at play within any given multicomputation: one essentially computational, and the other essentially multicomputational, in origin. The standard thermodynamic description of entropy is essentially a measure of non-injectivity of coarse-graining (i.e. how many distinct microstates get mapped to the same macrostate under the action of the coarse-graining function), and, due to Liouville's theorem,  and thus the one-to-one correspondence that exists between position/momentum values and possible evolution histories, in classical mechanics this definition is provably equivalent to a definition in terms of possible evolution trajectories (i.e. how many distinct branches of history would have resulted in the same coarse-grained macrostate). However, for more general multiway systems with more complex branching and merging structure, this correspondence does not necessarily hold, and so the two definitions of entropy diverge (although many computations of e.g. entanglement entropies in the context of quantum gravity may implicitly assume that they are equivalent\cite{shah}\cite{susskind}\cite{stanford}\cite{brown}). The manifold implications of this disambiguation remain another worthy topic for future study.

\section*{Acknowledgments}

The author would like to thank Mohamed Barakat, Nicolas Behr, Matteo Capucci, Bob Coecke, Fabrizio Genovese, Manojna Namuduri, Stephen Wolfram and Yorick Zeschke for various stimulating and insightful discussions, enjoyed at a variety of different gestational stages of the ideas presented within this work. The author would also like to acknowledge James Boyd for his christening of the term ``multicomputational irreducibility'' in an earlier blog post, and Juan Arturo Silva-Ordaz for forcing the author to think about the complexity of equivalence functions to a far greater extent than he would ever have chosen to do voluntarily.


\begin{thebibliography}{99}
\bibitem{wolfram}
\textsc{S. Wolfram} (1985), ``Undecidability and Intractability in Theoretical Physics'', \textit{Physical Review Letters} \textbf{54} (8): 735--738. \url{https://journals.aps.org/prl/abstract/10.1103/PhysRevLett.54.735}.

\bibitem{wolfram2}
\textsc{S. Wolfram} (2002), \textit{A New Kind of Science}. Champaign, IL: Wolfram Media, Inc. \url{https://www.wolframscience.com}.

\bibitem{gorard}
\textsc{J. Gorard} (2018), ``The Slowdown Theorem: A Lower Bound for Computational Irreducibility in Physical Systems'', \textit{Complex Systems} \textbf{27} (2): 177--185. \url{https://www.complex-systems.com/abstracts/v27_i02_a05/}.

\bibitem{turing}
\textsc{A. M. Turing} (1937), ``On Computable Numbers, with an Application to the Entscheidungsproblem'', \textit{Proceedings of the London Mathematical Society} \textbf{S2-42} (1): 250--265. \url{https://londmathsoc.onlinelibrary.wiley.com/doi/abs/10.1112/plms/s2-42.1.230}.

\bibitem{garey}
\textsc{M. R. Garey and D. S. Johnson} (1979), \textit{Computers and Intractability: A Guide to the Theory of NP-Completeness}. San Francisco, CA: W. H. Freeman. ISBN: 978-0716710455.

\bibitem{hopcroft}
\textsc{J. E. Hopcroft, R. Motwani and J. D. Ullman} (2013), \textit{Introduction to Automata Theory, Languages and Computation} (3rd Edition). London: Pearson. ISBN: 978-1292039053.

\bibitem{zenil}
\textsc{H. Zenil, F. Soler-Toscano and J. J. Joosten} (2012), ``Empirical Encounters with Computational Irreducibility and Unpredictability'', \textit{Minds and Machines} \textbf{22} (3): 149--165. \url{https://link.springer.com/article/10.1007/s11023-011-9262-y}.

\bibitem{zwirn}
\textsc{H. Zwirn and J-P. Delahaye} (2013), ``Unpredictability and Computational Irreducibility'', \textit{Irreducibility and Computational Equivalence: 10 Years after Wolfram's A New Kind of Science}, H. Zenil (ed): 273--295. Berlin, Heidelberg: Springer. \url{https://link.springer.com/chapter/10.1007/978-3-642-35482-3_19}.

\bibitem{reisinger}
\textsc{D. Reisinger, T. Martin, M. Blankenship, C. Harrison, J. Squires and A. Beavers} (2013), ``Exploring Wolfram's Notion of Computational Irreducibility with a Two-Dimensional Cellular Automaton'', \textit{Irreducibility and Computational Equivalence: 10 Years after Wolfram's A New Kind of Science,}, H. Zenil (ed): 263--272. Berlin, Heidelberg: Springer. \url{https://link.springer.com/chapter/10.1007/978-3-642-35482-3_18}.

\bibitem{maclane}
\textsc{S. Mac Lane} (1998), \textit{Categories for the Working Mathematician}, Graduate Texts in Mathematics \textbf{5} (2nd Edition). Berlin, Heidelberg: Springer. ISBN: 0-387-98403-8.

\bibitem{milnor}
\textsc{J. Milnor} (1962), ``A Survey of Cobordism Theory'', \textit{L'Enseignement Math\'ematique Revue International} \textbf{8}: 16--23. \url{https://www.e-periodica.ch/cntmng?pid=ens-001:1962:8::12}

\bibitem{adams}
\textsc{J. F. Adams} (1974), \textit{Stable Homotopy and Generalised Homology}. Chicago IL: University of Chicago Press. ISBN: 978-0202361406.

\bibitem{maclane2}
\textsc{S. Mac Lane} (1963), ``Natural Associativity and Commutativity'', \textit{Rice Institute Pamphlet - Rice University Studies} \textbf{49} (4): 28--46. \url{https://scholarship.rice.edu/handle/1911/62865}.

\bibitem{baez}
\textsc{J. C. Baez} (2006), ``Quantum Quandries: a Category-Theoretic Perspective'', \textit{The Structural Foundations of Quantum Gravity}, D. Rickles, S. French and J. T. Saatsi (eds): 240--366. Oxford: Oxford University Press. \url{https://arxiv.org/abs/0811.2280}.

\bibitem{atiyah}
\textsc{M. Atiyah} (1988), ``New Invariants of 3- and 4-Dimensional Manifolds'', \textit{The Mathematical Heritage of Hermann Weyl: Proceedings of Symposia in Pure Mathematics} \textbf{48}: 258--299. \url{https://archive.org/details/mathematicalheri0000symp/page/285/mode/2up}.

\bibitem{atiyah2}
\textsc{M. Atiyah} (1988), ``Topological quantum field theories'', \textit{Publications Math\'ematiques de l'Institut des Hautes \'Etudes Scientifiques} \textbf{68} (68): 175--186. \url{https://link.springer.com/article/10.1007/BF02698547}.

\bibitem{segal}
\textsc{G. B. Segal} (1988), ``The Definition of Conformal Field Theory'', \textit{Differential Geometrical Methods in Theoretical Physics}, K. Bleuler, M. Werner (eds) \textbf{250}: 165--171. \url{https://link.springer.com/chapter/10.1007/978-94-015-7809-7_9}.

\bibitem{goldreich}
\textsc{O. Goldreich} (2008), \textit{Computational Complexity: A Conceptual Perspective}. Cambridge: Cambridge University Press. ISBN: 978-0521884730.

\bibitem{papadimitriou}
\textsc{C. H. Papadimitriou} (1993), \textit{Computational Complexity}. London: Pearson. ISBN: 978-0201530827.

\bibitem{jacobson}
\textsc{N. Jacobson} (2009), \textit{Basic Algebra II} (2nd edition). Mineola, NY: Dover Publications, Inc. ISBN: 978-0486471877.

\bibitem{stong}
\textsc{R. E. Stong} (1968), \textit{Notes on Cobordism Theory}. Princeton NJ: Princeton University Press. ISBN: 978-0691649016.

\bibitem{wolfram3}
\textsc{S. Wolfram} (2020), ``A Class of Models with the Potential to Represent Fundamental Physics'', \textit{Complex Systems} \textbf{29} (2): 107--536. \url{https://www.complex-systems.com/abstracts/v29_i02_a01/}.

\bibitem{gorard2}
\textsc{J. Gorard} (2020), ``Some Relativistic and Gravitational Properties of the Wolfram Model'', \textit{Complex Systems} \textbf{29} (2): 599--654. \url{https://www.complex-systems.com/abstracts/v29_i02_a03/}.

\bibitem{gorard3}
\textsc{J. Gorard} (2020), ``Some Quantum Mechanical Properties of the Wolfram Model'', \textit{Complex Systems} \textbf{29} (2): 537--597. \url{https://www.complex-systems.com/abstracts/v29_i02_a02/}.

\bibitem{rabin}
\textsc{M. O. Rabin and D. Scott} (1959), ``Finite Automata and Their Decision Problems'', \textit{IBM Journal of Research and Development} \textbf{3} (2): 114--125. \url{https://ieeexplore.ieee.org/document/5392601}.

\bibitem{gorard4}
\textsc{J. Gorard, M. Namuduri and X. D. Arsiwalla} (2020), ``ZX-Calculus and Extended Hypergraph Rewriting Systems I: A Multiway Approach to Categorical Quantum Information Theory'', \textit{arXiv preprint}: \url{https://arxiv.org/abs/2010.02752}.

\bibitem{gorard5}
\textsc{J. Gorard, M. Namuduri and X. D. Arsiwalla} (2021), ``ZX-Calculus and Extended Wolfram Model Systems II: Fast Diagrammatic Reasoning with an Application to Quantum Circuit Simplification'', \textit{arXiv preprint}: \url{https://arxiv.org/abs/2103.15820}.

\bibitem{kelly}
\textsc{G. M. Kelly} (1964), ``On MacLane's conditions for coherence of natural associativities, commutativities, etc.'', \textit{Journal of Algebra} \textbf{1} (4): 397--402. \url{https://www.sciencedirect.com/science/article/pii/0021869364900183}.

\bibitem{kelly2}
\textsc{G. M. Kelly} (1982), \textit{Basic Concepts of Enriched Category Theory}, London Mathematical Society Lecture Notes Series. Cambridge: Cambridge University Press. ISBN: 978-0521287029.

\bibitem{baez2}
\textsc{J. C. Baez and M. Stay} (2011), ``Physics, Topology, Logic and Computation: A Rosetta Stone'', \textit{New Structures in Physics}, B Coecke (ed), Lecture Notes in Physics \textbf{813}: 95--172. \url{https://arxiv.org/abs/0903.0340}.

\bibitem{joyal}
\textsc{A. Joyal and R. Street} (1993), ``Braided Tensor Categories'', \textit{Advances in Mathematics} \textbf{102} (1): 20--78. \url{https://www.sciencedirect.com/science/article/pii/S0001870883710558}.

\bibitem{joyal2}
\textsc{A. Joyal and R. Street} (1986), ``Braided monoidal categories'', \textit{Macquarie Mathematics Reports} (860081). \url{http://web.science.mq.edu.au/~street/JS1.pdf}.

\bibitem{chari}
\textsc{V. chari and A. Pressley} (1995), \textit{A Guide to Quantum Groups}. Cambridge: Cambridge University Press. ISBN: 978-0521558846.

\bibitem{aguiar}
\textsc{M. Aguiar and S. Mahajan} (2010), \textit{Monoidal Functors, Species and Hopf Algebras}, CRM Monograph Series \textbf{29}. Ann Arbor, MI: American Mathematical Society. ISBN: 978-0821847763.

\bibitem{booth}
\textsc{K. S. Booth and C. J. Colbourn} (1977), ``Problems Polynomially Equivalent to Graph Isomorphism'', \textit{University of Waterloo Computer Science Department Technical Report} \textbf{CS-77-04}. \url{https://cs.uwaterloo.ca/research/tr/1977/CS-77-04.pdf}.

\bibitem{kobler}
\textsc{J. K\"obler, U. Sch\"oning and J. Tor\'an} (1993), \textit{The Graph Isomorphism Problem: Its Structural Complexity}. Basel: Birkh\"auser. ISBN: 978-0817636807.

\bibitem{gorard6}
\textsc{J. Gorard} (2016), ``Uniqueness Trees: A Possible Polynomial Approach to the Graph Isomorphism Problem'', \textit{arXiv preprint}: \url{https://arxiv.org/abs/1606.06399}.

\bibitem{ehrig}
\textsc{H. Ehrig, M. Pfender and H. J. Schneider} (1973), ``Graph-grammars: An algebraic approach'', \textit{IEEE Conference Record of 14th Annual Symposium on Switching and Automata Theory}: 167--180. \url{https://ieeexplore.ieee.org/document/4569741}.

\bibitem{ehrig2}
\textsc{H. Ehrig, K. Ehrig, U. Prange and G. Taentzer} (2006), \textit{Fundamentals of Algebraic Graph Transformation}, Monographs in Theoretical Computer Science. An EATCS Series. Berlin, Heidelberg: Springer. ISBN: 978-3642068317.

\bibitem{kissinger4}
\textsc{A. Kissinger} (2014), ``Finite matrices are complete for (dagger-)hypergraph categories'', \textit{arXiv preprint}: https://arxiv.org/abs/1406.5942.

\bibitem{fong}
\textsc{B. Fong} (2015), ``Decorated Cospans'', \textit{Theory and Applications of Categories} \textbf{30} (33): 1096--1120. \url{https://arxiv.org/abs/1502.00872}.

\bibitem{fong2}
\textsc{B. Fong} (2016), ``The Algebra of Open and Interconnected Systems'', \textit{DPhil Thesis, University of Oxford}. \url{https://arxiv.org/abs/1609.05382}.

\bibitem{habel}
\textsc{A. Habel, J. M\"uller and D. Plump} (2001), ``Double-pushout graph transformation revisited'', \textit{Mathematical Structures in Computer Science} \textbf{11} (5): 637--688. \url{https://www.cambridge.org/core/journals/mathematical-structures-in-computer-science/article/abs/doublepushout-graph-transformation-revisited/AF02050525390437E1DF746DE4459926}.

\bibitem{lack}
\textsc{S. Lack and P. Soboci\'nski} (2004), ``Adhesive Categories'', \textit{International Conference on Foundations of Software Science and Computation Structures}, Lecture Notes in Computer Science \textbf{2987}: 273--288. \url{https://link.springer.com/chapter/10.1007/978-3-540-24727-2_20}.

\bibitem{kissinger}
\textsc{A. Kissinger} (2011), ``Pictures of Processes: Automated Graph Rewriting for Monoidal Categories and Applications to Quantum Computing'', \textit{DPhil Thesis, University of Oxford}. \url{https://arxiv.org/abs/1203.0202}.

\bibitem{kissinger2}
\textsc{A. Kissinger} (2008), ``Graph rewrite systems for classical structures in dagger-symmetric monoidal categories'', \textit{MSc thesis, University of Oxford}.  \url{https://www.cs.ox.ac.uk/people/bob.coecke/Aleks.pdf}.

\bibitem{kissinger3}
\textsc{A. Kissinger} (2009), ``Exploring a Quantum Theory with Graph Rewriting and Computer Algebra'', \textit{Intelligent Computer Mathematics. CICM 2009}, J. Carette, L. Dixon, C. S. Coen, S. M. Watts (eds): 90--105. \url{https://link.springer.com/chapter/10.1007/978-3-642-02614-0_12}.

\bibitem{gorard7}
\textsc{J. Gorard, M. Namuduri and X. D. Arsiwalla} (2021), ``Fast Automated Reasoning over String Diagrams using Multiway Causal Structure'', \textit{arXiv preprint}: \url{https://arxiv.org/abs/2105.04057}.

\bibitem{abramsky}
\textsc{S. Abramsky and B. Coecke} (2004), ``A categorical semantics of quantum protocols'', \textit{Proceedings of the 19th IEEE Symposium on Logic in Computer Sceince}. Turku, Finland: 415--425. \url{https://ieeexplore.ieee.org/document/1319636}.

\bibitem{abramsky2}
\textsc{S. Abramsky and B. Coecke} (2008), ``Categorical quantum mechanics'', \textit{Handbook of Quantum Logic and Quantum Structures}, K. Engesser, D. M. Gabbay, D. Lehmann (eds): 261--323. Elsevier. \url{https://arxiv.org/abs/0808.1023}.

\bibitem{schreiber}
\textsc{U. Schreiber} (2009), ``AQFT from $n$-Functorial QFT'', \textit{Communications in Mathematical Physics} \textbf{291}: 357--401. \url{https://arxiv.org/abs/0806.1079}.

\bibitem{baez3}
\textsc{J. C. Baez and J. Dolan} (1995), ``Higher-dimensional Algebra and Topological Quantum Field Theory'', \textit{Journal of Mathematical Physics} \textbf{36}: 6073--6105. \url{https://arxiv.org/abs/q-alg/9503002}.

\bibitem{burgin}
\textsc{M. Burgin} (1970), ``Categories with involution, and correspondences in ${\gamma}$-categories'', \textit{Transactions of the Moscow Mathematical Society} \textbf{22}: 161--228. \url{http://www.mathnet.ru/php/archive.phtml?wshow=paper&jrnid=mmo&paperid=235}.

\bibitem{lambek}
\textsc{J. Lambek} (1999), ``Diagram chasing in ordered categories with involution'', \textit{Journal of Pure and Applied Algebra} \textbf{143} (1--3): 293--309. \url{https://www.sciencedirect.com/science/article/pii/S0022404998001157}.

\bibitem{kelly3}
\textsc{M. G. Kelly and M. L. Laplaza} (1980), ``Coherence for compact closed categories'', \textit{Journal of Pure and Applied Algebra} \textbf{19}: 193--213. \url{https://www.sciencedirect.com/science/article/pii/0022404980901012}.

\bibitem{doplicher}
\textsc{S. Doplicher and J. Roberts} (1989), ``A new duality theory for compact groups'', \textit{Inventiones Mathematicae} \textbf{98}: 157--218. \url{https://link.springer.com/article/10.1007/BF01388849}.

\bibitem{selinger}
\textsc{P. Selinger} (2011), ``Finite Dimensional Hilbert Spaces are Complete for Dagger Compact Closed Hilbert Spaces'', \textit{Electronic Notes in Theoretical Computer Science} \textbf{270} (1): 113-119. \url{https://www.sciencedirect.com/science/article/pii/S1571066111000119}.

\bibitem{hasegawa}
\textsc{M. Hasegawa, M. Hofmann and G. Plotkin} (2008), ``Finite Dimensional Vector Spaces are Complete for Traced Symmetric Monoidal Categories'', \textit{Pillars of Computer Science}, A. Avron, N. Dershowitz, A. Rabinovich (eds). Lecture Notes in Computer Science \textbf{4800}: 367--385. Berlin, Heidelberg: Springer. \url{https://link.springer.com/chapter/10.1007/978-3-540-78127-1_20}.

\bibitem{gorard8}
\textsc{J. Gorard} (2020), ``Algorithmic Causal Sets and the Wolfram Model'', \textit{arXiv preprint}: 
url{https://arxiv.org/abs/2011.12174}.

\bibitem{gorard9}
\textsc{J. Gorard} (2021), ``Hypergraph Discretization of the Cauchy Problem in General Relativity via Wolfram Model Evolution'', \textit{arXiv preprint}: \url{https://arxiv.org/abs/2102.09363}.

\bibitem{benabou}
\textsc{J. B\'enabou} (1967), ``Introduction to bicategories'', \textit{Reports of the Midwest Category Seminar}: 1--77. \url{https://link.springer.com/chapter/10.1007/BFb0074299}.

\bibitem{coecke}
\textsc{B. Coecke and R. Lal} (2013), ``Causal Categories: Relativistically Interacting Processes'', \textit{Foundations of Physics} \textbf{43} (1): 458--501. \url{https://arxiv.org/abs/1107.6019}.

\bibitem{johnson}
\textsc{N. Johnson and D. Yau} (2021), ``Bimonoidal Categories, ${E_n}$-Monoidal Categories, and Algebraic $K$-Theory'', \textit{arXiv preprint}: \url{https://arxiv.org/abs/2107.10526}.

\bibitem{arsiwalla}
\textsc{X. D. Arsiwalla and J. Gorard} (2021), ``Pregeometric Spaces from Wolfram Model Rewriting Systems as Homotopy Types'', \textit{arXiv preprint}: \url{https://arxiv.org/abs/2111.03460}.

\bibitem{arsiwalla2}
\textsc{X. D. Arsiwalla, J. Gorard and H. Elshatlawy} (2021), ``Homotopies in Multiway (Non-Deterministic) Rewriting Systems as $n$-Fold Categories'', \textit{arXiv preprint}: \url{https://arxiv.org/abs/2105.10822}.

\bibitem{shulman}
\textsc{M. Shulman} (2017), ``Homotopy type theory: the logic of space'', \textit{New Spaces in Mathematics and Physics}, G. Catren, M. Anel (eds). \url{https://arxiv.org/abs/1703.03007}.

\bibitem{shah}
\textsc{R. Shah and J. Gorard} (2019), ``Quantum Cellular Automata, Black Hole Thermodynamics and the Laws of Quantum Complexity'', \textit{Complex Systems} \textbf{28} (4): 393--410. \url{https://www.complex-systems.com/abstracts/v28_i04_a01/}.

\bibitem{susskind}
\textsc{L. Susskind} (2016), ``Computational Complexity and Black Hole Horizons'', \textit{Fortschritte der Physik} \textbf{64} (24): 44--48. \url{https://arxiv.org/abs/1402.5674}.

\bibitem{stanford}
\textsc{D. Stanford and L. Susskind} (2014), ``Complexity and Shock Wave Geometries'', \textit{Physical Review D} \textbf{90} (12): 126007. \url{https://arxiv.org/abs/1406.2678}.

\bibitem{brown}
\textsc{A. R. Brown, L. Susskind and Y. Zhao} (2017), ``Quantum Complexity and Negative Curvature'', \textit{Physical Review D} \textbf{95} (4): 045010. \url{https://arxiv.org/abs/1608.02612}.
\end{thebibliography}
\end{document}